\documentstyle[12pt]{article}

\begin{document}
\def\thebibliography#1{\section*{REFERENCES\markboth
 {REFERENCES}{REFERENCES}}\list
 {[\arabic{enumi}]}{\settowidth\labelwidth{[#1]}\leftmargin\labelwidth
 \advance\leftmargin\labelsep
 \usecounter{enumi}}
 \def\newblock{\hskip .11em plus .33em minus -.07em}
 \sloppy
 \sfcode`\.=1000\relax}
\let\endthebibliography=\endlist

\hoffset = -1truecm
\voffset = -2truecm


\title{\large\bf
QCD-Motivated BSE-SDE Framework For Quark-Dynamics \\
Under Markov-Yukawa Transversality
\begin{center}                        
\large A Unified View of $q{\bar q}$ And $qqq$ Systems - Part I
\end{center}
}
\author{
{\normalsize\bf
A.N.Mitra \thanks{e.mail: (1) ganmitra@nde.vsnl.net.in;
(2) anmitra@csec.ernet.in}
}\\
\normalsize 244 Tagore Park, Delhi-110009, India.}

\date{24 January 1999}

\maketitle

\begin{abstract}
 This article aims at an integrated formulation of BSE's for 2- and 3-quark 
hadrons under the Markov-Yukawa Transversality Principle ({\bf MYTP}) which
provides a deep interconnection between the 3D and 4D BSE forms, and hence 
offers a unified treatment of 3D spectroscopy with 4D quark-loop integrals 
for hadronic transitions. For the actual dynamics, an NJL-type realization
of $DB{\chi}S$ is achieved via the interplay of Bethe-Salpeter (BSE) and 
Schwinger-Dyson (SDE) equations, which are simultaneously derivable from a 
chiral Lagrangian with a gluonic (Vector-exchange) 4-fermion interaction of 
`current' $uds$ quarks, specifically addressing the non-perturbative regime. 
A prior critique of the literature on various aspects of the non-perturbative 
QCD problem, on the basis of some standard criteria, helps converge on a 
BSE-SDE framework with a 3D-4D interconnection based on {\bf MYTP}. This 
framework is then employed for a systematic self-contained presentation of 
2- and 3-quark dynamics on the lines of {\bf MYTP}-governed $DB{\chi}S$, 
with enough calculational details illustrating the techniques involved.
Specific topics include: 3D-4D interconnection of $q{\bar q}$ and $qqq$
wave functions by Green's Function methods; pion form factor; 3-hadron form 
factors with unequal mass loops; $SU(2)$ mass splittings; Vacuum condensates
(direct and induced); Complex H.O. techniques and $SO(2,1)$ algebras for 
Baryon Spectroscopy; plus others.  

\end{abstract}

\newpage

\tableofcontents                                                       

Appendix A: Derivation Of $F(k^2)$ And $N_H$ For P-meson                \\
Appendix B: Gauge Corrections To Kaon E.M. Mass                         \\
Appendix C: A 4D NJL-Faddeev Model                                      \\
            C.1:  $qq$ Bound State in NJL-Model                         \\
            C.2:  NLJ-$qqq$ Bound State Problem                         \\
            C.3:  Solution of the Bound State $qqq$ Eq.(C.8)            \\
            C.4:  Comparison of NJL-Faddeev with 3D-4D BSE              \\
Appendix D: $SU(6) \otimes O(3)$ Wave Fns In Complex Basis              \\
            D.1:  Construction of Orbital Functions in Complex Basis    \\
            D.2:  Normalization of Natural and Unnatural Parity Baryons \\
Bibliography

\vspace{1cm}
\section{Introduction: QCD-Type Confinement Models}
\vspace{1cm} 

        One of the biggest challenges in  physics to-day  is a viable
theory of strong interactions for which QCD is the leading candidate. 
Unfortunately, despite many of its extremely attractive features, this 
theory is not yet available in a sufficiently tractable form  so as to  
appeal instantly to all its practitioners in as universal a manner as, e.g.,
in QED. The bone of contention in this regard is the non-perturbative 
sector of QCD which shows up as the phenomenon of `confinement'  at low 
and moderate energies: As yet there is no visible evidence of a sort 
of minimum consensus on a common dynamical framework to incorporate 
this physical effect in QCD applications in the strong interaction sector 
in a sufficiently convincing yet doable manner. As a result, there exist a 
multiplicity of approaches which, while incorporating the QCD ideas in 
varying degrees of sophistication, nevertheless often need to resort to 
additional parametric assumptions to calculate various low energy hadronic 
properties. Some of the principal approaches are:
\par
        Bag models [1]; QCD-sum rules [2]; later adaptations [3] of the 
Nambu-Jona-lasino model [4]; QCD bosonization approaches [5]; Instanton 
methods [6]; Vacuum self-dual gluon fields [7]; Quark confinement models 
[8]; Schwinger-Dyson and Bethe-Salpeter models [9]; adaptation of BSE-SDE 
to a 3D-4D hybrid form [10-11] in a `two-tier' fashion (to incorporate 
the spectroscopy sector). To this list one should also add QCD-motivated 
`quarkonia' models which is one of the oldest types in existence, and whose 
state of the art may be found in a fairly recent collection [12]. Last not 
least, one must pay homage to ``Lattice QCD" which addresses confinement at 
a more fundamental level, and has grown into a self-contained field of study 
by itself. However its philosophy and methodology have so little in common 
with the less ambitious approaches listed above [1-12], that it does not 
fall within the scope of the present study.  
\vspace{1cm}
\subsection{A Short Critique of Models [1-12]} 
\vspace{1cm}
        If the pretence of implementing confinement through an exact 
solution of the QCD equations of motion is given up in favour of an
`effective confinement' programme, the central issue boils down to the 
extent to which the same  can be formulated in a manner which is both 
physically convincing as well as mathematically tractable enough to warrant
wide-ranging applications, all the way from low-energy spectroscopy [12] 
to deep inelastic processes amenable to perturbative QCD [2]. Such a 
philosophy is reminiscent of Bethe's ``Second Principle Theory" for 
effective nucleon-nucleon interactions, now reborn at the level 
of quark-quark interactions, with confinement addressed in a semi-
empirical manner which incorporates the main features of QCD structure.
Unlike Lattice-QCD, such  programmes are not meant to address confinement 
directly, but rather to take its role more or less  for granted in 
anticipation of future developments. It is from this angle that most of 
the approaches listed above may be viewed, from putting in the QCD 
feature by hand [1,8-12], to a conscious effort to `derive' its content  
more explicitly [5-7]. Of the last group, the model that comes closest to 
tackling confinement is perhaps [7], but its methodology has understandably 
more limitations for wider applicational purposes. A complementary role is 
that of [5] which is characterized by a `chiral perturbation' approach to the 
mechanism of formation of hadronic states in  QCD, and in the process gives 
rise to an effective chiral Lagrangian for low energy hadron physics. However 
the {\it perturbative} expansion in the momenta, deemed small in the low 
momentum limit, robs such a Lagrangian of a vital property: its capacity to 
predict the {\it bound} (confined) states of hadrons in the low momentum 
regime, due to the lack of a `closed form' approach. (A closed form approach 
is best exhibited by some sort of form factors characterized by a confinement 
scale, a feature that gets lost in any expansion in the momenta).
\par  
         A `two-tier' 3D-4D BSE approach like [10-11] is meant for a `more 
conscious' incorporation of the spectroscopy sector, i.e., an explicit 
recognition of the fact, often ignored in the more usual formulations of 
BSE-cum-SDE methods [9], that the observed hadronic spectra are O(3)-like 
[13], while a literal BSE formulation in euclidean form, with a standard 
4D support to the kernel leads to O(4)-like spectra [14]. In this respect, 
the two-tier strategy [11] invokes the Markov-Yukawa Transversality Principle 
({\bf MYTP}) [15] wherein the quark-quark interaction is in a hyperplane 
which is {\it transverse} to the 4-momentum $P_\mu$ of the composite hadron, 
so that the modified BSE has a (covariant) 3D support to its kernel. 
This feature in turn leads to an exact 3D reduction of the BSE from its 4D 
form, and an equally exact {\it reconstruction} of the 4D wave function in 
terms of 3D ingredients [16], thus implying an {\it {exact interconnection}} 
between the 3D and 4D BSE forms [16]. (A parallel formulation of {\bf MYTP} 
[15] by the Pervushin group [17] gave rise to the 3D reduction from the 4D 
BSE form, but the {\it inverse} connection (from 3D to 4D) was missing in 
their paper [17]). The 3D BSE form makes contact with the observed O(3)-
like spectra [13], while the reconstructed 4D BSE form provides a natural
language for evaluating transition amplitudes via quark loops [16,11]. 
\par    
        A  3D BSE form has its own logical basis which receives support from 
several independent angles in view of its crucial role in the understanding 
of physical processes in general and the theme of the present article in 
particular (see below). We list three supporting themes which have been
developed over the decades from entirely different premises, all converging
to a basically 3D picture for the effective $qq$/$q{\bar q}$ interaction:
\par 
i) It gives a physical meaning to the interaction of the quark 
constituents in their respective mass shells, consistently with the 
tenets of local field theory [18];
ii) It arises from the concept of instantaneous interaction [17,16] among 
the quark constituents in accordance with the Markov-Yukawa picture [15]
of transversality to the composite 4-momentum $P_\mu$;
iii) It is the only structure of the BSE kernel which makes this equation
compatible with a pair of Dirac equations for two particles under
their mutual interaction [19].     
\vspace{1cm}
\subsection{\bf {Bethe's `Second Principle' Criteria for Model Selection}}
\vspace{1cm}
Next we consider certain guiding principles (criteria) which form the basis 
of this study, before identifying the specific model/models for a more 
detailed and self-contained exposure. A clue is provided by the  observation 
that most of these approaches [1-12], irrespective of their individual 
theoretical premises, have a common characteristic: Applicability to hadronic 
processes viewed as quark composites, limited only by their individual
predictive powers, while a deeper  understanding of the underlying models 
themselves is best left to future investigations. Perhaps the painfully 
slow progress of Lattice QCD results gives an inkling of this scenario: The 
formidable dimensions of the quark-gluon strong interaction physics leaves 
little alternative to the respective practitioners of QCD but to settle for 
a less ambitious approach to the problem. Thus the different models [1-12] 
are best regarded as alternative strategies, each with its own methodology 
and parametric limitations, aimed at selected sectors of hadron physics that 
are suited to their structural budgets before the `final' theory unfolds 
itself, leaving their successes or otherwise to be judged in the interim by 
the depth and range of their respective predictions vis-a-vis the data. This 
is once again Bethe's `second principle' philosophy  in retrospect, which 
would presumably continue to operate perhaps as long as QCD remains a 
partially solved theory. Within this restricted philosophy,  some obvious 
criteria for theme selection from among the available candidates [1-12] 
could be the following partial shopping list (not mutually exclusive):
\par
A) Maximum number of mutually compatible as well as time-tested ideas 
that can be extracted from out of [1-12] in the sense of an ``HCF";
B) Close proximity to QCD as the ideal theory which stands on the
three pillars of Lorentz-, Gauge- and Chiral- invariance;
C) Formal capacity to address several sectors of physics simultaneously, 
all the way from hadron spectroscopy to quark loop integrals for different 
types of transition amplitudes within a common dynamical framework;  
D) Sufficient flexibility of the core dynamical framework to permit smooth
on-line incorporation of mutually compatible ideas, like Markov-Yukawa
transversality {\bf MYTP} [15], and Dynamical Breaking of Chiral Symmetry 
($DB{\chi}S$) [4], and other similar principles if need be, without causing 
any structural (or parametric) damage to the basic framework itself; 
E) Natural capacity of the {\it {conceptual premises}}  to include both 2- 
and 3-quark hadrons within a common dynamical framework, to give concrete 
shape to the (widely accepted) principle of meson-baryon duality;   
F) A built-in microcausality in the dynamical framework which takes in its
stride sensitive items like the structure of the vacuum in strong 
interaction physics, without the need for fresh ansatze/parametrizations. 
\par
        Although these criteria are neither exhaustive nor mutually exclusive,
their collective effect is nevertheless focussed enough to eliminate many 
prospective candidates in favour of a chosen few that would survive the
tests (A)-(F) for a reasonably self-contained account of strong interaction
physics within the tenets of the `Second Principle' philosophy. Thus the too 
simplistic premises of Quarkonia models [11] which played a crucial role in 
the early stages of QCD-motivated investigations, would not stand the tests 
of (B),(C) and (D). Bag models [1] which had also played a similar role 
in the early phases, would not qualify under (B), (C) and (F). QCD sum
rules [2] represent perhaps one of the most successful applications of
perturbative QCD by relating the high energy quark-gluon sector to the
low energy hadron sector through the principle of an FESR-like duality
discovered in the Sixties [20]. Even to-day it is extensively used for
many hadronic investigations. Yet it fails on count (C) mainly because 
of its failure to satisfy (F): The lack of microcausality in this model
can be traced to the `matching condition' between the quark-level and
hadron-level amplitudes whose solution is far from unique. Indeed,
while the borelisation technique suffices for the prediction of ground 
state hadron masses, that very mechanism also causes it to {\it lose} 
information on the spectra of the {\it excited} states, thus reducing
its predictability on this vital (low energy) front. 
\par
        As for models [5-8], the Quark Confinement model [8]  lacks
enough microcausality - condition (F) - which shows up through a 
relatively poor satisfaction of (C), since the spectroscopy sector is
badly neglected. QCD bosonization methods [5] have made very impressive 
strides in respect of {\it {transition amplitudes}} through the powerful 
technique of construction of effective Lagrangians in terms of the 
hadronic fields, which make them ideally suited for `tree-diagrams'. 
However such structures hide from view the {\it composite} character of 
the hadrons in the `soft' QCD regime, which is best exhibited in `closed' 
form via quark-hadron form factors - the vehicle for sensitivity to various
non-perturbative features of the theory. For the baryon dynamics too, the 
inadequacy of the formalism [5] to depict correctly the 3-quark form factors 
shows up through its excessive dependence on the quark-diquark description 
with a {\it rigid} diquark structure which results in an inevitable loss of 
information on the true 3-quark content of the baryon. Finally, while [6,7] 
satisfy the conditions (B),(C),(F) on separate counts, there is not enough 
published evidence of support from the other 3 quarters (A),(D),(E). Hence 
while their comparison with others is useful for a comparative discussion of 
different models, their claims to a primary `theme' status fall rather short 
of a good starting point. This leaves the BSE-SDE framework [9-11] for a 
more detailed scrutiny to follow.   
\vspace{1cm}
\subsection{BSE-SDE Framework:$DB{\chi}S$ and `Soft'-QCD}
\vspace{1cm}
The last group [9-11] is characterized by an interplay of BSE and SDE, 
both  derivable from a suitably chosem 4-fermion Lagrangian as input.
It has a very wide canvas and is fully attuned to Bethe's Second Principle 
Theory ($BSPT$ for short). Its general framework equips it with arms to meet 
most of the conditions (A)-(F), ranging from flexibility to wide-ranging  
predictivity, thus lending it some credibility within the broad premises 
of `BSPT'. In particular, its natural roots in field theory endow it with 
standard features like dynamical breaking of chiral symmetry ($DB{\chi}S$) 
via the non-trivial solution of the SDE, thus giving it the powers to
subsume the contents of the NJL-model [4]. Indeed, soon after the discovery 
of the NJL model [4], a field-theoretic understanding of its underlying 
idea was achieved in the form of a generalized $DB{\chi}S$ in the QED 
domain via the non-trivial solution of the SDE [21]. And after the advent of 
QCD [22], the same feature showed up through the  solution of the BSE for 
$q{\bar q}$ interaction via one-gluon-exchange [3]. This is a typical non-
perturbative effect, although it does not cover all its aspects. 
\par
        Subsequently the concept of $DB{\chi}S$ was generalized to show that 
this feature is shared by any extended {\it vector-type} 4-fermion coupling 
[23-24] which preserves the chiral symmetry of the Lagrangian  but the same 
gets {\it {broken dynamically}} through the non-trivial solution of the SDE, 
derivable from such a Lagrangian. Indeed the sheer generality of a Lagrangian-
based BSE-cum-SDE framework, by virtue of its firm roots in field theory, 
gives it a strong mandate, in terms of both predictivity and flexibility, 
to accommodate additional principles like Markov-Yukawa Transversality [15] 
({\bf MYTP}) while staying within a basically Lagrangian framework, so that   
criteria (A)-(F) are still satisfied. 
\par
        In particular, a basic proximity to QCD is ensured through a vector-
type interaction (condition (B)) [10,23], which while maintaining the 
correct o.g.e. structure in the perturbative region, may be fine-tuned to 
give any desired structure in the infrared domain as well. The latter part 
is admittedly empirical, but captures a good deal of physics in the 
non-perturbative domain while retaining a broad QCD orientation, and 
hence does not rule out a deeper understanding of the infrared part of the 
gluon propagator within the same framework. More importantly, the 
non-trivial solution of the SDE corresponding to this generalized gluon 
propagator [11] gives rise to a dynamical mass function $m(p)$ [11] as a 
result of $DB{\chi}S$, even while the input Lagrangian has chiral invariance 
due to the vector-type 4-fermion interaction [23,24] between almost massless 
$u-d$ quarks. These considerations further strengthen the case of a 
Lagrangian-based BSE-SDE framework for a theme choice.

\subsection{3D-4D BSE: From Spectra To Loop Integrals}
\par
        Now the canvas of a (second-principle) BSE-SDE framework is broad 
enough to accommodate a whole class of approaches, and facilitates further 
fine-tuning in response to the needs of major experimental findings such as 
the observed O(3)-like spectra [13], which essentially amounts to treating 
the time-like momenta separately from the space-like ones, as has long been 
known since the classic work of Feynman et al [25]. In this regard, the 
{\bf MYTP} constraint [15] seems to fit this bill, by imparting a 3D 
support to the pairwise BSE kernel [15-17], an ansatz which can be motivated 
from several different angles [16,18,19]. As to the `soft' non-perturbative
part of the gluonic propagator, it still remains empirical since orthodox QCD 
theory does not yet provide a closed form representation out of the infinite 
chain of equations that connect the successively higher order Green's 
functions in the standard fashion [26], thus necessitating parametric 
representations [27]. Parametrization is also compatible with {\bf MYTP} 
[15] (see [11]), wherein the key constants are attuned to the hadron spectra 
of both 2- [28] and 3- body [29] types, within a common framework. The 3D 
support ansatz of {\bf MYTP}[15] in turn gives a characteristic `two-tier' 
[16] structure to the entire BSE formalism, wherein the first stage (3D BSE) 
addresses the meson [28] and baryon spectra [29], while the reconstructed 
4D wave (vertex) functions [16] fit in naturally with the Feynman language 
of 4D quark loop diagrams for various types of transition amplitudes 
[11,30-33] in a unified fashion.        
\par 
        A BSE-SDE formulation [9] represents a 4D field-theoretic 
generalization of `potential models' [12], and is thus equipped to deal with 
a wider network of processes (e.g., high energy processes) not accessible 
to potential models [12]. In this way, the BSE-SDE approach occupies an 
intermediate position, sharing the off-shell feature with potential models 
[12], as well as the high energy flavour of QCD-SR [2], but its dynamical 
spirit is much nearer to [12] than to [2]). Indeed the role of the `potential'
[12] is played by the generalized 4-fermion kernel [23] (which is a paraphrase 
for the non-perturbative gluon propagator [11]). The 4D feature of BSE-SDE 
gives this framework a ready access to high energy amplitudes, as in 
QCD-SR [2] as well as in other models [5-8], while its `off-shell' feature 
gives it a natural access to hadronic spectra [13], in company with 
(potential-oriented) quarkonia models [12] . (In contrast, certain models 
[2,5-8] do not have a basic infrastructure to address spectroscopy). Now 
with this twin feature of off-shellness and Lorentz-covariance, the BSE-SDE 
framework formally overcomes the shortcomings of `potential' models [23] in 
obtaining numerically `correct' values for the various condensates which are 
employed as inputs in QCD-SR calculations [2]. This was indeed confirmed by  
a later derivation of similar results [30] in terms of a Lorentz-covariant 
formulation [11] of the BSE-SDE framework, which showed that vacuum 
condensates are calculable within a spectroscopy-rooted [28,29] framework.

\subsection{Off-Shellness in BSE: Parametric Links with QCD-SR}

The calculations [23,30] raise the interesting question of the
possibility of a basic connection among the input parameters of different
models, although conceived within very different premises. Thus in QCD-SR 
[2], the `free' parameters of the theory are the condensates themselves as 
input, while in the BSE-cum-SDE methods [9-11], the corresponding parameters 
are contained in the input structure of the infrared part of the gluon 
propagator [11]. Now since the condensate parameters of QCD-SR [2] are  
explicitly calculable in BSE-SDE models by quark loop techniques [23, 30],
using the gluonic parameters [11,30], this result at least settles the 
issue of a one-way connection: from BSE [11] to QCD-SR [2]. 
\par
        To pursue this question a bit further, let us compare the features
of potential models (of which BSE is a 4D generalization) with those of   
QCD-SR: Potential models are characterized by `off-shell' features, whose 
parameters (corresponding to given `potential' forms) are primarily attuned 
to low energy spectroscopy, so that their predictions tend to work upwards 
on the energy scale, starting from the low energy end. QCD sum rules on the 
other hand are attuned to the perturbative QCD regime, so that their  
predictions tend to work downwards on the energy scale, starting from the 
high energy end. The `softness' aspects of QCD-SR are typically simulated 
via the Wilson OPE expansion in inverse powers of 4-momentum $Q^2$ where 
the `twist' terms of successively higher dimensions are symbolized by the 
corresponding `vacuum condensates' which are thus the free parameters 
of the theory. Therefore prima facie it appears that the two methods are
largely complementary to each other. The former, by virtue of its low
energy/off-shell emphasis, is particularly successful on the spectroscopic 
front, but its techniques do not find easy access to transition amplitudes 
due to inadequate treatment of the high energy front (lack of covariance).
The latter (QCD-SR) on the other hand, is ideally suited to the high energy
regime, but does not find ready access to areas involving soft QCD physics,
especially the spectroscopic regime. This is at least partly attributable
to the methodology of QCD-SR [2] which makes use of the `quark-hadron 
duality' for `matching' the respective amplitudes [2,20]: Because of the 
relatively `macroscopic' nature of the `matching' which is effected with a
`Borelization' technique [2], the  predictions are reliable only for 
the hadronic {\it ground} states, but do not readily extend to the 
spectra of excited states.
\par
        Due to the complementary nature of the two descriptions, it is not
generally easy to relate the parameters of one to those of the other. 
However, the off-shellness feature of potential models gives it access to
information on the interaction of the quark pair with the environment; in 
particular they possess signatures on the structure of the 
{\it {degenerate vacuum}} in the form of `vacuum condensates'.  That the 
vacuum condensates of QCD-SR [2] can be be expressed in terms of potential 
[23] and BSE [30] models is a reflection of their crucial `off-shellness'
property. As to the converse question, there is no published evidence of 
a corresponding exercise in the {\it opposite} direction viz., a derivation 
of the parameters of the BSE kernel/gluon propagator in terms of the various 
vacuum condensates that characterize QCD-SR [2]. A possible reason may lie 
in the role of {\it microcausality} (condition (F)) which is well satisfied 
by potential models, but perhaps not by QCD-SR [2]. Thus it would appear 
that `microcausality' which underlies the `off-shellness' feature of the 
`potential models' enhances their predictive powers vis-a-vis those which 
do not possess this  crucial property. 
\par    
        Now the off-shell structures of all `potential-oriented' models 
[9-12] have a fairly direct connection with the `spectral' predictions, 
unlike other types of confinement models [2,5-8], which do not permit 
such predictions in an equally natural way. And for 3-quark states [29], 
the dichotomy seems to be even sharper, inasmuch as there is a strong 
tendency in the literature to simplify the 3-quark systems as quark-diquark 
systems [2,5-8], thus partly ``freezing'' some genuine 3-body d.o.f.'s 
and causing a loss of information on the spectra of L-excited states.     
\par
        The off-shell characteristics of the BSE-SDE framework [9-12] are 
perhaps the most important single feature responsible for extending their 
predictive powers  all the way from 3D spectra to 4D transition amplitudes 
(of diverse types) via 4D quark loop integrals, under one broad canvas. The 
key to this capacity lies in the vehicle of the BS wave (vertex) function 
which has at its command the entire `off-shell' information noted above. 
Here it is important to stress that this wave function is a {\it genuine} 
solution of the BS dynamics [11], so that it leaves {\it no} scope for any 
{\it free} parametrization beyond what is already contained in the (input) 
gluon propagator. (Potential models [12] also have this capacity in principle, 
but their 3D structure does not allow full play to the `loop' aspect).

\subsection{Markov-Yukawa Transversality on the Null Plane}
\par
        Covariant Instaneity Ansatz (CIA) on the BSE [16] is not the
only form of invoking {\bf MYTP} to achieve an exact interconnection
between the 3D and 4D structures of BSE. As will be found later (see Sec.4 
for details), the CIA which makes use of the {\it local} c.m. frame of 
the $q{\bar q}$ composite, has a disadvantage: The 4D loop integrals are 
ill-defined due to the presence of time-like momentum components in the 
exponential/gaussian factors (associated with the vertex functions) caused 
by a `Lorentz-mismatch' among the rest-frames of the participating hadrons.
This is especially so for triangle loops and above, such as the pion form 
factor, while 2-quark loops [32] just escape this pathology. This problem 
is probably absent if the null-plane ansatz (NPA) is invoked, as found in 
an earlier study of 4D triangle loop integrals [33], except for possible 
problems of covariance [34]. The CIA approach [16] which makes use of the 
TP [15], was an attempt to rectify the Lorentz covariance defect, but the 
presence of time-like components in the gaussian factors inside triangle 
loop integrals [31] impeded further progress on CIA lines. 
\par
        Is it possible to enjoy the best of both the worlds, i.e., ensure
a formal covariance without having to encounter the time-like components in 
the gaussian wave functions inside the 4D loop integrals ? Indeed the 
problem boils down to a covariant formulation of the null-plane approach.
Now the null-plane approach (NPA) itself has a long history [35], and it is
not in the scope of this article to dwell on this vast subject as such. 
Instead our concern is limited to the {\it covariance} aspects of NPA, a 
subject which is of relatively recent origin [18,36-38]. However in all
these approaches [38], the primary concern has been with the NP-dynamics
in 3D form only, as in the other familiar 3D BSE approaches [39] over the 
decades. On the other hand, the aspect of NPA which is of primary concern 
for this article, is on the possibility of invoking {\bf MYTP} for achieving
a  3D-4D BSE interconnection on the {\it covariant} Null Plane, on similar 
lines to Covariant Instantaneity (CIA) for the pairwise interaction [16]. 
Now it seems that a certain practical form of the null-plane formalism [33] 
had all along enjoyed {\it both} 3D-4D interconnection {\it and} a sort of
`pedagogical covariance'(albeit implicitly) [40]. This basic feature can be 
given a formal shape by merely extending the Transversality Principle [15]
from the covariant rest frame of the (hadron) composite [16], to a 
{\it {covariantly defined}} null-plane (NP) [41]. Because of its obvious
relevance, the subject of 3D-4D interlinkage on the covariant Null-Plane [41] 
will be covered in Sec.(4.2), with a parallel CIA treatment in Sec.(4.1).
\vspace{1cm}
\subsection{Scope of the Article: Outline of Contents}
\vspace{1cm}
        We now focus on a BSE-cum-SDE form of dynamics derivable from 
a chirally invariant Lagrangian with an effective gluon-exchange-like 
interaction (pairwise), as the central theme of this study for a reasonably 
self-contained presentation, under the further constraint of Markov-Yukawa
Transversality Principle (MYTP). The emphasis is on a {\it pedagogical} 
perspective on the problem of effective color confinement, converging on a 
vector exchange mediated Lagrangian whose chiral symmetry gets broken 
{\it dynamically}, after giving a bird's eye view of the main approaches to 
effective confinement [1-11]. Indeed the ($DB{\chi}S$) theme, although 
originating from the NJL-model [4] for contact pairwise interaction, admits 
a simple generalization to a (space-time extended) vector exchange 
$q{\bar q}$ / $qq$ interaction which exhibit chiral symmetry at the input 
Lagrangian level, but get broken dynamically via the solution of the 
Schwinger-Dyson Equation (SDE) [23-27].  A more explicit QCD motivation 
must be achieved by hand, e.g., identification of the pairwise interaction 
with the entire gluon propagator (perturbative and non-perturbative [28a]),
which in turn has several desirable consequences, such as the color effect 
which ensures that the strength of the $qq$ force is {\it half} that of 
$q{\bar q}$, within a common parametrization.   
\par
        The second item of emphasis concerns the remarkable facility of
an {\it exact} interconnection between the 3D and 4D BSE forms [16], that
is provided by {\bf MYTP}, a facility that other 3D approaches to BSE [39],
or (basically 3D) Null-Plane approaches [35-38], do {\it not} seem to possess.
This property allows the exposiition of the BSE-cum-SDE techniques in a very
simple way, so as to provide the reader with a quick working knowledge
of their applications to a wide class of problems which may be broadly 
classified in a {\it two-tier} form: A) Mass spectra; B) Quark-Loop diagrams.  
Such a division is natural since investigations of types (A) and (B)
are mainly governed by the 3D and 4D aspects of the BSE respectively.
Therefore after an introductory phase on the general BSE-SDE formulation,
an early specialization to its {\bf MYTP}-governed 3D-4D form (from Sec.4 
onwards) will form the basis for this (application oriented) article.   
\par
        A third item of emphasis is on the {\it second} stage of the 3D-4D 
BSE framework, viz., techniques of 4D quark-loop amplitudes, with a 
comparative study of CIA [16] vs CNPA [41], to bring out their relative 
(strong/weak) features.
\par
        This article has been built on the infrastructure of one with a 
similar theme [40] written about a decade ago; it incorporates major 
advances through the present decade on the 3D-4D BSE front [42a-b], viz., 
the Covariant Instaneity Ansatz (CIA) and its more recent Null-Plane 
counterpart (CNPA)[41], both under the umbrella of {\bf MYTP} [15]. The 
background of ref.[40] will be freely used, but the details on (3D) spectra 
on which CIA [28] and CNPA [41] have similar predictions, will now be 
omitted, except for drawing attention to their structural similarities. 
Instead more attention will be paid to the structure of 4D quark loop 
integrals of selected types to bring out the applicational potential of 
this {\bf MYTP}-governed formalism [41-2]. These types include i) certain 
hadronic form factors built out of triangle loops; ii) typical self-energy 
problems dealing with  $SU(2)$-mass splittings among hadrons; iii) vacuum 
condensates which are inputs in QCD-SR [2], but calculable in the 3D-4D 
BSE-SDE formalism [30].
\par            
        While giving the details of this article, we repeat at the outset
that, except for the contents of Sections 1-3, it is {\it not} intended as 
a conventional `review' of the BSE-SDE framework such as [9]. Nor are 
conventional 3D BSE approaches [39], or the conventional NPA formalisms 
[35-38] the subjects of our detailed description. Aspects of {\it contact} 
NJL-type 4- and 6-fermion couplings (often employed in the`nuclear' field), 
are also not of interest here. 
\par
        As to the actual details, the Table of Contents, preceding the
Introduction (Section 1) gives a fair cross section of the included items:
{\bf Sect.2} gives a panoramic view of the NJL-Model [4] and its aftermath. 
{\bf Sect.3} gives a general derivation of BSE and SDE in an interlinked 
fashion, with a gluon-like (Vector-exchange) propagator whose mass function 
$m(p)$ stems via $DB{\chi}S$ from a spatially extended 4-fermion interaction 
in the input Lagrangian. With this general background of SDE-BSE as well
as of $DB{\chi}S$, the rest (Sect.4-11) deals with different facets of the
3D-4D BSE-SDE framework under the Markov-Yukawa Transversality Principle 
[15] at {\it two} distinct levels of operation, viz., CIA [16, 17] which 
has been around for some time, and CNPA [41] which is formally a new 
proposal, although in effective (practical) use for quite some time [33,40]. 
\par
        Of the subsequent Sections, {\bf Sect.4} collects the background 
for interlinked 3D-4D BSE techniques for $q{\bar q}$ hadrons. For the 
fermionic BSE, we have preferred to stick to its `Gordon-reduced' version 
[10a-b] {\it adapted} to the {\it off-shell} constituents [10a]. This is a 
conscious departure [10b] from the standard BSE-form [26] to make the BSE 
more tractable for wider applications, as in other BSE approaches [10c-d], 
and does not violate the `Bethe Second Principle' spirit, since the input 
4-fermion coupling is an effective description of the pairwise interaction).
\par
        {\bf Sects.(5-8)} deal with some selected applications of triangle loops
(form factors), two-loops (self-energy), and one-loop (vacuum condensates) 
techniques respectively. These include, among other things, a technique to
include QED gauge insertions in arbitrary momentum-dependent vertex functions 
for the e.m. self energy and form factors. Wherever possible, a parallel 
treatment is provided for CIA and CNPA for a comparative view of the two 
distinct {\bf MYTP}-governed BSE formalisms, but some technical problems
with CIA [16] often lead to a preference for CNPA. Some calculational
details on the form factor plus normalization are given in Appendix A. 
\par
        {\bf Sect.6} gives a general method for triangular quark-loop integrals 
applicable to a large class of transition amplitudes for 3-hadron coupling 
[31], to bring out a major simplifying feature of the resulting structure 
arising out of a `cancellation' mechanism between the 4D quark propagators 
and the 3D $D$-functions in the hadron-quark vertices of the two-tier BS 
formalism [16]. This prevents free propagation of quarks by eliminating the 
Landau-Cutkowsky (overlapping) singularities [16,31]. 
\par
        {\bf {Sects.7,8}} give results for self-energy diagrams [32] and of vacuum 
condensates [11,30], requiring two and one $S_F$-functions respectively. The 
self-energy calculations in Sect.7 are illustrated with SU(2) mass splittings 
of pseudoscalar mesons [32b]. A general method to deal with QED gauge 
corrections to the e.m. mass differences is outlined in Appendix B. For the 
vacuum condensates [11,30], Sect.8 offers a new gauge invariant technique for
loop integrations, on the lines of Schwinger [43]. We reiterate that such 
predictions are intimately linked with spectroscopy via the infrared 
structure of the gluon propagator [11,30]. 
\par
        The third part {\bf Sects.9-11} deals with the BSE formalism for a 
3-quark baryon, with emphasis on the $qqq$ structure taking into view that 
in most approaches, including other BSE models [9b], the dynamical 
treatment has often relied heavily on the quark-diquark approximation 
[5b,8b,9b,11], which amounts to a ``freezing'' of the 3-body degrees of 
freedom. It has also been recognized in the literature that with a 3-body 
BSE treatment to the baryon, there are some technical problems associated 
with the status of the spectator [44]. In the Two-tier BSE model this problem 
has been regularly addressed at various stages of its development [10,40,29]. 
The 3-quark dynamics is described in 3 Sections (9-11) as follows.  
\par
        {\bf Sect.9}: A panoramic view of the baryon dynamics as a general 3-body 
problem with full permutation symmetries [45] in all the relevant d.o.f.'s 
incorporated; a detailed correspondence with the quark-diquark model; 
Complex HO techniques for the $qqq$ problem [46]; problems of 3D reduction 
and 4D reconstruction for $qqq$ BSE [47]; and fermionic BSE with gluonic 
interactions in pairs [29]. 
\par
        {\bf Sect.10}: Green's function techniques for 3D reduction of the BSE, and 
reconstruction of the 4D $qqq$ wave function [47]; see Table of Contents. 
\par  
        {\bf Sect.11}: A summary of the relativistic fermionic $qqq$ BSE with 
the same gluonic propagator as employed for the $q{\bar q}$ problem;  the 
3D reduction [29] of the $qqq$ BSE is on closely parallel lines to the 
two-body case [28]. The derivation of an explicit mass formula is greatly
facilitated by  taking a complex HO basis [46]. However loop techniques for 
baryonic amplitudes are not included for explicit presentation.
         
\section{ NJL Model: Recent Developments (Nambu)}

The precursor of the NJL-model [4] was the `Nambu-Goldstone' picture of the 
pion as a zero mass particle arising from the chiral non-invariance of the 
vacuum [48]. This view of the pion received quantitative shape at the hands
of Gell-Mann and Levy [49a] who started with a $SU(2) \times SU(2)$  symmetry 
of the Lagrangian (termed SU(2) $\sigma$-model) involving an $I=1$
pseudoscalar ${\bf \pi}$ and an $I=0$ scalar $\sigma$ field. Due to 
spontaneous symmetry breaking of the vacuum, the $\sigma$-field shifts to 
a minimum $<\sigma>=-f_\pi \neq 0$, while the pion field remains unshifted
($<{\bf \pi}>=0$) and stays at zero mass.        
\par
        The Gell-Mann Levy $\sigma$-model set the stage for modern chiral 
theories, stimulated by an important paper due to Skyrme [49b], to describe
pseudoscalar mesons and baryons through a solitonic picture wherein baryons 
are generated as bound states of weakly interacting mesons. These models 
were developed in the QCD context wherein, in the large $N$ limit, QCD 
becomes equivalent to a non-linear meson theory. The underlying logic is
that although the QCD Lagrangian has chiral symmetry for massless quarks,
this symmetry is spontaneously broken, giving rise to massless pions, etc.
These methods give rise to effective Lagrangian descriptions at the tree
level [5], but will not concern us any further in this article.     
\par
        The NJL-model [4] on the other hand, which is the very raison 
d'etre of this article, is characterized by chiral symmetry breaking in a 
{\it dynamical} fashion, and allows a formally {\it composite} structure of 
the pion, in company with other hadrons. This is a distinct advance over the 
elementary field picture of the pion [48-49], and facilitates a more natural 
understanding of many of its observational properties (form factor, 
$L$-excited states, etc). A short summary of the NJL model follows. 

\subsection{Outline of NJL Model}
\setcounter{equation}{0}
\renewcommand{\theequation}{2.\arabic{equation}}   
The NJL Lagrangian [4] may be written in two different ways:
\begin{equation}
L_{NJL} = L_0+L_i = (L_0 +L_s) + (L_i-L_s) \equiv L_0' + L_i';
\end{equation}
where $L_0 = -{\bar \psi}\gamma.\partial \psi$ and $L_i$ (see below) are 
chirally invariant, but  $L_s = -m{\bar \psi} \psi$ which stands for the
{\it observed} fermion, is not, and represents the symmetry breaking effect.
The interaction term $L_i$ is given by 
\begin{equation}
L_i = g_0 [({\bar \psi}{\psi})^2 - ({\bar \psi}\gamma_5{\psi})^2]
= -g_0 [({\bar \psi}i\gamma_\mu{\psi})^2 - 
({\bar \psi}i\gamma_\mu\gamma_5{\psi})^2]/2
\end{equation} 
The rearrangement in (2.1) is meant to diagonalize $L_0'$, and treat $L_i'$
as a perturbation; this implies a redefinition of the vacuum by introducing 
a complete set of `quasi-particle' states which are eigen states of $L_0'$.
The $L_s$ is now determined from the requirement that $L_i'$ shall not yield
additional self-energy effects. This gives the standard Schwinger-Dyson 
Equation (SDE) for $m$ in terms of the loop self energy
\begin{equation}
m = \Sigma |_{(i\gamma.p+m =0)} = -8img_0 \int (2\pi)^{-4} d^4p 
{F(p,\Lambda) \over {m^2+p^2-i\epsilon}}
\end{equation}
where $F(p,\Lambda)$ is a cut-off factor. The trivial solution $m=0$ 
corresponds to the usual chiral invariant vacuum characteristic of 
perturbation theory. The non-trivial solution $m=m_{NJL}$ is found from
\begin{equation}
(2\pi)^4 = -8i g_0 \int d^4 p (m^2 + p^2 - i\epsilon)^{-1} F(p,\Lambda)
\end{equation}
in terms of $g_0$ and $\Lambda$. It is alos called the `gap' equation, and
is based on a shifted vacuum $\Omega_m$ which is chiral non-invariant. 
With a fixed Lorentz-invariant cut-off $\Lambda$ in Euclidean space (and
$F=1$), eq.(2.4) reduces to 
\begin{equation}
2\pi^2/g_0 = \Lambda^2 - m^2 ln(\Lambda^2/m^2 + 1); \quad 
0 < 2\pi^2 {g_0}^{-1} {\Lambda}^{-2} < 1
\end{equation}
The two vacuua $\Omega_0$ (chiral invariant) and $\Omega_m$ (non-invariant)
are fully orthogonal to each other, and correspond to two different worlds.
$\Omega_m$, with the lower energy, is the true ground state. The chirality
operator defined as $\chi$ = $\int {\bar \psi}\gamma_4\gamma_5 \psi d^3$  
 commutes with the original hamiltonian $H_0$ with vacuum $\Omega_0$, but
not with $H_m$ with vacuum $\Omega_m$. However $\chi$ has no matrix elements 
connecting the two worlds $\Omega_0$ and $\Omega_m$, a sort of superselection
rule. Now the following paradox arises: The $\chi$-conservation in the 
$\Omega_0$ basis implies the existence of a conserved current
\begin{equation}
j_{\mu 5}= i{\bar \psi}\gamma_\mu\gamma_5 \psi; \quad \partial_\mu j_{\mu 5}=0
\end{equation}
On the other hand, for a massive Dirac particle in the $\Omega_m$ basis
\begin{equation}
\partial_\mu ({\bar \psi}\gamma_\mu \gamma_5 \psi) 
= 2m{\bar \psi}\gamma_5 \psi \neq 0
\end{equation}    
To reconcile these two statements, the $\chi$-current operator between 
{\it physical} states suffers radiative corrections w.r.t. the simple term
$i\gamma_\mu \gamma_5$ so that, on grounds of Lorentz invariance
\begin{equation}
<p'|j_{\mu 5}|p> = F(k^2) {\bar u}(p')[i\gamma_\mu \gamma_5 + 
{{2m\gamma_5 k_\mu} \over k^2}] u(p); \quad (k=p-p')
\end{equation}
Thus the real fermion (quark) is {\it not} a point particle since its $\chi$-
current has an anomalous $\gamma_5$ term. This in turn implies a pole at
$k^2 = 0$ for the $\gamma_5$ term, corresponding to a {\it {zero mass}}
pseudoscalar, whose natural identification is the pion. 
\par
        The pion which arises here as the lowest $q{\bar q}$ bound state,
has clearly the nature of a {\it {collective excotation}}, thus also
implying the existence of higher excitations in the same package, something 
which the elementary field model [48-49] could not provide. Indeed the BS
amplitude $\Psi$ for the bound state composite is
\begin{equation}
\Psi(x,y) = <0|T(\psi(x) {\bar \psi}(y))|B>
\end{equation}
which is related to the vertex function $\Gamma$ in momentum space as
\begin{equation}
\Psi(p_1,p_2) = S_F(q+P/2) \Gamma(q,P) S_F(q-P/2)
\end{equation}
where the individual quark momenta $p_{1,2}=P/2 \pm q$ in terms of the total
$(P)$ and relative $(q)$ 4-momenta. For the pseudoscalar state in question,
the BSE for $\Gamma(p_1,p_2)$, viz.,
\begin{equation} 
(2\pi)^4 \Gamma(q,P) = 2i g_0 \gamma_5 \int d^4 q' Tr [\gamma_5 S_F(q'+P/2)
\Gamma(q',P)S_F(q'-P/2)]
\end{equation}   
which for $P_\mu =0$, has a self-consistent solution $\Gamma = C\gamma_5$, 
$C$ being a constant, {\it provided} $g_0$ satisfies eq.(2.4), which is just
the `gap' equation (SDE) for the mass $m$. 
\par
        This crucial result of the NJL model, which shows that in the chiral
limit $P_\mu = 0$, the BSE and the SDE are {\it identical}, is a direct
consequence of the $\gamma_5$-invariance of the input Lagrangian. It also
tells us that in the $P_\mu = 0$ limit, the $q{\bar q}$ vertex function and
the quark mass function $m$ have the same (constant) structure. The 
constancy of each is of course a consequence of the contact interaction, but
the basic equality of these two quantities is also valid for an extended
4-fermion chirally invariant Lagrangian such as a vector mediated one [24].
\par
            The true significance of NJL was realized in the QCD context 
[22], through the study of non-perturbative solutions of SDE as a 
$DB{\chi}S$ mechanism for more general 4-fermion couplings [3,8,23,24],
including reformulations of the bag model [8a], and renormalization group
equations [24d]. And it was eventually subsumed in the generalized BSE-SDE 
formalism [9-11], which is of course the subject of this review. 

\subsection{BCS Mechanism, Mass Relations, SYSY, Etc}

        During the last decade, Nambu [50] has abstracted the findings of a 
new symmetry from BCS-type theories of dynamical symmetry breaking (due to
short-range attraction), resulting in a new vacuum state. The residual
symmetry in question is a remarkably simple relation among the fermion mass
$m_f$ and the composite boson $({\bf \pi}, \sigma)$ masses as low energy 
modes in the new vacuum, viz.,  $(M_\pi: m_f: M_\sigma)$ = $(0:1:2)$. In
more complex the fermion mass $m_f$ and the (composite) boson masses 
$(M_1,M_2)$  obey the generalized relation $M_1^2 +M_2^2 = 4m_f^2$. The
low energy properties of the system can be represented by an effective 
Hamiltonian like in the $\sigma$-model [49a] where the coupling constants
are so related as to yield such mass relations automatically.   
\par
        Coming to the SUSY aspects, the essential thrust of Nambu's
discovery [50b] is a hidden SUSY in the BCS mechanism, manifesting via
two physical scenarios: i) a cascading chain of symmetry breaking (tumbling);
ii) a bootstrap mechanism in which the symmetry sustains itself among a set
of effective fields without the need to refer to a substructure. The main
ideas are the following.
\par
        A BCS mechanism has two energy scales: i) The high energy scale 
corresponds to the force responsible for the formation of Cooper pairs;
its analogue in particle physics is the pion decay constant $f_{\pi}$. ii)
The low energy scale is the pairing energy, and one of its manifestations
is the quasi-fermion mass $m_f$ which corresponds to the constituent quark
mass $m_q = m_{NJL}$. More explicitly, there are both fermionic and bosonic 
excitations in the low energy scale: the quasi-fermion $(m_f)$, the 
Goldstone boson (pion) and the Higgs (sigma) boson. In the simplest BCS 
(NJL) mechanism, their masses are in the ratio $m_f:m_{\pi}:m_{\sigma}$
= $1:0:2$. This low energy picture can also be articulated by an effective 
Ginzburg-Landau-Gell Mann-Levy Hamiltonian involving these fermion and 
boson fields with Yukawa couplings and a Higgs potential. Their 
characteristic parameters are the `high-energy' sigma-condensate 
$c \sim f_{\pi}$, and the `low-energy' dimensionless Yukawa coupling
constant $G=m_f/c$. To satisfy the mass ratio constraints, the Higgs self-
coupling must be equal to $G^2$. The non-relativistic analogue of the
condensate $c$ is ${\sqrt N}/2$ where $N$ is the density of states of the
constituents at the fermi surface. The origin of the mass scales is a more
dynamical question depending on the SUSY Hamiltonian structure, for a
derivation of which the interested reader is referred to [50 a-d].            
\par
        The physical scenario envisaged by Nambu for this broken SUSY
structure is two-fold. The first is a cascading hierarchy of symmetry 
breaking (`tumbling') which in particle physics [51a] means something
like the following. Suppose a symmetry breaking at a high energy level
gives rise to a $\sigma$-boson at the low energy scale. The latter, being
a scalar, will induce attraction between the quasi-fermions, which in turn
may generate a second generation symmetry-breaking, and so on. According 
to Nambu, a similar example of tumbling also exists in nuclear physics. 
Thus the $\sigma$-boson, which is a fall-out of chiral symmetry breaking
and quark-mass generation in the bulk of nuclear binding, also causes
nuclear pairing which can be estimated quite accurately [50c]. 
\par
        The second scenario [50e] is the theoretical possibility of a
(Chew-like) bootstrap, not at the hadron, but at the quark-lepton level,
on the assumption that the  t'Hooft self-consistency condition [51b] is
satisfied between these two levels. This leads to the following bounds 
on the $t$-quark and Higgs masses: $m_t > 120 GeV$; $m_H > 200 GeV$. 
This and other details may be found in [50e].             

\section{Gauge Theoretic Formulation of SDE-BSE} 

As seen in Sec.2, the simple NJL-model [4] succinctly articulates the 
$DB{\chi}S$ mechanism which gives rise to dynamical quark-mass generation
on the one hand, and a Nambu-Goldstone [48] realization of the massless
pion on the other. Another result is the formal identity of the mass-gap
equation (SDE) with the homogeneous BSE for the vertex function for a
{\it massless} pseudoscalar $q{\bar q}$ composite. We are now in a 
position to pursue the same logic to give a formal theoretical basis to a 
gluon-exchange (vector)-like 4-fermion interaction (to simulate QCD effects) 
in the input Lagrangian by deriving from it an interlinked BSE-SDE 
framawork [9-12] which is the backbone of this article. In this respect
we shall skip an alternative non-perturbative treatment of the BCS-NJL
pairing mechanism by the Bogoliubov-Valatin method [23] which is not easy
to adapt to a Lorentz-invariant formulation. 

\setcounter{equation}{0}
\renewcommand{\theequation}{3.\arabic{equation}}

\subsection{Minimal Effective Action: SDE $\&$ BSE}        

We outline a treatment due to Munczek [52] on the derivation of the 
equations of motion for composite fields. Consider an action functional
\begin{equation}
S= \int dx [{\bar \psi}(-\gamma.\partial-M)\psi+{\bar \psi}\lambda(x)+ h.c.]            
-{1 \over 2} \int \int dx dy \Sigma_s {G_s(x-y)J_s(x)J_s(y)}
\end{equation}
where $\lambda(x)$ is an external source, $G_s$ is the propagator of the
exchanged boson, and $J_s(x)$ =${\bar \psi}(x) \Gamma_s \psi(x)$ is the 
current function. This form is approximately derivable from the standard
generating function for non-abelian QCD with $\Gamma_s$ = 
$i\gamma_\mu \lambda/2$, when $G_s$ becomes the gluon propagator. The
NJL-type contact interaction corresponds to $G_s \equiv \delta^4 (x-y)$,
but the treatment is more generally valid for non-local interactions too.
The standard approach is to introduce bilocal boson fields [53] which for 
several types of spin excitations has the form [52]
\begin{equation}
\eta(x,y) = \Sigma_s \Gamma_s \psi (x) \Gamma_s{\bar \psi}(y) G_s (x-y)
\end{equation}        
where $\eta$ has a $4 \times 4$ matrix form. With a second auxiliary field
$B(x,y)$ [52], one gets the following generating functional
\begin{equation}
Z = N^{-1} \int D{\psi} D{\bar \psi} D{\eta} D{B} exp [iS(\psi, {\bar \psi},
B, \eta) + i \int dx ({\bar \psi}\lambda+ {\bar \lambda}\psi)]
\end{equation}
\begin{eqnarray}
S    &=& \int dx {\bar \psi}(-\gamma.\partial-M)\psi - Tr \int dx dy \eta(x,y)
[B(y,x)-\psi(y){\bar \psi}(x)] \\ \nonumber
     & & + {1 \over 2} Tr \int \int dx dy \sum_s G_s(x-y)
B(x-y)\Gamma_s B(y,x)\Gamma_s
\end{eqnarray}
When the functional integration is carried out over $\eta(x,y)$, it gives
a $\delta$-function $\delta[B(y,x)-\psi(y){\bar \psi}(x)]$. Subsequent
integration over $B$ gives eq.(3.1). After this check, the order of 
integration may be reversed so as to integrate out over $\psi$ and 
${\bar \psi}$, and yield the effective action
\begin{equation}
S=Tr[-iln(-\gamma.\partial-\eta)-\eta B +{\bar B}B/2]; \quad
{\bar B}(x,y)= \Sigma_s G_s(x-y)\Gamma_s B(x,y)\Gamma_s
\end{equation}   
Here $\eta$, $B$, $(\gamma.\partial+M)$, are matrices in spinor, internal
symmetry, and configuration space indices, so that
\begin{equation}
\eta B = \int dz \eta(x,z)B(z,y) \equiv <x|\eta B|y>; \quad
Tr [\eta B] = Tr \int dx <x|\eta B|x>
\end{equation}
Varying $S$ w.r.t. $B$ and $\eta$ gives
\begin{equation}
\eta(x,y) = {\bar B}(x,y); \quad B = i(-\gamma.\partial-M-\eta)^{-1}
= i(-\gamma.\partial-M-B)^{-1}
\end{equation}
Replacing $B$ in (3.7) by the vacuum expectation value $<B>$ = $i S_F$,
gives the SDE
\begin{equation}
S_F = (-\gamma.\partial - M -iS_F)^{-1} = \Sigma_s G_s(x-y)\Gamma_s 
S_F(x-y) \Gamma_s
\end{equation}
whose detailed form is
\begin{equation}
(-\gamma.\partial-M)S_F(x-y)-i \int dz \Sigma_s G_s(x-z) \Gamma_s S_F(x-z)
\Gamma_s S_F(z-y) = \delta^4(x-y)
\end{equation}
Next, for the quantum corrections to $B$, write
\begin{equation}
B(x,y) = iS_F(x-y) + \phi(x,y)
\end{equation}
and obtain the homogeneous equation
\begin{equation}
i \sum_1^{\inf} S_F({\bar \phi} S_F)^n = iS_F{\bar \phi} S_F +
S_F {\bar \phi} \phi; \quad {\bar \phi}(x,y) = \Sigma_s G_s(x-y) \Gamma_s
\phi(x,y) \Gamma_s
\end{equation}
If the non-linear term in $\phi$ in (3.11) is neglected, the result is the 
homogeneous BSE
\begin{equation}
\phi(x,y) = i \int \int dz dt S_F(x-z) \Sigma_s G_s(z-t) \Gamma_s \phi(z-t)
\Gamma_s S_F(t-y)
\end{equation}
which must be solved along with the SDE (3.8) for the propagator. Note that 
the kernel of the BSE is $G_s$, i.e., the {\it same} form factor as appears
in the input Lagrangian itself. This is the basic logic of the interplay of
the SDE with the BSE. Next we describe this interplay in momentum space for
the case $\Gamma_s$ = $i\gamma_\mu \lambda_a/2$, to bring out the Nambu-
Goldstone nature of a pseudoscalar state ($\phi$ proportional to $\gamma_5$),
one in which the Ward identity plays a crucial role.

\subsection{Self-Energy vs Vertex Fn in Chiral Limit}

The formal equivalence of the mass-gap equation (SDE) and the BSE  for a 
pseudoscalar meson in the chiral limit [24] will now be demonstrated for
an arbitrary confining form $D(k)$ (not just the perturbative form $k^{-2}$).
Denoting the mass operator by $\Sigma(p)$ and the vertex function by 
$\Gamma_H$, the SDE after replacing the color factor $\lambda_1.\lambda_2/4$
by its Casimir value $4/3$, reads as
\begin{equation}
\Sigma(p) = {4 \over 3}i (2\pi)^{-4} \int d^4k D_{\mu\nu}(k) \gamma_\mu
S_F'(p-k)\gamma_\nu; \quad D_{\mu\nu}(k) = (\delta_{\mu\nu}-
k_\mu k_\nu/k^2) D(k)
\end{equation} 
$S_F'$ is the full propagator related to the mass operator $\Sigma(p)$ by
\begin{equation}
\Sigma(p) +i \gamma.p = S_F^{-1}(p) = A(p^2)[i\gamma.p + m(p^2)]
\end{equation}
thus defining the mass function $m(p^2)$ in the chiral limit $m_c =0$.
In the same way the vertex function $\Gamma_H(q,P)$ for a $q{\bar q}$
hadron $(H)$ of 4-momentum $P_\mu$ made up of quark 4-momenta 
$p_{1,2}= P/2 \pm q$ satisfies the BSE 
\begin{equation}
\Gamma_H(q,P)= -{4 \over 3}i(2\pi)^{-4} \int d^4q'D_{\mu\nu}(q-q')\gamma_\mu
S_F(q'+ P/2) \Gamma_H(q',P) S_F(q'-P/2) \gamma_\nu
\end{equation}
The complete equivalence of (3.13) and (3.15) for the pion case in the 
chiral limit $P_\mu \rightarrow 0$ is easily established. Indeed, with 
the self-consistent ansatz $\Gamma_H$ =$\gamma_5 \Gamma(q)$, eq.(3.15)
simplifies to
\begin{equation}
\Gamma(q)= {4 \over 3}i (2\pi)^{-4} \int d^4k \gamma_\mu S_F'(k-q) \Gamma(q-k)
S_F'(q-k)\gamma_\nu
\end{equation}
where the replacement $q'= q-k$ has been made. Substitution for $S_F'$ from
(3.14) in (3.16) gives
\begin{equation}
\Gamma(p) = -{4 \over 3}i (2\pi)^{-4} \int d^4k {{D(k) \Gamma(p-k)} \over 
{A^2(p-k) (m^2((p-k)^2) + (p-k)^2)}}
\end{equation}
where we have relabelled $q \rightarrow p$. On the other hand substituting
for $S_F'$ (3.14) in (3.13) gives for the mass term of $\Sigma(p)$ the result
\begin{equation}
A(p^2)m(p^2)= -{4 \over 3}i (2\pi)^{-4} \int d^4k {{D(k)A(q')m(q'^2)} \over
 {A^2(q')(m^2(q'^2) + q'^2)}}
\end{equation}
where $q'=p-k$. A comparison of (3.17) and (3.18) shows their equivalence
with the identification $\Gamma(q)=A(q)m(q^2)$, i.e. the identity of the
vertex and mass functions in the chiral limit, provided $A=1$. this last
is a consequence of the Landau gauge for $D_{\mu\nu}$ in eq.(3.13), since in
this gauge, the function $A(p)$ does not undergo renormalization [54], 
so that it may be set equal to unity. Note that this result is more 
general than in the contact type NJL model, since both quantities are 
now functions of momentum due to the extended nature of the 4-fermion 
coupling caused by the gluonic propagator $D(k)$. 

\subsection{$\Sigma(p)$ vs $\Gamma(q,P)$ via Ward Identities}

The connection between $\Sigma(p)$ and $\Gamma(q,P)$ away from the chiral 
limit $(P_\mu =0)$ is achieved via a systematic use of the Ward identities
for vector and axial vector types. The following derivation due to [24a] may
be instructive for applications. Consider some approximation scheme (based
on a BSE with a specified kernel) to determine $\Sigma(p)$ via eq.(3.13),
so as to obey the Ward-Takahashi identities. E.g., the quark-gluon vertex
function $\Gamma_\lambda$ satisfies the {\it inhomogeneous} equation
\begin{equation}
\Gamma_\lambda=\gamma_\lambda - {4 \over 3}i(2\pi)^{-4} \int d^4q'\gamma_\nu
S_F'(q'+P/2) \Gamma_\lambda S_F'(q'-P/2)\gamma_\mu D_{\mu\nu}(q-q')
\end{equation}
Multiplying (3.19) by $P_\lambda$ and using the WT-identity 
\begin{equation}
P_\lambda \Gamma_\lambda(q,P) = S_F'^{-1}(q+P/2) - S_F'^{-1}(q-P/2)
\end{equation}
gives the result
\begin{eqnarray}
\lefteqn{{1 \over S_F'(P/2+q)}-{1 \over S_F'(q-P/2)} = } \\ \nonumber
   &  &  \gamma.P -{{4i} \over 3} \int {{d^4q'}\over {(2\pi)^4}} 
D_{\mu\nu}(q-q')\gamma_\nu [S_F(q'-P/2)-S_F(q'+P/2)]\gamma_\mu
\end{eqnarray}
which is entirely consistent with (3.13) when one uses the definition (3.14) 
for $\Sigma(p)$. In a similar way, for the axial vector $\Gamma_{\mu 5}$,
the corresponding BSE obeying chiral symmetry is
\begin{equation}
\Gamma_{\lambda 5}(q,P)= i\gamma_\lambda \gamma_5 - {4 \over 3}i(2\pi)^{-4}
\int d^4q'D_{\mu\nu}(q-q')\gamma_\nu S_F(q'+P/2) \Gamma_{\lambda 5}(q',P)
S_F(q'-P/2)\gamma_\mu
\end{equation}
It is again consistent with eq.(3.15) and the definition (3.14) for 
$\Sigma(p)$ if one uses the axial WT identity
\begin{equation}
-iP_\lambda \Gamma_{\lambda 5}(q,P)= S_F'^{-1}(q+P/2)\gamma_5 +\gamma_5
S_F'^{-1}(q-P/2)
\end{equation}
The LHS of (3.23) must now be identified with the pseudoscalar vertex
function $\Gamma_5(q.P)$, so that the corresponding RHS gives its full 
structure that is consistent with gauge invariance, viz., 
\begin{eqnarray}
\lefteqn{\Gamma_5(q,P)\gamma_5 = }  \\  \nonumber
 & &  i\gamma.(q+P/2)A(q+P/2) - i\gamma.(q-P/2)A(q-P/2)+ B(q+P/2)+B(q-P/2); \\ \nonumber
 & &  B(p) = A(p) m(p^2)
\end{eqnarray}
This equation checks with (3.18), in the Landau gauge $(A=1)$, in the 
chiral limit $P_\mu = 0$, but now provides the corrections for 
$P_\mu \neq 0$ as well. In the Landau gauge (3.24) simplifies to
\begin{equation}
\Gamma_5\gamma_5 = i\gamma.P + m(q+P/2) + m(q-P/2)
\end{equation}
In recent years, the determination of vertex functions via WT identities
has become a fairly standard practice, although it is not always the most
convenient method in practice for incorporating gauge-invariance within a
given (semi-phenomenoligical) framework. For the present report, we shall 
have occasion to incorporate QED gauge invariance in arbitrary momentum-
dependent form factors, and the method will be explained in Sec.(5), and
in more detail in Appendix B, in connection with the $P$-meson e.m. self-
energy calculations  to be given in Sect.7 

\section{3D-4D SDE-BSE Formalism Under MYTP} 

        As per the programme outlined in Sect.1, we shall from now on 
specialize to a more practical form of SDE-BSE framework born out of 3D 
support (defined covariantly) to a vector-exchange mediated 4-fermion 
coupling at the input Lagrangian level with `current' (almost massless)
quarks. The vector exchange simulates the effect of a gluonic propagator,
encompassing both the perturbative and non-perturbative regimes, and thus
preserves the chiral character of the input coupling. The derived SDE and
BSE, a la Chap 3, automatically incorporates  $DB{\chi}S$ and hence 
generates the dynamical mass function $m(p)$ whose low momentum limit $m(0)$ 
gives the bulk contribution to the {it constituent} mass $m_{cons}$, while 
the {\it current} mass $m_{curr}$ for $uds$ quarks (that enter the input 
Lagrangian) gives a small effect. This last is in keeping with Politzer's 
Additivity principle [55], viz., $m_{cons}$ = $m_{curr}+m(0)$, providing a 
rationale for the quark masses usually employed in potential models [12].   
\par
        Now to implement the covariant 3D constraint of {\bf MYTP} [15] on 
the BSE kernel (which stems from one on the input Lagrangian), we shall 
consider two methods in parallel for a direct comparison: i) Covariant 
Instantaneity Ansatz (CIA) [16-17]; ii) Covariant Null-Plane Ansatz (CNPA) 
[41]. The latter [41] gives a formal `covariance structure' to an earlier 
pragmatic formulation with essentially the same content [40], while the 
former [16] is already covariant as it is. We shall now outline a connected 
account of the 3D BSE reduction for both CIA and CNPA types (with scalar 
followed by fermion quarks), to bring out the structural identity of the 
resulting BSE's for a $q{\bar q}$ system. This will be followed by a 
reconstruction of the 4D BS vertex functions for both types [16, 41] 
which will serve as the basic framework for 4D quark loop calculations 
in the subsequent chapters.             

\setcounter{equation}{0}
\renewcommand{\theequation}{4.\arabic{equation}}

\subsection{3D-4D BSE Under CIA: Spinless Quarks}
 
To keep the contents fairly self-contained, we start with a few definitions
for unequal mass kinematics in the notation of [16,10b]. Let the quark 
constituents of masses $m_{1,2}$ and 4-momenta $p_{1,2}$ interact to produce
a composite hadron of mass $M$ and 4-momentum $P_\mu$. The internal 4-momentum 
$q_\mu$ is related to these by
\begin{equation}
p_{1,2} = {\hat m}_{1,2}P \pm q; \quad P^2=-M^2; \quad 
2{\hat m}_{1,2}= 1 \pm (m_1^2 - m_2^2)/M^2
\end{equation}
These Wightman-Garding definitions [56] of the fractional momenta 
${\hat m}_{1,2}$ ensure that $q.P=0$ on the mass shells $m_i^2+ p_i^2=0$
of the constituents, though not off-shell. Now define ${\hat q}_\mu$ =
$q_\mu - q.PP_\mu/P^2$ as the relative momentum {\it transverse} to the
hadron 4-momentum $P_\mu$ which automatically gives ${\hat q}.P \equiv 0$,
for all values of ${\hat q}_\mu$. If the BSE kernel $K$ for the 2 quarks
is a function of only these transverse relative momenta, viz. 
$K= K({\hat q}, {\hat q}')$, this is called the ``Cov. Inst.Ansatz (CIA)'' 
[16] which accords with {\bf MYTP} [15]. For two scalar quarks with 
inverse propagators $\Delta_{1,2}$, this ansatz gives rise to the following
BSE for the wave fn $\Phi(q,P)$ [16, 10b]:
\begin{equation}
i(2\pi)^4 \Delta_1\Delta_2 \Phi(q,P)= \int d^4q'K({\hat q}, {\hat q}')          
\Phi(q',P); \quad  \Delta_{1,2}= m_{1,2}^2 + p_{1,2}^2
\end{equation} 
The quantities $m_{1,2}$ are the `constituent' masses which are strictly
momentum dependent since they contain the mass function $m(p)$ [55], but
may be regarded as almost constant for low energy phenomena $m(p) \cong m(0)$.
Further, under CIA, $m(p)= m({\hat p})$, a momentum-dependence which is 
governed by the $DB{\chi}S$ condition [4] (see below). 
\par
        To make a 3D reduction of eq.(4.2), define the 3D wave function 
$\phi({\hat q})$ in terms of the longitudinal momentum $M \sigma$ as
\begin{equation}
\phi({\hat q}) = \int Md\sigma \Phi(q,P); \quad M\sigma = M q.P/P^2
\end{equation}
using which, eq.(4.2) may be recast as
\begin{equation}
i(2\pi)^4 \Delta_1 \Delta_2 \Phi(q,P)= \int d^3{\hat q}' K({\hat q},{\hat q}')
\phi({\hat q}'); \quad d^4q' \equiv d^3{\hat q}' M d{\sigma}'
\end{equation}
Next, divide out by $\Delta_1\Delta_2$ in (4.4) and use once again (4.3) to
reduce the 4D BSE form (4.4) to the 3D form 
\begin{equation}
(2\pi)^3 D({\hat q})\phi({\hat q}) = \int d^3 {\hat q}' K({\hat q},{\hat q}')
\phi({\hat q}'); \quad {{2i\pi}\over {D({\hat q})}} \equiv 
\int {{M d\sigma} \over {\Delta_1\Delta_2}}
\end{equation}
Here $D({\hat q})$ is the 3D denominator function associated with the like
wave function $\phi({\hat q})$. The integration over ${d\sigma}$ is carried 
out by noting pole positions of $\Delta_{1,2}$ in the $\sigma$-plane, where
\begin{equation}
\Delta_{1,2} = {\omega_{1,2}}^2 - M^2 ({\hat m}_{1,2} \pm \sigma)^2; \quad
{\omega_{1,2}}^2 = m_{1,2}^2 + {\hat q}^2
\end{equation}
The pole positions are given for $\Delta_{1,2}=0$ respectively by
\begin{equation}
M(\sigma + {\hat m}_1) = \pm \omega_1 \mp i\epsilon; \quad
M(\sigma - {\hat m}_2) = \pm \omega_2 \mp i\epsilon
\end{equation}
where the $(\pm)$ indices refer to the lower/upper halves of the $\sigma$-
plane. The final result for $D({\hat q})$ is expressible symmetrically [16]:
\begin{equation}
D({\hat q}) = M_{\omega} D_0({\hat q}); \quad {{2} \over {M_{\omega}}} =
{{{\hat m}_1} \over {\omega_1}} + {{{\hat m}_2} \over {\omega_2}} 
\end{equation}
\begin{equation}
{{1} \over {2}} D_0({\hat q}) = {\hat q}^2 - {{\lambda(m_1^2, m_2^2, M^2)}
 \over {4M^2}}; \quad \lambda = M^4 -2M^2(m_1^2 + m_2^2)+(m_1^2-m_2^2)^2
\end{equation}
The crucial thing for the {\bf MYTP} is now to observe the {\it equality} of 
the RHS of eqs (4.4) and (4.5), thus leading to an {\it {exact 
interconnection}} between the 3D and 4D BS wave functions:
\begin{equation}
\Gamma({\hat q}) \equiv \Delta_1\Delta_2 \Phi(q,P) = 
{{D({\hat q}\phi({\hat q})} \over {2i\pi}} 
\end{equation}
Eq.(4.10) determines the hadron-quark vertex function $\Gamma({\hat q})$ as 
a product $D\phi$ of the 3D denominator and wave functions, satisfying a 
relativistic 3D Schroedinger-like equation (4.5).               
\par
        Some comments on the entire BSE structure are now in order. The 
`two-tier' character of the formalism is seen from the simultaneous
appearance of the 3D form (4.5) and the 4D form (4.4), leading to their
interconnection (4.10). The 3D form (4.5) gives the basis for making contact
with the 3D spectra [13], while the reconstructed 4D wave (vertex) function
(4.10) in terms of 3D ingredients $D$ and $\phi$ enables the evaluation of
4D quark-loop integrals in the standard Feynman fashion [40]. Note that the
vertex function $\Gamma=D\phi/(2i\pi)$ has quite a general structure, and 
independent of the details of the input kernel $K$. Further, the $D$-
function, eq.(4.8), is universal and well-defined off the mass shell of 
either quark. The 3D wave function $\phi$ is admittedly model-dependent, but
together with $D({\hat q})$, it controls the 3D spectra via (4.5), so as to
offer a direct experimental check on its structure. Both functions depend on 
the single 3D Lorentz-covariant quantity ${\hat q}^2$ whose most important
property is its positive definiteness for time-lke hadron momenta ($M^2 >0$).

\subsection{CNPA for 3D-4D BSE: Spinless Quarks}

As a preliminary to defining a 3D support to the BS kernel on the null-plane 
(NP), on the lines of CIA [16], a covariant NP orientation [41] may be 
represented by the 4-vector $n_\mu$, as well as its dual ${\tilde n}_\mu$, 
obeying the normalizations $n^2 = {\tilde n}^2 =0$ and $n.{\tilde n} = 1$.
In the standard NP scheme (in euclidean notation), these quantities 
are $n=(001;-i)/\sqrt{2}$ and ${\tilde n}=(001;i)/\sqrt{2}$, while the two
other perpendicular directions are collectively denoted by the subscript
$\perp$ on the concerned momenta. We shall try to maintain the $n$-dependence 
of various momenta to ensure explicit covariance; and to keep track of
the old NP notation $p_{\pm} = p_0 \pm p_3$, our covariant notation is 
normalized to the latter as  $p_+ = n.p \sqrt{2}$; 
$p_- = -{\tilde n}.p \sqrt{2}$, while the perpendicular components continue 
to be denoted by $p_{\perp}$ in both notations.
\par 
        In the same nitation as for CIA [16],  the 4th component of the 
relative momentum $q={\hat m}_2 p_1-{\hat m}_1 p_2$,  that should be 
eliminated for obtaining a 3D equation, is now proportional 
to $q_n \equiv {\tilde n}.q$, as the NP analogue [40] of $P.qP/P^2$ in  
CIA [16], where $P=p_1+p_2$ is the total 4-momentum of the hadron. However 
the quantity $q - q_n n$ is still only $q_\perp$, since its square is 
$q^2 - 2 n.q{\tilde n}.q$, as befits $q_\perp^2$ (readily checked against the 
`special' NP frame). We still need a third component $p_3$, for which a
first guess is $zP$, where $z=n.q/n.P$. And for calculational convenience
we shall need to (temporarily) invoke the `collinear frame' which amounts 
to $P_\perp.q_\perp = 0$, a restriction which will be removed later by a 
simple prescription of `Lorentz completion'. Unfortunately the definition 
${\hat q}_\mu$ = $(q_{\perp\mu}, zP_\mu)$ does not quite fit the bill for a 
covariant 3-vector, since a short calculation shows again that ${\hat q}^2$ 
=$q_\perp^2$. The correct definition is seen as $q_{3\mu} = z P_n n_\mu$, 
where $P_n$=$P.{\tilde n}$, giving ${\hat q}^2$ = $q_\perp^2 + z^2M^2$, as 
required. We now collect the following definitions/results:
\begin{eqnarray}
 q_\perp &=& q-q_n n; \quad {\hat q}=q_\perp+ x P_n n; \quad x=q.n/P.n;
 \quad P^2 = -M^2; \\ \nonumber
 q_n,P_n &=& {\tilde n}.(q,P);{\hat q}.n = q.n; \quad {\hat q}.{\tilde n} = 0; 
\quad P_\perp.q_\perp = 0;   \\ \nonumber 
     P.q &=& P_n q.n + P.n q_n; \quad P.{\hat q} = P_n q.n; \quad
{\hat q}^2 = q_\perp^2 + M^2 x^2
\end{eqnarray}        
Now in analogy to CIA, the reduced 3D BSE (wave-fn $\phi$) may be derived
from the 4D BSE (4.2) for spinless quarks (wave-fn $\Phi$) when its kernel 
$K$ is {\it decreed} to be independent of the component $q_n$, i.e., 
$K=K({\hat q},{\hat q}')$, with ${\hat q}$ = $(q_\perp,  P_n n)$, 
in accordance with the TP [15] condition imposed on the null-plane (NP), 
so that $d^4 q$ = $d^2 q_\perp dq_3 dq_n$. Now define a 3D wave-fn 
$\phi({\hat q})$ = $ \int  d{q_n}\Phi(q)$, as the CNPA  counterpart of the
CIA definition (4.3), and use this result on the RHS of (4.2) to give 
\begin{equation}
i(2\pi)^4 \Phi(q) = {\Delta_1}^{-1} {\Delta_2}^{-1} 
\int d^3{\hat q}' K({\hat q},{\hat q}')\phi({\hat q}')  
\end{equation}
which is formally the same as eq.(4.4) for CIA above. Now integrate both 
sides of eq.(4.12) w.r.t. $dq_n$ to give a 3D BSE in the variable ${\hat q}$:
\begin{equation}  
(2\pi)^3 D_n({\hat q}) \phi({\hat q}) =  \int d^2{q_\perp}'d{q_3}'  
K({\hat q},{\hat q}') \phi({\hat q}')
\end{equation}
which again corresponds to the CIA eq.(4.5), except that the function 
$D_n(\hat q)$ is now defined by 
\begin{equation}
\int d{q_n}{\Delta_1}^{-1} {\Delta_2}^{-1} = 2{\pi}i D_n^{-1}({\hat q})
\end{equation}
and may be obtained by standard NP techniques [40] (Chaps 5-7) as follows. 
In the $q_n$ plane, the poles of $\Delta_{1,2}$ lie on opposite sides of 
the real axis, so that only {\it one} pole will contribute at a time. Taking
the $\Delta_2$-pole, which gives 
\begin{equation}
2q_n = -{\sqrt 2} q_- = {{m_2^2 + (q_\perp-{\hat m}_2P)^2} 
 \over {{\hat m}_2 P.n - q.n}}
\end{equation}
the residue of $\Delta_1$ works out, after a routine simplification, to 
just $2P.q = 2P.n q_n+2P_n q.n$, after using the collinearity condition 
$P_\perp.q_\perp = 0$ from (4.11). And when the value (4.15) of $q_n$ is 
substituted in (4.14), one obtains (with $P_n P.n = -M^2/2$): 
\begin{equation}
D_n({\hat q}) = 2P.n ({\hat q}^2 -{{\lambda(M^2, m_1^2, m_2^2)} \over {4M^2}});
\quad {\hat q}^2 = q_\perp^2 + M^2 x^2; \quad x = q.n/P.n      
\end{equation}
Now a comparison of (4.12) with (4.13) relates the 4D and 3D wave-fns: 
\begin{equation}  
2{\pi}i \Phi(q) = D_n({\hat q}){\Delta_1}^{-1}{\Delta_2}^{-1} \phi({\hat q})
\end{equation}
as the CIA counterpart of (4.10) which is valid near the bound state pole. 
The BS vertex function now becomes $\Gamma = D_n \times \phi/(2{\pi}i)$. This
result, though dependent on the NP orientation, is nevertheless formally
{\it covariant}, and closely corresponds to the pedagogical result of the
old NPA formulation [40], with $D_n \Leftrightarrow D_+$.    
\par
        A 3D equation similar to the covariant eq.(4.13) above, also obtains 
in alternative NP formulations such as in Kadychevsky-Karmanov [38] (see 
their eq.(3.48)). Both are `covariantly' dependent on the orientation 
$n_\mu$ of the NP, i.e., have certain $n$-dependent 3-scalars, in addition to 
genuine 4-scalars. However the {\it independent} 4-vector ${\tilde n}_\mu$ 
which has a dual interplay with $n_\mu$ in the above CNPA formulation, does
not seem to have a counterpart in [38]. Secondly this manifestly covariant 
4D formulation needs no 3-vector like ${\bf n}$, or explicit Lorentz 
transformations, as in such alternative NP formulations [38]. As to the 
`angular condition', a question first raised by Leutwyler-Stern [35d], 
no special effort has been made to satisfy this requirement, since the very 
appearance of the `effective' 3-vector ${\hat q}_\mu$ in the 3D BSE in a 
rotationally invariant manner is an automatic guarantee (in the sense of a 
`proof of the pudding') of the satisfaction of this condition [35d] without 
further assumptions.    
\par
        A second aspect of the above 3D-4D BSE under CNPA (which allows for 
off-shell momenta) is that it has no further need for `spurions' [38] (to 
make up for energy-momentum balance due to on-shellness of the momenta 
in such formulations [38]), so that normal 4D Feynman techniques suffice, as
in the old-fashioned NPA formulation [33,40]. However, to rid the 
{\it physical} amplitudes of $n_\mu$-dependent terms in the external (hadron) 
momenta, after integration over the internal loop momenta, one still needs to
employ a simple technique of `Lorentz-completion' (to be illustrated in 
Sec.5 for the pion form factor calculation) as an alternative to other 
NP prescriptions [37,38] to remove ${\bf n}$-dependent terms.     
\par
        A more succinct comparison with other null-plane approaches concerns 
the inverse process of {\it reconstruction} of the 4D hadron-quark vertex, 
eq.(4.17)), which has no counterpart in them [37,38], as these are basically 
3D oriented. Thus in [38], the nearest analogue is to express the 3D NP wave 
function in terms of the 4D BS wave function (see eq.(3.58) of [38]), but 
{\it {not vice versa}}. This problem of `loss of Hilbert space information' 
inherent in such a process of reconstruction, has been discussed recently 
in the context of the $qqq$ problem [47]; (see Section 10 for details).                 

\subsection{Fermion Quarks with QCD-Motivated BSE}

We are now in a position to give a corresponding description of the 3D-4D
BSE for fermion quarks, for both CIA and CNPA cases taken together, just as
for spinless quarks above. The 4D BSE for fermion quarks under a gluonic 
(vector-type) interaction kernel with 3D support has the form [10 a-b]:  
\begin{equation}  
i(2\pi)^4 \Psi(P,q) = S_{F1}(p_1) S_{F2}(p_2) \int d^4 q' 
K({\hat q}, {\hat q}') \Psi(P,q'); \quad K= F_{12} i\gamma_\mu^{(1)}
i\gamma_\mu^{(2)} V({\hat q},{\hat q}')
\end{equation}
where $F_{12}$ is the color factor $\lambda_1.\lambda_2/4$ and the $V$- 
function expresses the scalar structure of the gluon propagator in the 
perturbative (o.g.e.) plus non-perturbative regimes. The `hat' notation
on the momenta covers both CIA and CNPA cases simultaneously, where the
longitudinal component ${\hat q}_3$ is defined for the CNPA case as 
$q_{3\mu} = z P_n n_\mu$, with  $P_n$=$P.{\tilde n}$. The full structure of
$V$ (used in actual calculations [28,40]) is collected as under, using 
the simplified notations $k$ for $q-q'$, and $V({\hat k})$ for the $V$ fn: 
\begin{equation}
V({\hat k})=4\pi\alpha_s/{\hat k}^2+{3 \over 4}\omega_{q{\bar q}}^2 
 \int d{\bf r} [r^2 {(1+4A_0 {\hat m}_1{\hat m}_2 {M_>}^2 r^2)}^{-1/2}
 -C_0/\omega_0^2] e^{i{\hat k}.{\bf r}};
\end{equation}
\begin{equation}
\omega_{q{\bar q}}^2 = 4M_> {\hat m}_1 {\hat m}_2 \omega_0^2 \alpha_s({M_>}^2);
\quad \alpha_s(Q^2)={{6\pi} \over {33-2n_f}} {\ln (M_>/\Lambda)}^{-1};
\end{equation}
\begin{equation}
{\hat m}_{1,2} = [1 \pm (m_1^2 - m_2^2)/M^2]/2; \quad 
M_> = Max (M, m_1+m_2); \quad C_0 = 0.27;\quad A_0=0.0283  
\end{equation}
And the values of the basic constants (all in $MeV$) are [28,40]
\begin{equation}
\omega_0= 158 ;\quad m_{ud}=265 ;\quad m_s=415 ; \quad m_c=1530 ; 
\quad m_b = 4900.
\end{equation}     
\par
        The BSE form (4.18) is however not the most convenient one for wider
applications in practice, since the Dirac matrices entail several coupled 
integral equations. Indeed, as noted long ago [10 a-b], a considerable 
simplification is effected by expressing them in `Gordon-reduced' form, 
(permissible on the quark mass shells, or better on the surface $P.q = 0$), 
a step which may be regarded as a fresh starting point of our dynamics, in the 
sense of an `analytic continuation' of the $\gamma$- matrices to `off-shell' 
regions (i.e., away from the surface $P.q =0$). Admittedly this constitutes
a conscious departure from the original BSE structure (4.18), but such 
technical modifications are not unknown in the BS literature [10c-d] in the 
interest of greater manoeuvreability, without giving up the essentials, in
view of the  "effective" nature of the BS kernel (see Chap 1 sec.6).  
\par
        The `Gordon-reduced' BSE form of (4.18) is given by [10a-b]
\begin{equation}
\Delta_1 \Delta_2 \Phi(P,q) = -i(2\pi)^{-4} F_{12} \int d^4q' V_\mu^{(1)} 
V_\mu^{(2)} V({\hat q},{\hat q}') \Phi(P,q');
\end{equation}
where the connection between the $\Psi$- and $\Phi$- functions is 
\begin{equation} 
\Psi(P,q) = (m_1-i\gamma^{(1)}.p_1)(m_2+i\gamma^{(2)}.p_2) \Phi(P,q); \quad
 p_{1,2} = {\hat m}_{1,2} P \pm q
\end{equation}
\begin{equation}
V_\mu^{(1,2)}= \pm 2m_{1,2} \gamma_\mu^{(1,2)}; \quad 
V_\mu^{(i)}=p_{i\mu}+p_{i\mu}'+i \sigma_{\mu\nu}^{(i)}(p_{i\nu}-p_{i\nu}')    
\end{equation}
\par
        Now to implement the Transversality Condition [15] for the entire 
kernel of eq.(4.23), all time-like components $\sigma, \sigma'$ in the 
product $V^{(1)}.V^{(2)}$ must first be replaced by their on-shell values. 
Substituting from (4.25) and simplifying  gives 
\begin{equation}
(p_1+p_1').(p_2+p_2') = 4{\hat m}_1{\hat m}_2 P^2 -({\hat q}+{\hat q}')^2
-2({\hat m}_1-{\hat m}_2) P.(q+q') + ``spin-Terms'';
\end{equation}
\begin{equation}
``Spin Terms'' = -i(2{\hat m}_1P+{\hat q}+{\hat q}')_\mu\sigma_{\mu\nu}^{(2)}
{\hat k}_\nu+i(2{\hat m}_2P-{\hat q}-{\hat q}')_\mu \sigma_{\mu\nu}^{(1)}
{\hat k}_\nu + \sigma_{\mu\nu}^{(1)} \sigma_{\mu\nu}^{(2)}
\end{equation}
This is identical to eq.(7.1.9) of Ref.[40], via the correspondence  
${\hat q}$ $\Leftrightarrow$ $q_x,q_y, q_3(= q_+/P_+)$, so that both CIA [16]
and CNPA [41] have formally the same structures as the `old-fashioned' NPA 
[40], and hence give identical predictions on the 2-body spectra [28].  
\par
     The 3D reduction of eq.(4.23) now goes through exactly as in the spin-0 
case, eqs.(4.2-8), so that without further ado, the full structure of the 
3D BSE  can be literally taken over from Ref.[40]-Chap 7 (derived under 
old-fashioned NPA). In particular, for harmonic confinement,  obtained by
dropping the $A_0$ term in the `potential' $U(r)$ of (4.23) (a very good 
approximation for {\it light} ($ud$) quarks), the 3D BSE works out as
\begin{equation}
D({\hat q}) \phi({\hat q})=\omega_{q{\bar q}}^2 {\tilde D}({\hat q})
\phi({\hat q});
\end{equation}
\begin{equation}
D_n({\hat q}) \phi({\hat q})={P_n \over M}\omega_{q{\bar q}}^2 
{\tilde D}({\hat q}) \phi({\hat q}); 
\end{equation}
for the CIA and CNPA cases respectively, where $D_n$ is given by (4.16) 
and $D({\hat q})$ by (4.8). The other quantities retain the same meaning
for both. Thus 
\begin{equation}
{\tilde D}({\hat q})= 4{\hat m}_1{\hat m}_2 M^2 ({\nabla}^2+C_0/\omega_0^2) 
+ 4{\hat q}^2{\nabla}^2+8{\hat q}.{\nabla}+18
- 8{\bf J}.{\bf S}+(4C_0/\omega_0^2){\hat q}^2
\end{equation}         
For the spectroscopic predictions on $q{\bar q}$ hadrons, vis-a-vis data, 
the reader is referred to [28]. For algebraic completeness however, 
the (gaussian) parameter $\beta$ of the 3D wave function $\phi({\hat q})$ = 
$exp(-{\hat q}^2/{2 \beta^2})$, which is the solution of (4.28-29) for a 
ground state hadron [40,28] is: 
\begin{equation}
\beta^4 = {{8{\hat m}_1^2{\hat m}_2^2 M_>^2 \omega_0^2 \alpha(M_>^2)} 
\over {[1 - 8 C_0 {\hat m}_1 {\hat m}_2 \alpha_s(M_>^2)] <\sigma>}};\quad 
<\sigma>^2 = 1 + 24 A_0 ({\hat m}_1{\hat m}_2 M_>)^2 /\beta^2       
\end{equation}
Note that $\beta$ is a 4D invariant quantity, independent of $n\mu$, etc.     
(For an $L$-excited hadron wave function, see [40]). The full 4D BS wave 
function $\Psi(P,q)$ in a 4x4 matrix form [40] is then reconstructed from 
(4.23-24) as in the scalar case, eq.(4.10), viz., [40,16,44]  
\begin{equation}
\Psi(P,q)= S_F(p_1) \Gamma({\hat q}) \gamma_D S_F(-p_2); \quad
\Gamma({\hat q}) = N_H [1; P_n/M] D({\hat q})\phi({\hat q})/{2i\pi}
\end{equation}   
where $\gamma_D$ is a Dirac matrix which equals $\gamma_5$ for a P-meson, 
$i\gamma_\mu$ for a V-meson, $i\gamma_\mu \gamma_5$ for an A-meson, etc.
The factors in square brackets stand for CIA  and CNPA  values respectively. 
$N_H$ represents the hadron normalization given by (see Appendix A):
\begin{equation}
N_H^{-2}=2(2\pi)^3 \int d^3{\hat q} [M_{\omega}; M] \phi^2({\hat q})
[(1+{\delta m}^2/M^2)({\hat q}^2-\lambda/{4M^2})
+2{\hat m}_1{\hat m}_2 (M^2-{\delta m}^2)]           
\end{equation}
where $M_{\omega}$ is given in (4.8), and ${\delta m}$ = $m_1-m_2$, and
again the factors in square brackets represent CIA/CNPA values.    

\subsection{Dynamical Mass As $DB{\chi}S$ Solution of SDE}

We end this Section with the definition of the `dynamical' mass 
function of the quark in the chiral limit ($M_\pi=0$) of the pion-quark
vertex function $\Gamma({\hat q})$, in the 3D-4D BSE framework [11]. 
The logic of this follows from the BSE-SDE formalism outlined in Sec.3,
eqs.(3.12-18), for the connection between eq.(3.18) for $m(p)$ and
eq.(3.17) for $\Gamma(q)$ in the limit of zero mass of the pseudoscalar. So,
setting $M=0$ in (4.28-29), the scalar part of the (unnormalized) vertex
function may be identified with the mass function $m({\hat p})$, in the
limit $P_\mu=0$, where $p_\mu$ is the 4-momentum of either quark;
(note the appearance of the `hatted' momentum). The result is [11]
\begin{equation}
m({\hat p}) = [\omega({\hat p}); {\sqrt 2}p.n]
{{m_q^2 + {\hat p}^2} \over {m_q^2}} \phi({\hat p})
\end{equation}
under CIA and CNPA respectively. The normalization is such that  in the 
low momentum limit, the constituent $ud$ mass $m_q$ is restored under CIA
[11], while the corresponding `mass' under CNPA is $p_+$ [35c].   
\par
        A more important aspect of the `dynamical' mass function is its
appearance as the non-trivial of the SDE under $DB{\chi}S$ [9,24]. We now
give a derivation of the 3D-4D counterpart [11] of this basic result [24].
To that end we start with the non-perturbative part of the gluon propagator 
$D_{\mu\nu}(k)$ = $D(k)[\delta_{\mu\nu}-k_\mu k_\nu/k^2]$ for the (harmonic) 
interaction of $ud$ quarks (forming the `pion'structure), where the scalar 
factor $D(k)$ has the form [11]
\begin{equation}
D(k)= {3 \over 4}(2\pi)^3 \omega_0^2 2m_q {\alpha_s(4m_q^2)}[\nabla_{\hat k}^2
+C_0/\omega^2] {\delta^3({\hat k})}
\end{equation}
This form is immediately derivable from the structure of the `potential'
function $V{\hat k})$ of eq.(4.19-20), with the $A_0$-term set equal to
zero, and taking $M_>=2m_q$ for the `pion' case. Note that $D({\hat k})$ 
has a directional dependence $n_\mu=P_\mu/P^2$ on the pion 4-momentum $P_\mu$,
so that  ${\hat k}^2 >0$ over all 4D space; it also possesses a well-defined 
limit for $P_\mu \rightarrow 0$. This structure may now be substituted in 
the `gap-equation' (3.17/18) foa a self-consistent solution in the low 
momentum limit. This exercise has been carried out in [11], wherein the
SDE (3.18) in the Landau gauge $A(p^2)=1$ reduces to the form
\begin{equation}
m(p^2)= {{3i} \over \pi} \int d^3{\hat k}dk_0 m_q \alpha_s \times
[\omega_0^2 {\nabla_{\hat k}}^2 +C_0] \delta^3({\hat k}) 
{m(p'^2) \over {(p'2+m^2(p'^2)}}
\end{equation}
where $p'=p-k$ is 4D, and (${\hat k}$, $k_0$) are (3D,1D) respectively.
The integration is essentially over the time-like $k_0$, with the `pole' 
position at ${p_0}'=m(p_0') \equiv m_{NJL}$, leading finally to [11]
\begin{equation}
m_{NJL}={{3m_q \alpha_s} \over {m_{NJL}}^2} [3\omega_0^2 - C_0 m_{NJL}^2];
\quad \alpha_s = {{6\pi} \over {29 ln(10 m_q)}}
\end{equation}
after substituting the value $\Lambda =200 MeV$ for the QCD constant.   
The further identification of $m_q$ with $m_{NJL}$ in this equation, yields
an independent self-consistent estimate $m_{NJL} \sim 300 MeV$, which may be 
compared to the input value $265 MeV$, eq.(4.22), employed for spectra [28].
This analysis so far ignores the Politzer relation [55] $m_{ud}=m_c+m_{NJL}$,
for the constituent mass $m_q$ away from the chiral limit. The derivation 
of the pion and $\sigma$-meson masses away from the chiral limit, may be
found in [11]. 

\vspace{1cm}

\section{CNPA Applications: Gauge-Inv Pion F.F.} 

The first example of our applications of the 3D-4D BSE structure developed
in Chap 4 is to 4D triangle loop integrals. This example has been chosen to 
illustrate the difficulties of CIA (as noted in Chap 1.4) in tackling their 
ill-defined nature as a result of acquiring time-like momentum components 
in the exponential/gaussian factors associated with the vertex functions 
(4.10) due to a `Lorentz-mismatch' among the rest-frames of the concerned 
hadronic composites, for triangle loops and above, such as the pion form 
factor, while 2-quark loops [32] just escape this pathology. This problem 
was not explicitly encountered in the old-fashioned NPA  treatment [33]
of the pion form factor, except for lack of explicit covariance. The CIA 
approach [16] to {\bf MYTP} [15] enjoys covariance, but its application
to triangle loop integrals causes other problems such as {\it complexity} 
of the corresponding amplitudes [57], apparently without good reason. On 
the other hand, the apparent success of the old-fashioned NPA [33] in 
circumventing this (complexity) problem [57], gives the hope that with its 
`covariant' formulation [41], CNPA, the powers of this method should stand
a better chance of testing via the form factor problem.
\par 
        To recall a short background, the pion form factor has through the 
ages been a good laboratory for subjecting theoretical models and ideas on 
strong interactions to observational test. Among the crucial parameters are 
the squared radius $<r_{exp}^2$ = $0.43 \pm .014 fm^2$ [58a], and the scaled 
form factor at high $k^2$, viz., $k^2 F(k^2) \approx 0.5 \pm 0.1 GeV^2$ [58b] 
that represent important check points for theoretical candidates such as 
QCD-sum rules [2b], Finite Energy sum rules [59], perturbative QCD [60], 
covariant null-plane approaches [37,38], Euclidean SDE [61], etc. The issue 
interface of perturbative and non-perturbative QCD regimes has been studied
in terms of the relative importance longitudinal vs transverse components 
[63], but this is the subject of a full-fledged dynamical theory (such as 
[9-12]), and not of some intuitive ansatze [62]. 
\par                            
        To that end, we outline a calculation of the P-meson form factor for 
unequal mass kinematics with full gauge invariance, including correction terms 
arising from QED gauge invariance, and illustrate the techniques of 
`Lorentz-completion' to obtain an explicitly Lorentz invariant quantity. 
As a check on the consistency of the formalism, the expected $k^{-2}$ 
behaviour of the pion form factor at high $k^2$ is realized.      
Some calculational details on the triangle-loop integral for the P-meson
form factor are given in Appendix 5.A.

\setcounter{equation}{0}
\renewcommand{\theequation}{5.\arabic{equation}}

\subsection{P-Meson Form Factor $F(k^2)$ for Unequal Masses} 

        Using the two diagrams (figs.1a and 1b) of ref.[33c], and in the
same notation, the Feynman amplitude for the $h \rightarrow h'+ \gamma$
transition contributed by fig.1a ( quark $2$ as spectator is given by [33c]
\begin{equation}                
2{\bar P}_\mu F(k^2)= 4(2\pi)^4 N_n(P)N_n(P') e{\hat m}_1 \int d^4 T_\mu^{(1)}
{{D_n({\hat q})\phi({\hat q})D_n({\hat q}')\phi({\hat q}')} \over 
{\Delta_1 \Delta_1' \Delta_2}} + [1 \Rightarrow 2];
\end{equation}   
\begin{equation}
4T_\mu^{(1)} = Tr [\gamma_5 (m_1-i\gamma.p_1) i\gamma_\mu (m_1-i\gamma.p_1')
\gamma_5 (m_2 + i\gamma.p_2)]; \quad \Delta_i = m_i^2 + p_i^2; 
\end{equation}
\begin{equation}
p_{1,2} = {\hat m}_{1,2}P \pm q; \quad p_{1,2}'={\hat m}_{1,2}P' \pm q'
\quad p_2 = p_2'; \quad P-P'=p_1-p_1'=k; \quad 2{\bar P} = P+P'.
\end{equation}
After evaluating the traces and simplifying via (5.2-3), $T_\mu$ becomes
\begin{equation}
T_\mu^{(1)}= (p_{2\mu}-{\bar P}_\mu)[{\delta m}^2-M^2-\Delta_2] -k^2p_{2\mu}/2
+(\Delta_1-\Delta_1')k_\mu/4
\end{equation} 
The last term in (5.4) is non-gauge invariant, but it does not survive the
integration in (5.1), since the coefficient of $k_\mu$, viz., 
$\Delta_1-\Delta_1'$ is antisymmetric in $p_1$ and $p_1'$, while the rest 
of the integrand in (5.1) is symmetric in these two variables. Next, to bring
out the proportionality of the integral (4.1) to ${\bar P}_\mu$, it is
necessary to resolve $p_2$ into the mutually perpendicular components 
$p_{2\perp}$, $(p_2.k/k^2)k$ and $(p_2.{\bar P}/{\bar P}^2){\bar P}$, of
which the first two will again not survive the integration, the first due
to the angular integration, and the second due to the antisymmetry of
$k=p_1-p_1'$ in $p_1$ and $p_1'$, just as in the last term of (5.4). The third
term is explicitly proportional to ${\bar P}_\mu$, and is of course 
gauge invariant since ${\bar P}.k =0$. (This fact had been anticipated while 
writing the LHS of (5.4)). Now with the help of the results
\begin{equation}
p_2.{\bar P}= -{\hat m}_2 M^2 -\Delta_1/4 - \Delta_1'/4; \quad
2{\hat m}_2 = 1-(m_1^2-m_2^2)/M^2; \quad {\bar P}^2=-M^2-k^2/4,
\end{equation} 
it is a simple matter to integrate (5.1), on the lines of Sec.4, noting 
that terms proportional to $\Delta_1 \Delta_2$ and $\Delta_1' \Delta_2$
will give zero, while the non-vanishing terms will get contributions 
only from the residues of the $\Delta_2$-pole, eq.(4.15). Before collecting
the various pieces, note that the 3D gaussian wave functions $\phi,\phi'$,
as well as the 3D denominator functions $D_n, D_n'$, do {\it not} depend 
on the time-like components $p_{2n}$, so that no further pole contributions
accrue from these sources. (It is this problem of time-like components of
the internal 4-momenta inside the gaussian $\phi$-functions under the CIA
approach [16], that had plagued a earlier CIA study of triangle diagrams 
[57]). To proceed further, it is now convenient 
to define the quantity ${\bar q}.n = p_2.n -{\hat m}_2 {\bar P}.n$ to 
simplify the $\phi$- and $D_n$- functions. To that end define the symbols: 
\begin{equation}
(q,q') = {\bar q} \pm {\hat m}_2 k/2; \quad  z_2 = {\bar q}.n/{{\bar P}.n}; 
\quad {\hat k}= k.n/{{\bar P}.n}; \quad (\theta_k, \eta_k)=1 \pm {\hat k}^2/4
\end{equation}
and note the following results of pole integration w.r.t. $p_{2n}$ [40]:
\begin{equation}
\int dp_{2n}{1 \over {\Delta_2}} [1/{\Delta_1}; 1/{\Delta_1'};
1/(\Delta_1 \Delta_1')] = [1/D_n; 1/D_n'; 2p_2.n/(D_nD_n')]
\end{equation}          
The details of further calculation of the form factor are given in 
Appendix A. An essential result is the normalizer $N_n(P)$ of the hadron, 
obtained by setting $k_\mu=0$, and demanding that $F(0)=1$. The reduced 
normalizer $N_H= N_n(P)P.n/M$, which is Lorentz-invariant, is given via
eq.(A.9) by:   
\begin{equation}
N_H^{-2}=2M (2\pi)^3 \int d^3{\hat q}
e^{-{\hat q}^2/\beta^2} [(1+{\delta m}^2/M^2)({\hat q}^2-\lambda/{4M^2})
+2{\hat m}_1{\hat m}_2 (M^2-{\delta m}^2)]           
\end{equation}
where the internal momentum ${\hat q}=(q_\perp,Mz_2)$ is formally a 3-vector,
in conformity with the `angular condition' [35d]. The corresponding 
expression for the form factor is (see Appendix A): 
\begin{equation}
F(k^2)= 2M N_H^2 (2\pi)^3 exp[-{(M{\hat m}_2{\hat k}/\beta)^2/{4\theta_k}}] 
(\pi \beta^2)^{3/2} {\eta_k \over {\sqrt \theta_k}}{\hat m}_1 G({\hat k}) +
[1 \Rightarrow 2]            
\end{equation}
where $G({\hat k})$ is defined by eqs.(A.12-13) of Appendix A.

\subsection{`Lorentz Completion' for $F(k^2)$}

The expression (5.9) for $F(k^2)$ still depends on the null-plane
orientation $n_\mu$ via the dimensionless quantity ${\hat k}$ =
$k.n/P.n$ which while having simple Lorentz transformation properties, 
is nevertheless {\it not} Lorentz invariant by itself. To make it explicitly 
Lorentz invariant, we shall employ a simple method of `Lorentz completion' 
which is merely an extension of the `collinearity trick' empolyed at the
quark level, viz., $P_\perp.q_\perp$ = 0; see eq.(4.11). Note that this
collinearity ansatz has already become reduntant at the level of the 
Normalizer $N_H$, eq.(5.8), which owes its Lorentz invariance to the 
integrating out of the null-plane dependent quantity $z_2$ in (5.8). This
is of course because $N_H$ depends only on one 4-momentum (that of a
{\it {single hadron}}), so that the collinearity assumption is exactly
valid. However the form factor $F(k^2)$ depends on {\it {two independent}}
4-momenta $P,P'$, for which the collinearity assumption is non-trivial,
since the existence of the perpendicular components cannot be wished away!
Actually the quark-level assumption $P_\perp.q_\perp$ = 0 has, so to say, got 
transferred, via the ${\hat q}$-integration in eq.(5.9), to the {\it {hadron
level}}, as evidenced from the ${\hat k}$-dependence of $F(k^2)$; therefore
an obvious logical inference is to suppose this ${\hat k}$-dependence  
to be the result of the  collinearity ansatz $P_\perp.P_\perp'$ = 0 at the
hadron level. Now, under the collinearity condition, one has
\begin{equation}
P.P' = P_\perp.P_\perp'+ P.n P'.{\tilde n} + P'.n P.{\tilde n}
 = P.n P_n'+ P'.n P_n ; \quad P.{\tilde n} \equiv P_n.
\end{equation}
Therefore `Lorentz completion'(the opposite of the collinearity ansatz) 
merely amounts to reversing the direction of the above equation by supplying
the (zero term) $P_\perp.P_\perp'$ to a 3-scalar product to render it a 
4-scalar! Indeed the process is quite unique for 3-point functions such as 
the form factor under study, although for more involved cases (e.g., 
4-point functions), further assumptions may be needed. 
\par
        In the present case, the prescription of Lorentz completion is 
relatively simple, being already contained in eq.(5.10). Thus since
$P,P'$= ${\bar P} \pm k/2$, a simple application of (5.10) gives 
\begin{equation}
k.n k_n = +k^2; \quad {\bar P}.n {\bar P}_n = -M^2 -k^2/4;
\quad {\hat k}^2= {{4k^2} \over {4M^2+k^2}}= 4\theta_k-4=4-4\eta_k 
\end{equation}    
This simple prescription for ${\hat k}$ automatically ensures the 4D 
(Lorentz) invariance of $F(k^2)$ at the hadron level. (It may be instructive
to compare this to the Cov. LF prescription [38] of `recognizing' the $n$-
dependent terms (unphysical) of $F(k^2)$ and then dropping them). For
more involved amplitudes (e.g., 4-point functions) too, this prescription
works fairly unambiguously, if their diagrams can be analyzed in terms of
more elementary 3-point vertices (which is often possible). We hasten to add 
however that strictly speaking, a `Lorentz completion' goes beyond the 
original premises of restricting the (pairwise $qq$) interaction to the 
covariant null-plane (in accordance with {\bf MYTP} [15]),
but such `analytic continuations' are not unwarranted, since in Cov. LF 
theories too [38], implementing the angular condition [35d] involves the 
introduction of `derivative' terms, implying a tacit enlargement of the 
Hilbert space beyond the null-plane (see Chap 2 of [38]).   

\subsection{QED Gauge Corrections to $F(k^2)$} 
\par
        While the `kinematic' gauge invariance of $F(k^2)$ has already been 
ensured in Sec.5.1 above, there are additional contributions to the
triangle loops - figs.1a and 1b of [33c] - obtained by  inserting the photon 
lines at each of the two vertex blobs instead of on the quark lines themselves.
These terms arise from the demands of QED gauge invariance, as pointed out 
by Kisslinger and Li (KL) [63] in the context of two-point functions, and are
simulated by inserting exponential phase integrals with the e.m. currents.
However, this method (which works ideally for {\it point} interactions) is 
not amenable to {\it extended} (momentum-dependent) vertex functions, and
an alternative strategy is needed, which is described below.     
\par
        The way to an effective QED gauge invariance lies in the simple-minded 
substitution $p_i-e_iA(x_i)$ for each 4-momentum $p_i$ (in a mixed $p,x$ 
representation) occurring in the structure of the vertex function. This 
amounts to replacing each ${\hat q}_\mu$ occurring in $\Gamma({\hat q})$ = 
$D({\hat q})\phi({\hat q})$, by ${\hat q}_\mu -e_q {\hat A}_\mu$, where 
$e_q = {\hat m}_2 e_1 -{\hat m}_1 e_2$, and keeping only first order terms
in $A_\mu$ after due expansion. Now the first order correction to 
${\hat q}^2$ is $-e_q{\hat q}.{\hat A}- e_q{\hat A}.{\hat q}$, which 
simplifies on substitution from  eq.(5.11) to 
\begin{equation}
-2e_q \tilde q.A \equiv -2e_q A_\mu [{\hat q}_\mu -{\hat q}.n {\tilde n}_\mu 
+ P.{\tilde n}{\hat q}.n n_\mu /P.n]
\end{equation}    
The net result is a first order correction to $\Gamma({\hat q})$ of amount 
$e_q j({\hat q}).A$ where  
\begin{equation}
j({\hat q})_\mu = -4M_>{\tilde q}_\mu\phi({\hat q}) (1-({\hat q}^2 -
\lambda/{4M^2})/{2\beta^2})
\end{equation}
The contribution to the P-meson form factor from this hadron-quark-photon 
vertex (4-point) now gives the QED gauge correction to the triangle loops, 
figs.(1a,1b) of [33c], to the main term $F(k^2)$, eq.(5.1), of an amount
which, after a simple trace evaluation (and anticipating the vanishing 
of all $\Delta$-terms remaining in the trace, as a result of contour 
integration over $q_n$)  simplifies to ($\phi=\phi({\hat q})$, etc)
\begin{equation}
F_1(k^2) = 4(2\pi)^4 N_H^2 e_q {\hat m}_1 M_>^2
\int d^4q  
(M^2-{\delta m}^2)\phi\phi'[{{D_n'{\tilde q}.{\bar P}} \over 
{\Delta_1' \Delta_2 P'.n}} + {{D_n {\tilde q}'.{\bar P}} \over 
{\Delta_1 \Delta_2 P.n}}]+ [1 \Rightarrow 2];    
\end{equation}
In writing down this term, the proportionality of the current to 
$2{\bar P}_\mu$ has been incorporated on both sides, on identical lines 
to that of (5.1), using results from (5.2-5.7) as well as from Appendix A.
Note that $e_q$ is {\it antisymmetric} in `1' and `2', signifying a change
of sign when the second term $[1 \Rightarrow 2]$ is added to the first. 
The term ${\tilde q}.{\bar P}/{\bar P}^2$ simplifies to
$2q.n (1-{\hat k}/2)/P.n$, after extracting the proportionality to 
${\bar P}_\mu$. Next, after the pole integrations over $q_n, q_n'$ in 
accordance with (5.7), it is useful to club together the results of photon 
insertions on {\it both} blobs for either index (`1' or `2'); this step 
generates two independent combinations for the `1' terms (and similarly 
for `2' terms): 
\begin{equation} 
A_n = q.n(1-{\hat k}/2); \quad 
B_n = q.n(1-{\hat k}/2)({\hat q}^2 -\lambda/{4M_>^2})/{2\beta^2} 
\end{equation}
Collecting all these contributions the result of $q_n$- integration is
\begin{equation}
F_1(k^2)= 8(2\pi)^3 N_H^2 e_q {\hat m}_1 M_>^2 \int d^3{\hat q}   
(M^2-{\delta m}^2)\phi\phi'[{{A_n +A_n'-B_n-B_n'} \over {\eta_k \times 
({\bar P}.n)^2}}] + [1 \Rightarrow 2]
\end{equation} 
The rest of the calculation is routine and follows closely the steps of
Appendix A for the (main) $F(k^2)$ term, including the translation 
$z_2 \rightarrow z_2 + {\hat m}_2 {\hat k}^2/{2\theta_k}$, and is omitted 
for brevity. The final result for $F_1(k^2)$ is
\begin{equation}
F_1(k^2) = -e_q{\hat m}_1{\hat m}_2 (3\eta_k+{\hat k}^2)[{{(M_>^2-{\delta m}^2)
\eta_k{\hat k}^4}(M_> {\hat m}_2)^2
\over {8G(0) \theta_k^{7/2} \beta^2}}]  + [1 \Rightarrow 2]
\end{equation}  
where we have dropped some terms which vanish on including the 
$[1 \Rightarrow 2]$ terms, noting the $(1,2)$ antisymmetry of $e_q$.
 
\subsection{Large and Small $k^2$ Limits of $F(k^2)$} 
\par
        We close this section with the large and small $k^2$ limits of 
the form factors $F(k^2)$ and $F_1(k^2)$. For large $k^2$, eq.(5.11) gives 
${\hat k}$= 2, $\theta_k$ = 2, and $\eta_k$ = $4M^2/k^2$, so that
\begin{equation}
F(k^2)= 2M N_H^2 (2\pi)^3 {\hat m}_1 {{4M^2} \over k^2} (\pi \beta^2/2)^{3/2}
G(\inf) exp[-{(M{\hat m}_2/\beta)^2/2}]   +[1 \Rightarrow 2] 
\end{equation}  
where, from eqs.(A.11-12), 
\begin{equation}
G(\inf) = (1+{\delta m}^2/M^2)(\beta^2- \lambda/{4M^2}+M^2{\hat m}_2^2) 
+(M^2-{\delta m}^2){\hat m}_2 -2{\hat m}_2^2 M^2 
\end{equation}  
Similarly from eq.(5.14), the large $k^2$ limit of $F_1(k^2)$ is
\begin{equation}
F_1(k^2) = 2{\sqrt 2}M^2k^{-2} e_q{\hat m}_1{\hat m}_2 (M_>^2-{\delta m}^2)
[{{M_>^2({\hat m}_1-{\hat m}_2)} \over {\beta^2 G(0)}}] 
\end{equation}
where we have taken account of the $(1,2)$ antisymmetry of $e_q$ in 
simplifying the effect of the $[1 \Rightarrow 2]$ term on the RHS. As a 
check, both $F(k^2)$ and $F_1(k^2)$ are seen to satisfy the `scaling' 
requirement of a $k^{-2}$ variation for large $k^2$. This result can be 
traced to the input dynamics of the (non-perturbative) gluonic interaction, 
eq.(4.19), on the structure of the vertex function, eq.(4.32). Perturbative 
QCD of course gives a $k^{-2}$ behaviour [61]. The covariant NP(LF) approach  
[38] also gives a similar behaviour, but extracted in a somewhat 
different way from the present `Lorentz completion' treatment. Note that
for the pion case the QED gauge correction term $F_1(k^2)$ gives zero
contribution in the large $k^2$ limit.   
\par        
        For small $k^2$, on the other hand, we have from eq.(5.11)
\begin{equation}
{\hat k}^2 = k^2/M^2; \quad [\theta_k, \eta_k] = 1 \pm  k^2/{4M^2} 
\end{equation}          
In this limit, the form factor, after substituting for $N_H$ from (5.8), 
and summing over the `1' and `2' terms, works out as
\begin{equation}   
F(k^2)= (1-3k^2/8M^2)[1- {\hat m}_1{\hat m}_2({{k^2} \over {4\beta^2}}- 
{{k^2{\delta m}^2} \over {M^2 G(0)}}) -{{3 k^2 \beta^2 (1+{\delta m}^2/M^2)}
 \over {8M^2G(0)}}]
\end{equation} 
where $G(0)$ is formally given by eq.(A.10), except for the replacement of
${\hat q}^2$ by $3\beta^2/2$. As a check, $F(k^2)$ is symmetrical in `1' 
`2', as well as satisfies the consistency condition $F(0)$ = 1. Similarly
the small $k^2$ value of $F_1(k^2)$, after taking account of the $(1,2)$
antisymmetry of $e_q$, is of minimum order $k^4$, so that it contributes 
neither to the normalization ($F_1(0)=0$), nor to the P-meson radius.    
\par
        For completeness we record some numerical results for large and
small $k^2$ limits. For the pion case, in the large $k^2$ limit, 
eqs.(5.12-13) yield after a little simplification the simple result  
\begin{equation}
F(k^2) = C/k^2; \quad C \equiv  2{\sqrt 2}{M_>^2 \over G(0)} (\beta^2+m_q^2) 
e^{-M_>^2/{8\beta^2}} 
\end{equation}
where $m_q= 265 MeV$ stands for $m_1=m_2$; and $M_>$ stands for the 
bigger of $m_1+m_2$ and $M$. 
Substituting for $\beta^2$=$0.0603 GeV^2$ [32] and $G(0)$=$0.166 GeV^2$, 
yields the result $C$=$0.35 GeV^2$, vs the expt value of $0.50 \pm 0.10$ 
[58b]. For comparison, we also list the perturbative QCD value [60] of 
$8\pi\alpha_s {f_\pi}^2$ = $0.296 GeV^2$, with $f_\pi$=$133MeV$, and the
argument $Q^2$ of $\alpha_s$ taken as $M_>^2$. 
\par
        For low $k^2$, eqs.(5.14-15) yield values of the pion and kaon
radii, in accordance with the relation $<R^2>$ = $-{\nabla_k}^2F(k^2)$ in
the $k^2$=$0$ limit. Substitution of numerical values from (4.21-22) yields
\begin{equation}
R_K = 0.629 fm (vs: 0.53- expt [58a]); \quad 
R_\pi = 0.661 fm (vs: 0.656- expt [58a])
\end{equation}
\par
        We end this Section with the remark that a simple-minded, 
conventional NP approach [33, 40] to BS dynamics had already produced 
most of the results of this form factor calculation, but had been  
critized [34] on grounds of `non-covariance'. The CNPA with an explicit 
formulation of the Transversality Principle (TP) [10-8] on a {\it covariant} 
null plane (NP), hopefully, keeps both the advantages, since the 4D loop 
integrals are now not only perfectly well-defined, but a major part of the 
$n_\mu$ dependence has got eliminated in the process of ${\hat q}$ 
integration, while the remaining NP orientation dependence has been 
transferred to the external (hadron) 4-momenta. In this regard the present 
approach is already in the company of a wider NP(LF) community [37-38] which 
has also to contend with some $n_\mu$ dependence. The solution offered 
here to overcome this problem is a simple prescription of `Lorentz completion'
wherein a `collinear frame' ansatz $P_\perp.q_\perp = 0$ is lifted on the 
external hadron momenta $P,P'$ etc, {\it after} doing the internal ${\hat q}$
integration, so as to yield an explicitly Lorentz-invariant result. The 
prescription, though different from other LF approaches [38], is nevertheless 
self-consistent, at least for 3-point hadron vertices, (and amenable to 
extension to higher-point vertices provided the latter can be expressed as a 
combination of simpler 3-point vertices). (It may be added parenthetically
that the old-fashioned NP treatment [33c] had yielded a slightly better curve
for the pion form factor, but this was due to the use of the ``half-off-
shell'' form of the NP wave function [40], which however did not come out
naturally from the present `covariant' treatment).      

\section{Three-Hadron Couplings Via Triangle-Loops}

For a large class of hadronic processes like $H \rightarrow H'+H''$ and
$H \rightarrow H'+\gamma$, the quark triangle loop [31] represents the 
lowest order ``tree'' diagram for their evaluation. Criss-cross gluonic 
exchanges inside the triangle (see Fig.1 of [31] ) are not important for
this kind of description in which the hadron-quark vertices, as well as 
the quark propagators are {\it {both non-perturbative}}, and thus take up
a lion's share of non-perturbative effects. This is somewhat similar to 
the ``dynamical perturbation theory'' of Pagels-Stokar [64]  in which such
criss-cross diagrams are neglected. 
\par
        In this Section we shall give an outline of the calculational 
techniques for such diagrams for the most general case of unequal mass
kinematics $m_1 \neq m_2 \neq m_3$, but with spinless quarks only,
since the `spin' d.o.f. does not introduce any new singularities over
the spin-0 case. In this we shall closely follow the method of ref.[31], 
which is a 3-hadron generalization of Sec.5 for the e.m. form factor of a 
pseudoscalar meson. However, as already noted therein, the CIA form [16] 
of 3D-4D BSE is fraught with problems of ill-defined integrals (and hence
complexity of amplitudes) due to the presence of time-like momentum 
components [25] in the (gaussian) wave functions of the participating hadrons.
So we shall work only with CNPA [41] structures, as derived in Sec.4.2.
\setcounter{equation}{0}
\renewcommand{\theequation}{6.\arabic{equation}}

\subsection{Kinematical Preliminaries}

According to Fig.1 of ref [31], and in the same notation, the 3 hadrons with
all incoming 4-momenta $P_i$, with masses $M_i$, interact via the quark 
triangle loop wherein $P_k$ dissociates into the quark pair with 4-momenta 
$(-p_i, p_j)$ and masses $(m_i, m_j)$ respectively, so that 
$P_k = -p_i + p_j$, and $P_k+P_i+P_j=0$. Thus [31]:
\begin{equation}
-P_k= P_i+P_j \equiv P_{ij}; \quad P_k^2= -M_k^2; \quad P_k=-P_{ij}=p_j-p_i
\end{equation} 
where $(i,j,k)$ are cyclic permutations of $1,2,3)$. Similarly the relative  
4-momenta $q_k$ between quarks $i,j$ corresponding to the break-up 
$P_k = p_j-p_i$, and $Q_k$ between hadrons $i,j$ for the break-up 
$P_k = -P_i- P_j$ are:
\begin{equation}
q_k = {\hat \mu}_{ij} p_j + {\hat \mu}_{ji} p_i; \quad 
Q_k = {\hat m}_{ij} P_j - {\hat m}_{ji} P_i; \quad -P_k =P_i+P_j
\end{equation}
The fractional momenta ${\hat \mu}_{ij}$ at the quark level, and 
${\hat m}_{ij}$ at the hadron level, are given by the Wightman-Gaerding 
[56] definitions
\begin{equation}
2{\hat \mu}_{ij} = 1 + {{m_i^2-m_j^2} \over {M_k^2}}; \quad 
2{\hat m}_{ij} = 1 + {{M_i^2-M_j^2} \over {M_k^2}}
\end{equation}    
The relative signs are determined by the phase convention of Fig.1 of [31]. 
\par
        Now to define the `hatted' relative 3-momenta ${\hat q}_k$ and 
${\hat Q}_k$, we must follow the CNPA procedure [41] instead of CIA [31].
Further, since the content of CNPA is for all practical purposes identical
with that of the old-fashioned NPA [40], considerable simplification is
achieved by adopting the latter notation [40], which is what is already done
in [31], albeit with CIA  content. Indeed, with  the collinear ansatz (Sec.4)
the NPA values of ${\hat q}_i$ are simply $q_{i\perp}, q_{i3}$, where 
$q_{i3} = M_i q_{i+}/P_{i+}$, etc [40]; and $Q_{i3} = M_i Q_{i+}/P_{i+}$.
However, since the $q_i$'s are not all independent, it is necessary to take
a basis momentum (say $p_2$) in terms of which to express others. Now in a
fixed $p_i$ basis, we have
\begin{equation}
q_k= p_i+{\hat \mu}_{ij}P_k; \quad q_j= p_j - {\hat \mu}_{ik}P_j; \quad
q_i = p_i - {\hat \mu}_{jk}P_j + {\hat \mu}_{kj}P_k
\end{equation}
For later purposes we shall consider a $p_2$ basis, for which
\begin{equation}
q_1=p_2+{\hat \mu}_{23}P_1; \quad q_3=p_2-{\hat \mu}_{21}P_3
\end{equation}
We also record some useful results for the kinematics of external particles, 
if they are on-shell $(Q_i.P_i=0)$, under the collinearity condition [31]:
\begin{equation}
P_{i\pm}= \pm {M_i \over Q_i}Q_{i\pm}; \quad M_i^2 = P_{i+}P_{i+}; \quad
Q_i^2=-Q_{i+}Q_{i-} = {{\lambda(M_1^2,M_2^2,M_3^2)} \over {4M_i^2}}
\end{equation}
which lead to the further symmetry relations
\begin{equation}
Q_1M_1=Q_2M_2=Q_3M_3= {\sqrt {\lambda(M_1^2,M_2^2,M_3^2)/4}}
\end{equation}
Further we can define a $3 \times 3$ matrix structure $n_{ij}$, with
$n_{ij} \equiv P_{i+}/P_{j+}$, which satisfy the relations
\begin{equation}
n_{ij}n_{ji}=1; \quad n_{12}n_{23}n_{31}=1; \quad \Sigma_i n_{i1}=0
\end{equation}
showing a definite phase relation among these quantities, which are more
explicitly expressed by the matrix structure
\begin{eqnarray}
[n_{ij}] &=&    1;  -{\hat m}_{13} \pm Q_2/M_2; -{\hat m}_{12} \mp Q_3/M_3 \\       
         & & -{\hat m}_{23} \mp Q_1/M_1;    1; -{\hat m}_{21} \pm Q_3/M_3 \\
         & & -{\hat m}_{32} \pm Q_1/M_1; -{\hat m}_{31} \mp Q_2/M_2;    1
\end{eqnarray}  
with a two-fold sign ambiguity  expressed by the statement that only the
upper, or only the lower signs, must be taken. It is easily verified that
eqs (6.8) (hence phase relations) are satisfied by the matrix (6.9). 
                   
\subsection{Structure of $HHH$ Form Factor}

The full structure of the 3-hadron amplitude due to Fig.1 of [31] is
\begin{equation}
A(3H)= {{2i} \over {\sqrt 3}} (2\pi)^8 \int d^4p_i 
\Pi_{123}{{\Gamma_i({\hat q}_i)} \over {\Delta_i(p_i)}} 
\end{equation}
exhibiting cyclic symmetry, where the normalized vertex function $\Gamma_i$ 
in CNPA [41] is given from Sec.4-5 as
\begin{equation}
\Gamma_i({\hat q}_i)= N_i (2\pi)^{-5/2}
D_i({\hat q}_i)\phi_i({\hat q}_i); \quad D_i= 2M_i({\hat q}_i^2
-{{\lambda(M_i^2, m_j^2,m_k^2)} \over {4M_i^2}})
\end{equation}
where we have defined the `reduced' denominator function $D_i$ as 
$D_{i+}M_i/P_{i+}$ and written the (invariant) normalizer $N_{iH}$ as $N_i$.  
The color factor and the effect of reversing the loop direction are given by
$2/{\sqrt 3}$, while $(2\pi)^8$ is the overall BS normalizer [40]. 
$\Delta_i$ = $m_i^2+p_i^2$ =$\omega_{i\perp}^2 -p_{i+}p_{i-}$. Spin and 
flavour d.o.f. will give rise to a standard `trace' factor [31] $[TR]$
which is skipped here for simplicity.
\par
        To evaluate (6.10), we first write the cyclically invariant measure:
\begin{equation}
d^4p_i = d^p_{\perp} {1 \over 2} d(x_i^2) M_i^2 dy_i; \quad 
x_i = p_{i+}/P_{i+}; \quad y_i = p_{i-}/P_{i-}
\end{equation}   
The cyclic invariance of this quantity ensures that it is enough to take any
index, say $2$, and first do the pole integration w.r.t. the $y_2$ variable
which has a pole at $y_2=\xi_2$$ \equiv $$\omega_{2\perp}^2/(M_2^2 x_2)$.
The process can be repeated, by turn, over all the indices and the results
added. Note that the $\phi$-functions do {\it not} include the time-like
$y_i$ variables under CNPA [41], so that the residues from the poles arise
from only the propagators. The crucial thing to note is that the denominator
functions $D_1$ and $D_3$ sitting at the opposite ends of the $p_2$-line in
Fig.1 of [31] will {\it {cancel out}} the residues from the complementary 
(inverse) propagators $\Delta_3$ and $\Delta_1$ respectively. Indeed by
substituting the pole value $y_2=\xi_2$, in $\Delta_{1,3}$, the corresponding
residues in an obvious notation work out as:
\begin{equation}
\Delta_{1;2}=\xi_2 n_{32}M_2^2 + x_2 n_{23}M_3^2 - 2{\hat \mu}_{21}M_3^2; \\\\       
\Delta_{3;2}=-\xi_2 n_{12}M_2^2 - x_2 n_{21}M_1^2 - 2{\hat \mu}_{23}M_1^2
\end{equation}
It is then found, with a short calculation using (6.5), that
\begin{equation}
{{D_3({\hat q}_3)} \over {\Delta_{1;2}}} = 2M_3x_2n_{23};\quad
{{D_1({\hat q}_1)} \over {\Delta_{3;2}}} = 2M_1x_2 n_{21} 
\end{equation}
which shows the precise cancellation mechanism between the $D_i$-functions 
and the residues of the propagators $\Delta_i$ at the $\Delta_2$ pole. This
mechanism thus eliminates [16] the (overlapping) Landau-Cutkowsky poles that 
would otherwise have caused free propagation of quarks in the loops. 
The same procedure is then repeated cyclically for the other two terms 
arising from the $\Delta_{3,1}$ poles. Collecting the factors, the result
of all the 3 contributions is compactly expressible as (c.f. [31]):
\begin{equation}
A(3H)= 8{\sqrt {{2\pi} \over 3}} \Sigma_{123} \int \int M_2 {n_{23} n_{21}}
\pi^2 dx_2 d\xi_2 x_2^2 [TR]_2 D_{2}({\hat q}_2) {\Pi_{123}{M_iN_i\phi_i}}   
\end{equation}
where the limits of integration for both variables are 
$-\inf<(\xi_2,x_2)<+\inf$, since these are governed, not by the on-shell 
dynamics of standard LF methods [37-38], but by off-shell 3D-4D BSE.  
The difference from [31] (under CIA [16]) arises from using CNPA [41] here. 

\subsection{Discussion on Applicability}

Eq.(6.14) is the central result of this Section. Its general nature stems
from the use of unequal mass kinematics at both the quark and hadron levels,
which greatly enhances its applicability to a wide class of problems which
involve 3-hadron couplings, either as complete process by themselves (such as
in decay processes) or as part of bigger diagrams in which 3-hadron couplings
serve as basic building blocks. What makes the formula particularly useful
for general applications is its explicit Lorentz invariance which has been 
achieved through the simple method of `Lorentz Completion' on the lines of 
Sec.5 for the e.m. form factor of P-mesons (pion). 
\par
        How much of this derivation is model independent, except for the 
use of the {\bf MYTP} [15] ? The answer lies in the structure 
$\Gamma_H$ = $D \times \phi$ for the hadron-quark vertex function, which is
a direct consequence of the 3D support ansatz which in turn receives support
from several angles [17-19], although this specific form [16] does not seem 
to have been used elsewhere. Its factorable structure in which the 
denominator function $D$ is quite universal and depends only on kinematics,
has helped reduce the 4D loop-integral to a 3D form, and in so doing, has
succeeded in eliminating the Landau-Cutkowsky (overlapping) singularities
in a very simple and transparent manner, thus preventing the free propagation 
of quarks in the intermediate (loop) stages. (The only model dependent entity
in the 3D wave function $\phi$, but it has been related to the (observable)
spectra [13]).     
\par
        In the spirit of this generality, this article was not intended for
specific allpications per se, but some possibilities are readily listed. The
simplest class is that of strong decay of a resonance ($H_1$) into two 
lighter hadrons ($H_1,H_2$) under kinematically allowed conditions (whose
signature is carried by the external variables $Q_i,M_i$ inside the integral
(6.14)). The amplitude $A(3H)$ can also be adapted, via Sce.5, to include 
e.m. or semi-leptonic processes, expressed by $H_1 \rightarrow H_2 + \gamma$,
(where $\gamma$ is real or virtual), where the signature of virtuality is 
carried by $M_3^2 \equiv t$. Non-leptonic weak decays (see Fig.2 of [31])
are also amenable to this treatment. As an example one may cite the 
experimental discrepancy [65] of the vector form factors in the semi-leptonic
process $D \rightarrow K^*e{\bar \nu}$ with theoretical models that prefer 
to represent the intermediate states through effective meson propagators [66].
On the other hand, a crucial role of the appropriate quark-triangle, with a
considerable effect of the unequal masses of the participating quarks, seemed
to be strongly indicated in resolving the discrepancy [67]. Other 
applications include the so-called Sullivan process [68], details of which the 
interested reader may find in [69].

\section{Two-Quark Loops: SU(2)-Breaking Problems}

To illustrate other applications of the 3D-4D BSE formalism, we now turn to 
the (simpler) problem of two-quark loops which are useful for estimating
SU(2)-breaking effects in phenomena like i) mass-splittings in P-mesons 
[18a], and ii) $\rho-\omega$ mixing in meson-exchange forces [18b]. Simpler
2-quark loops, such as those involved in the weak and e.m. decay constants
of hadrons, are already available in previous studies of this formalism [40],
and  will not be the subject of this semi-review. Further, its scope does 
not include detailed analysis of these 2-loop phenomena [18], but only 
their essential physics, and a quick derivation of their core structures, 
leaving the reader to ref [18]  for numerical results plus more references.
               
\setcounter{equation}{0}
\renewcommand{\theequation}{7.\arabic{equation}}

\subsection{Strong SU(2)-Breaking in P-meson Multiplets}
 
To recapitulate the essential physics of hadronic mass splittings within 
SU(2)multiplets $(I= 0.5, 1.0)$, these were for long thought to be of e.m. 
origin, until the advent of QCD [22] when the possibility of strong 
breaking of SU(2)  due  to  the  intrinsic  $u-d$  mass  difference 
started being taken  seriously. (This  was  despite   the  prior 
existence of the GMOR- mechanism [70] which had sought  to  relate 
the pseudoscalar masses to  the  current  quark  masses  and  the 
vacuum condensates ). In this respect the trend was  set  largely 
by Weinberg's analysis [71], characterized by the `Weinberg ratios'     
$m_u/m_d$ = $0.55$, and $m_s/m_{ud}$ = $20.1$, confirmed by a recent 
analysis [72]. A conservative estimate of the u-d mass difference 
is believed to be $d-u$ = $3-4 MeV$ [71-72].  On the other hand 
the absolute values of the current masses are not as well  known, but  
the SU(2) mass splittings [13] among the known  pseudoascalar multiplets
$(\pi,K,D,B)$ is a useful mathematical laboratory to determine the $d-u$ 
mass difference from the corresponding `constituent mass' difference,
 via Politzer additivity [55]. The issue is basically a dynamical one 
(in view of the sensitive nature of this labiratory), necessitating  a 
high degree of parametric control on the strong vertex functions involved 
in the concerned Feynmam diagrams (Figs.1(a,b,c) of ref.[18a]).  
The problem clearly goes beyond mere additivity in the quark masses, as 
the observed pattern of mass splittings [13] seems to suggest a 
basically  decreasing trend  from  the  lightest  (pion)  to  the 
heaviest (beauty) flavour, tapering off almost to zero for the 
$B_0 -B_+$  mass difference.
\par
        For the meson self-energy, there are 3 basic contributions,
a la Figs.1(a,b,c) of [18a]: i) In Fig.1a, a 2-point ${\delta m_{ud}}$ vertex
inserted at each propagator by turn represents the principal source of
strong SU(2) breaking; ii) Fig.1b simulates the e.m. breaking effect by 
joining the two quark lines internally by a photon propagator; iii) Finally 
Fig.1c simulates the effect of the difference of the quark condensates 
$<u{\bar u}>$ and $<d{\bar d}>$ on the strong SU(2) breaking of hadron 
masses. Figs.1(a,c) are one loop diagrams, while Fig.1b represents a 
{\it two-loop} process, which is moreover sensitive to QED gauge 
constraints [63], as in the e.m. form factor case (see Sec.5). 
\par
        Using the dynamical framework collected in Sec.4.(2-3), it is
fairly straightforward to write down the integrals accruing from these
diagrams. Both CIA [16] and CNPA [41] are valid mechanisms for evaluating
these diagrams, but in view of a prior exposure [18a] of CIA for this
problem, it may be instructive to adopt the CNPA alternative here. Thus
Fig.1a gives in terms of the results of the previous sections,
\begin{equation}    
\Pi_a(M^2)= i(2\pi)^{-1} N_H^2 \int d^4q D^2({\hat q}) \phi^2({\hat q})
Tr [\gamma_5 S_F({\hat m}_1P+ q) \gamma_5 S_F(-{\hat m}_2 P+q)]
\end{equation}
where we have used the representation of the normalized vertex function
given in eq.(6.11), and $D$ is the reduced denominator fn in CNPA.   
After evaluating the traces this expression simplifies to
\begin{equation}
\Pi_a(M^2) =-2i(2\pi)^{-1} N_H^2 \int d^4 q D^2({\hat q}) \phi^2({\hat q}) 
{{\Delta_1+\Delta_2 +M^2-{\delta m}^2} \over {\Delta_1 \Delta_2}}
\end{equation}
where  ${\delta m}= m_1 - m_2$, and all kinematical quantities are as
defined in Sec.4 The integration over the time-like component $q_-$ is
carried  out very simply, using the result (5.7). Note that the vertex 
function $D \times \phi$ does not involve this variable, and  also the 
$D$-fn exactly cancels out the residues arising from the propagators, as
shown generally in Sec.6. The  resultant 3D integration over $d^3{\hat q}$
is expressible simply as 
\begin{equation}
\Pi_a(M^2) = 2 N_H^2 \int d^3 {\hat q} \phi^2({\hat q})D({\hat q}) 
[{D({\hat q}) \over {4Mx_2}}+{D({\hat q}) \over {4Mx_1}}+M^2-{\delta m}^2]  
\end{equation}                            
where $x_i$= $p_{i+}/P_+$ = ${\hat m}_i \pm x$ for i=1,2 respectively. The 
third component  $q_3$ of CNPA [40-41] is simply $Mx$, so that ${\hat q}$ =
$q_{\perp}, q_3$. The normalizer $N_H$ is given by eq.(5.8). The parallel
CIA result is [18a]
\begin{equation}
\Pi_a(M^2) =2 N_H^2 \int d^3{\hat q}[D^2({\hat q})({1 \over {2\omega_1}} +
{1 \over {2\omega_2}}) +D(({\hat q})(M_2 - {\delta m}^2)]
\end{equation}  
\par 
        A comparison between the CNPA and CIA forms of $\Pi_a$ is now in
order. In CIA, eq.(7.4), the $D^2$ term is well defined and is amenable to
simple quardature. On the other hand, the CNPA form, eq.(7.3), encounters
singularities at $x_{1,2}=0$, on integration w.r.t. $x$, taking account of 
the relations $x_{1,2}$=${\hat m}_{1,2}-x$, and ${\hat q}^2$ = 
$q_{\perp}^2 + M^2 x^2$. The final results are quite similar for both cases.            
\par
        The formulae (7.3-4) for $\Pi_a$ and (5.8) for $N_H$, show explicit  
dependence  on the masses $m_{1,2}$, and facilitate the evaluation of mass 
splittings within the SU(2) isospin multiplets as follows:  For the 
$K,D,B$ mesons, take $m_2$  as the mass of the $ud$-quark  with $m_1>m_2$,
and  while  differentiating w.r.t. $m_2$, consider the increment $\delta_c$.
A little reflection then shows (by virtue of the Politzer [55]  additivity  
relation) that this quantity  may  be directly identified with the difference 
$m_{dc} - m_{uc}$  between the current $d$- and $u$ -masses provided  the 
hadron mass with the  $u$-quark  gets subtracted from that with the d-quark 
(e.g.,$K_0 - K_-$ , etc). Of course the normal rules of  differentiation  
apply, viz., ${\delta f(m_2)}$ = $ f'(m_2) \delta_c$, where the  argument  of 
$f'(m_2)$  must  use the average `constituent mass' of $ud$-quark, viz.,  
$265 MeV$, eq(4.22). 
\par
        For the pion case, some extra care is necessary  since  
both the constituents are now $u/d$-quarks so that both $m_1$ and $m_2$
must be subjected to differentiation in turn . On the other hand these two 
contributions come with just  equal but opposite signs, so that they cancel 
out exactly, giving a net vanishing contribution, as seen more directly from 
the fact that the  ${\Delta I}= 1$ field $u{\bar u} - d{\bar d}$ in the 
Lagrangian cannot contribute to $\pi_+ - \pi_0$  anyway. For the details
of numerical results on ${\delta {\Pi(M^2)}}$, see [18a]. 

\subsection{E.M. Contribution to Self-Energy}

The e.m. contribution to the hadron self-energy is given by a 2-loop 
diagram (Fig 1b of [18a]) in a slightly simplified notation as follows:
\begin{equation} 
\Pi_b (M^2) = N_H^2 e^2 e_1 e_2 \int \int d^4 q d^4 q' 
{{D\phi D'\phi'} \over {(2\pi)^5 k^2}} Tr[\gamma_5 S_F(p_1)i\gamma_\mu
S_F(p_1')\gamma_5 S_F(-p_2')i\gamma_\mu S_F(-p_2)]
\end{equation} 
where $k$=$q-q'$=$p_1 -p_1'$=$p_2'-p_2$  is the exchanged quantum ; 
$e_1$ and $e_2$  are the charges of the quarks involved ( in  units  of 
$e$ ) and $q,q'$ are the internal 4-momenta of the LHS and RHS  
hadrons respectively. This integral involves simultaneous  (pole)  
integrations over the time-like components of $q$ and $q'$ which  do 
not figure in the respective vertex functions and therefore  can 
be carried out exactly. However the rest of the 3D integrations 
(two sets) do not quite factor out, so they  need  some  strategy   
before they can be  carried out without much  tears.To  that  end 
a simple device that suggests itself naturally is based  on   the 
following observation : By the very topology of the  diagram   it 
is  fairly  clear  that the time-like  components  of  both  the 
4-vectors $q$ and $q'$ are quantitatively similar,  so  that  their 
effects largely "cancel out" in the factor $k^{-2}$ in eq.(7.5). 
As a result the quantity $k$ =$q - q'$   effectively reduces to the 
space-like quantitty $({\hat q}-{\hat q'})^2$ which can be manipulated 
to desired numerical accuracy in the 3D integrations over ${\hat q}$ and 
${\hat q'}$. We list both CIA [18a] and CNPA (new) results in the form 
of 3D integrals in ${\hat {q,q'}}$ jointly as follows. 
\begin{equation}
\Pi_b(M^2) = 4N_H^2 e^2 e_1 e_2 (2\pi)^{-3} \int \int d^3{\hat q}d^3{hat q}'
[...] {{\phi({\hat q})\phi({\hat q}')} \over {({\hat q}-{\hat q}')^2}};
\end{equation}
the quantity $[...]$ is first listed for CIA as follows [18a]: 
\begin{eqnarray}
[...]_{CIA} &=& (M^2-{\delta m}^2)^2-2m_1m_2(M^2-m_{12}^2)-{\delta m}^2
({\hat q}-{\hat q}')^2+  \\   \nonumber 
            & &  ({1 \over {2m_1}}+{1 \over {2m_2}})  
({1 \over {2m_1}'}+{1 \over {2m_2}'}) D({\hat q})D({\hat q}')   \\ \nonumber
            & & + (M^2-m_1^2-2m_2^2+m_1m_2)D({\hat q})/\omega_2
+ (M^2-m_2^2-2m_1^2+m_1m_2)D({\hat q})/\omega_1;
\end{eqnarray}   
The dual quantity $[...]_{CNPA}$ may be simply read from the above merely
by the replacements $\omega_{1,2}$$ \rightarrow $$2(Mx_1,Mx_2)$ respectively,
where $x_{1,2}$=${\hat m}_{1,2} \pm x$. 
\par
        To convert the mass shifts from quadratic to linear, it is of
course necessary to divide both $\Pi_{a,b}$ in the avobe equations by
$2M$. In the e.m. case, no further `differentiation' w.r.t. $m_2$ is 
necessary, since (7.6) is already of second order in $e$. As regards the 
factor $e_1e_2$, its differential ${\delta (e_1e_2)}$ is easily found as    
$+(1/2)$, $-(1/3)$, $+(2/3)$ and $-(1/3)$ for the  differences $\pi_+-\pi_0$, 
$K^0-K^+$,$D^+-D^0$, and $B^0-B^-$ in {\it this} order .It turns out [18a]
that this alternating sign pattern is of  great  help in  reinforcing and 
reducing the net differences within the $K,D,B$ multiplets (after taking 
account of the strong breaking effects, Figs 1a and 1c), towards a good 
pattern of accord [18a] with the data [13]. (For results, see [18a]). 
\par
        We now  consider QED gauge corrections [63] to the e.m. value, 
eq.(7.6), arising from Fig.1b of [32a], on the lines of corresponding 
corrections to the e.m. form factor derived in Sec.5. This correction is 
sketched in Appendix B for P-mesons, using diagrams listed in ref.[63], and 
in their notation for the contributing figures. The resulting QED correction
for the kaon e.m. mass difference turns out to be nearly a $60$ percent 
increase over the CIA result $1.032 MeV$ [18a] arising from the main term 
(7.6). We omit the corresponding CNPA treatment for brevity.

\subsection{Effect of Quark Condensates}

Another source of mass splittings arises from  the difference between the 
$u/d$-quark condensates, in accordance with Fig.(1c) of [18a]. Indeed some
recent  calculations via QCD-sum rules have used this as the principal 
mechanism [73] for the mass splittings, with much less contribution from
Fig.(1a). Indeed the value of ${\delta <q{\bar q}>}$ in itselh has been the 
subject of separate investigations in chiral perturbation theory [74] as well 
as in QCD-sum rules [74]. On the other hand, the BSE-SDE formalism [11,23]
provides a `direct' ab initio estimate [11] of this condensate (as well as 
others [30]). 
\par
        To recapitulate the logic of the condensate calculation by the
`direct' method [11], in terms of the quark's {\it non-perturbative} mass 
function, $m(p)$, note that the latter is the chiral ($M_\pi =0$) limit
of the pion-quark vertex function $\Gamma({\hat q})$, given by eq.(4.35),
and must be used in the expression of the full propagator, $S_F(p)$, Sec.7,
which appears in the formal definition of the condensate as follows [11]: 
\begin{equation}
<{\bar q}q> =  {{iN_c N_f} \over {(2\pi)^4}} Tr [\int d^4 p S_F(p)]  \\
= - {3 \over {4{\pi}^3}} \int d^3{\hat p}{m({\hat p}) \over {\sqrt 
{{\hat p}^2 + m^2({\hat p})}}}
\end{equation}
after doing the pole-integration over the time-like component of $p_\mu$. 
Here $N_c=3$, and $N_f=1$ (since each separate flavour $(u/d)$ is counted).  
Now to evaluate the 3D integral (7.8), substitute the CIA structure (4.35)
for $m({\hat p})$, with $\phi({\hat p})$=$exp(-{\hat p}^2/{2\beta^2})$. This
integral formula has an analytic structure in terms of the constituent mass 
$m_q$ of the $u/d$-quark, so that it is now a matter of simple 
differentiation to give an explicit form of its increment w.r.t. 
${\delta m_q}$ which equals ${\delta_c}$.  The final formula is [18a]:
\begin{equation}
{\delta {<{\bar q}q>}} = {{-3{\delta c}} \over {\pi^2 m_q}} \int dk k^4
\phi_\pi(k) \omega(k) [1- {{2k^2} \over {m_q^2}}] (m^2(k)+k^2)^{-3/2}
\end{equation}                                                             
Using the inputs from (4.21-22) gives $\beta^2$ = $0.0603$, and the final
results under CIA are 
\begin{equation}
<{\bar q}q> = -(266 MeV)^3 ; \quad {\delta <{\bar q}q>} = + 0.0664 {\delta c}
\end{equation} 
These values are fully rooted in spectroscopy but are otherwise free from 
adjustable parameters, except for the quantity ${\delta_c}$. They have 
a fair overlap with QCD-SR determinations [76].  
\par
        For completeness we now give the condensate results under CNPA,
substituting the CNPA mass function (4.35) in (7.8). This gives
\begin{equation} 
<{\bar q}q> =  {{12i{\sqrt 2}} \over {(2\pi)^4}} \int d^3 {\hat p} dp_n 
{{p.n [1+{{\hat p}^2 \over {m_q^2}}] \phi({\hat p})} \over 
{m_q^2 +p_\perp^2 -2 p.n p_n}}
\end{equation} 
The integration over $p_n$ is trivial and yields
\begin{equation}
<{\bar q}q> ={{-3{\sqrt 2}} \over {(2\pi)^3}} \int d^3{\hat p}
[1+{{\hat p}^2 \over {m_q^2}}] \phi({\hat p}) = -(242 MeV)^3
\end{equation}
Substituting the gaussian form (as above) for $\phi$ and integrating, yields
an analytic structure useful for calculating ${\delta <{\bar q}q>}$:
\begin{equation}
<{\bar q}q> = -3{\sqrt 2}(\beta^2/{2\pi})^{3/2}[1+3\beta^2/m_q^2] 
= - (242 MeV)^3 
\end{equation}  
a value which seems to be even closer to the estimate $-(240)^3$ of 
QCD-SR [2] than the CIA result $-(266)^3$ of [18a].  
\par 

        As to the contribution of ${\delta <{\bar q}q>}$ to the strong SU(2)
mass splittings, a la Fig.1c of [18a], we skip the detailed derivation in 
favour of [18a], since it turns out to be rather small within this BSE 
framework. This is in sharp contrast to the QCD-SR findings [73] wherein
the condensate contribution seems to dominate. This is not too surprising  
since within a BSE-cum-SDE framework,  most  of  the  non-perturbative 
effects are already contained in the hadron-quark vertex function, 
with a correspondingly smaller role  for  the  condensates. On the other
hand in QCD-SR [2] these represent major non-perturbative effects 
when seen from the high energy perturbative QCD end. 
\par
        A few comments on the main results of this exercise are in order. 
The e.m. contributions alternate in sign in the mass splittings between the
charged and neutral components in the sequence $\pi,K,D,B$. The condensate
contribution to strong SU(2)-breaking being small, the sensitivity to the
$d-u$ mass difference comes almost entirely from Fig.1(a) of [18a]. Next,
the feature of unequal mass kinematics has played a big role in the 
formalism, being mainly responsible for a sustematic decrease in mass 
splittings as one goes up on the mass scale. This aspect has come about
mainly from the properties of the $D$-functions (mostly model independent) 
The numerical values show a good overall pattern of agreement with data [13],
(within less than half MeV), for the parameter $\delta_c$ in the range
$(3.5- 4.0) MeV $ for all the 3  cases $K,D,B$. This value of $\delta_c$ 
appears well within the phenomenological limits of acceptability [72]. 
However, as the results of Appendix B on QED gauge corrections indicate,
inclusion of these tends to {\it decrease} the effective value of $\delta_c$.  
Finally the calculational technique seems to conform to the spirit of 
`Dynamical Perturbation Theory' of  Pagels-Stokar [64] (neglect of 
`criss-cross' diagrams) which must be carefully distinguished from a naive 
interpration of  perturbative QCD. 

\subsection{Off-Shell $\rho-\omega$ Mixing}
        
Before concluding this Section, we shall briefly draw attention to  
a similar SU(2)-breaking phenomenon which has proved to be of considerable 
interest for the understanding of certain anomalies in nuclear forces [77]: 
off-shell $\rho-\omega$ mixing. Although nuclear topics are not of direct 
concern for this article, the basic logic of charge-symmetry-breaking (CSB) 
to explain the Nolen-Schiffer anomaly [77] via $\rho-\omega$ mixing [78], 
stimulated by new experiments [79] on polarized $n-p$ scattering, comes 
directly under the theme of this Section. Indeed, the sensitivity of 
$\rho-\omega$ mixing to the $d-u$ mass difference $\delta_c$, especially 
{\it off-shell} [78a,18b], is as strong as that of $P$-meson masses [18a].
\par
        To recall the basic logic, the small difference between the proton
vs neutron analyzing powers at an angle $\theta_0$ corresponding to the 
vanishing of the average analyzing power [79], is proportional to the CSB
potential $V_{CSB}$ whose contribution from $\rho-\omega$ mixing may be
schematically expressed as [78a]
\begin{equation}
V_{CSB}^{\rho-\omega} = <NN| H_{int}|NN\omega>G_0<\omega|H_{CSB}
|\rho^0>G_0<\rho^0NN|H_{int}|NN>+(\rho^0 \Leftrightarrow \omega)
\end{equation}
Here $G_0$ is the appropriate $V$-meson propagator, and 
$<\omega|H_{CSB}|\rho^0>$ gets its dominant theoretical contribution from
the $d-u$ mass difference $\delta_c$, with $H_{CSB}$ = 
$\rho.\omega {\delta_c}^2$, and a partial contribution from the e.m. chain
$\rho \Rightarrow \gamma \Rightarrow \omega$ via vector dominance and/or
2-quark loops. Alternatively, the matrix element can be estimated from the
experimental $e^+e^- \Rightarrow \pi^+\pi^-$ amplitude at the $\omega$-pole,
which gives the {\it on-shell} value $\theta(M^2)$ of the $\rho-\omega$ 
mixing amplitude [80]. On the other hand, it is its {\it off-shell} value 
$\theta(q^2)$ which is relevant to the CSB potential, eq.(7.14), for the 
$V$-meson exchange in a space-like region where its effect on $V_{CSB}$ has 
been claimed to be greatly suppressed [78a]. This question in turn requires 
a theoretical model for the necessary extrapolation which can be defined in
terms of a dimensionless parameter $\lambda$ as [18e]:
\begin{equation}
\theta(q^2) =\theta(M^2)[1 - (1+ q^2/M^2)\lambda]
\end{equation}          
\par
        A calculation of this parameter $\lambda$ is the central issue of
any investigation of the CSB effect, wherein its value has been variously 
estimated to be within the $(0-1)$ range [78]. In particular, the function 
$\theta(q^2)$ is also amenable to the 3D-4D formalism [16], using the self-
energy techniques [18b] outlined in this Section. Its on-shell value 
$\theta (M^2)$ [18b] agrees with the data [80], while the {\it off-shell} 
parameter $\lambda$ comes close to unity, signifying a change of sign for
$\theta(q^2)$ in the transition region between the space-like and time-like
momenta, in agreement with a `nucleonic' self-energy calculation [78d].     

\section{QCD Parameters from Hadron Spectroscopy}

In this Section we outline a simple method of calculation [30] of QCD 
Condensates in terms of the (spectroscopy-oriented) parameters of the 3D-4D 
BSE framework. These parameters of QCD simulate non-perturbative effects as 
coefficients in Wilson's operator product expansions ($OPE$) [81,55]. The 
method of QCD sum rules represented the first practical attempt [2a] to
relate these quark-gluon quantities to hadronic amplitudes by employing a 
duality principle [20] between the quark-gluon and meson-baryon pictures. 
Basically the idea is to find a $Q^2$ region ($\approx 1\, GeV^2$) where 
one may incorporate nonperturbative physics, generated via $OPE$ [81], into 
the perturbative QCD treatment of physical processes involving hadrons. 
The QCD-SR ansatz [2] for the evaluation of a certain correlation function 
$\Pi (p)$, is to replace the free quark (or gluon) propagator by one more
suitable for the nontrivial vacuum, and on the other hand to express, via 
dispersion relations, the same correlation function in terms of the 
variables of the meson-baryon picture. The two results are then equated to 
yield sum rules connecting the variables of the two physical descriptions.

\subsection{Field-Theoretic Definition of Condensates}
\setcounter{equation}{0}
\renewcommand{\theequation}{8.\arabic{equation}}

While QCD-SR per se [2,82] is not the subject of this review, its basic 
building blocks- the condensates, are the main concern of this Section. 
These may be defined in terms of quark- and gluon- fields [82,30]:

\begin{equation}
<{\bar q} {\cal O}_i q> = \sum_{a,j} <0|:{\bar q}_j^a(0) {\cal O}_i
      q_j^a(0):|0> 
      =-\int {d^4 p\over (2\pi)^4} Tr {{\tilde S}^A_F(p) {\cal O}_i},
\end{equation} 
where ${\cal O}_i$ is an operator representing the nature of condensate, the
index $A$ represents the effect of a background field, and ${\tilde S}^A_F(p)$
is the quark propagator with the perturbative part suitably subtracted. At 
this stage, we must distinguish between the gluonic background field and other
external ones (electromagnetic, axial, etc.): The latter can be taken
perturbatively, but the former, with its characteristic problem of color gauge
invariance, must be addressed more fully, a subject on 
which there exists a vast literature [83]. However it is possible to 
incorporate in practice a major fraction of this effect through the simple
device of changing the variable of integration in eq.(8.1) from $p_\mu$ to
$\Pi_\mu = p_\mu - {1\over 2} g_s \lambda^a G^a_\mu$, where $G^a_\mu$ is the 
gluon field. This would in general not be possible if one were to evaluate 
complicated integrals involving more propagators and vertex functions, but
since the integral in (8.1) ``sees'' only one such quantity, the trick should
work, especially since we are mainly interested in a constant background
$G_{\mu\nu}-$field, i.e. $G^a_\mu(x) = -{1\over 2} x_\mu G^a_{\mu\nu}$. This 
is basically a non-abelian adaptation of the famous Schwinger method [43] to
the present situation but the details of the available methods [83] are not
necessary for justifying this step. With this understanding, we shall
not use any additional subscript or superscript in (8.1) to specify the
gluonic background, but rather take the integration variable $p_\mu$ to
represent $\Pi_\mu = p_\mu -{1\over 2} g_s \lambda^a G^a_\mu$. 
\par
        The principal quark condensate  $<\bar q q>_0$ corresponds to
${\cal O}_i =1$ and $A=0$. The corresponding gluon condensate is defined as
\begin{equation} 
<g^2_s G^2> = Tr (\nabla_\mu \nabla_\nu -\delta_{\mu\nu} \nabla^2) g_s^2
D_{\mu\nu}(0), 
\end{equation}
where $\nabla_\mu$ is the gauge covariant derivative and $D_{\mu\nu}(x)$ is 
the non-perturbative part [11] of the gluon propagator.  
These quantities which are free parameters in QCD-SR, provide access to the
non-perturbative domain of QCD, but except for the two principal condensates 
$<{\bar q}q>_0$ and $<g_s^2 G^2>_0$, which are amenable to cross checks 
against many data, the determination of the higher order ones often
leave ambiguities. A partial list is [82] 
\begin{equation}
<{\bar q} i \gamma_\mu \gamma_5 q>_A,\quad
<{\bar q} \sigma_{\mu\nu} q>_F,\quad
<{\bar q} {1\over 2} \lambda \, \sigma \cdot G q>_0,\quad 
<{\bar q} {1\over 2} \lambda G_{\mu\nu} q>_F.
\end{equation}
In the method of QCD-SR  [2,82], there is no intrinsic mechanism to evaluate 
them from first principles but only an extrinsic `matching' between the two 
sides of the duality relation with the help of suitable parameters. And
for condensates of still higher dimensions, additional assumptions, such as
factorization, are needed. The BSE-SDE framework [9-12] on the other hand,
has a more microscopic structure which gives it simultaneous access to both
high and low energy phenomena under one umbrella. Thus the condensates (8.3)
as well as others, are calculable within such a framework with as much ease
as the (low energy) spectroscopy is accessible to it (See Sec.1.4 for  
discussions thereof). The same facility also holds for its 3D-4D adaptation
which provides a two-tier structure, with the 3D sector specifically 
attuned to spectroscopy, while the 4D structure is good for loop integrals, 
thus naturally giving rise to a spectroscopic linkage between the high and 
low energy descriptions of hadrons via QCD. To that end eqs (3.12-18) of the
BSE-SDE interplay [82], adapted to its 3D-4D form [11], are collected in 
Sec.4.4: i) an explicit expression (4.33) for the mass function $m(p)$
derived from the condition that it is the pion-quark vertex function in
the chiral limit of $M_\pi$=$0$; ii) the non-perturbative gluon propagator 
$D({\hat k})$, eq.(4.34); iii) its more general form $V({\hat k})$, 
eqs.(4.19-20); iv) and the formula (4.30) for the inverse range $\beta$ of 
the 3d wave function $\phi$. These are the main ingredients needed for the 
condensate calculations in this Section. 
 
\subsection{The Gluon Condensate in 3D-4D Formalism}

We start by rewriting the gluon propagator in a more general form than (4.34)
by making use of the more complete  $V$-function, eqs.(4.19-20), as under:
\begin{equation}
{\cal D}_{\mu\nu}^{ab}(k) = \delta^{ab} (\delta_{\mu\nu} - {k_\mu k_\nu \over
k^2}) {\cal D}({\hat k}), 
\end{equation}
where $a,\, b$ are the color indices in the adjoint representation. 
The logic of the connection between the ${\cal D}_{\mu\nu}$-fn, and the $V$-fn
Eqs.(4.19-20) comes about from its relation with the fermionic kernel $K$:
\begin{equation} 
K(q, q') \Leftrightarrow \gamma_\mu {\cal D}_{\mu\nu}(q-q') \gamma_\nu.
\end{equation}, 
where the scalar part ${\cal D}(\hat k)$  in the infrared region may be
identified with the confining part of the $K$-function as [11,28a]: 
\begin{equation}
{\cal D}(\hat k) = {3\over 4} (2\pi)^3 \omega_0^2 2m_q \alpha_c(4 m_q^2)
\lbrack {\nabla_{\hat k}^2 \over \sqrt{1 - A_0 m_q^2 \nabla_{\hat k}^2}}
+{C_0\over \omega_0^2}\rbrack \delta^3(\hat k),
\end{equation}
using the full $q-{\bar q}$ potential $V({\hat k})$ which fits the
spectroscopy for all flavors (light and heavy) [28], but specialized to the
equal mass ($m_q$) case. The constants $C_0$, $\omega_0$, $A_0$ are given by
eqs.(4.21-22), while the QCD coupling constant $\alpha_s$ is given by [28]:
\begin{equation}
\alpha_s(Q^2) = {4\pi\over 11-{2\over 3}N_f}\cdot {1\over ln{Q^2\over
\Lambda_s^2}}; \qquad \Lambda_s=200\, MeV.
\end{equation}
The coordinate representation ${\cal D}(\hat R)$ of the gluon propagator 
(8.6) is 
\begin{equation}
{\cal D}(\hat R) ={3\over 4}\omega_0^2 \cdot 2m_q \alpha_s(Q^2)
[{C_0\over \omega^2_0} - {{\hat R}^2\over \sqrt{1 +A_0 m_q^2 {\hat R}^2}}]. 
\end{equation}
Note the, in the ${\hat R}\to \infty$ limit, ${\cal D}(\hat R)$ is linear in
${\hat R}$ as well as flavor independent (the $m_q-$factor cancels out), 
except for the $\alpha_s(Q^2)$ effect. Thus the structure (8.8), despite its
empiricity, respects the standard QCD constraint, but only in the strict
confining region. On the other hand, the smallness of $A_0$ $(=0.0283)$ ensures
that for light flavors its structure is dominated by the harmonic form, which
amounts to setting $A_0=0$. This is an excellent approximation for the
pion-vertex function in the chiral limit ($M_\pi=0$), and hence for the quark
mass function given by (4.33) [11], and normalized to $m(0) = m_q$ and 
identified as the constituent mass for the $ud$-quarks only (ignoring
their small `current' values). The 3D wave function $\phi({\hat q})$ is a
gaussian with inverse range $\beta$ given by (4.30), which for equal masses
simplifies to
\begin{equation}
\phi(\hat q) = exp(-{1\over 2}{\hat q}^2/\beta^2); \qquad \beta^4={2
m_q^2\omega_0^2\alpha_s(4m_q^2) \over {1 -2\alpha_s(4m_q^2)C_0}}.
\end{equation}
For the inputs (4.21-22), $\beta^2$ works out as $(0.060) GeV^2$.
\par
        We shall use the mass function (4.33)/(8.9) in  the next subsection 
for the quark condensates. Here we indicate briefly a derivation of the gluon
condensate, eq.(8.2), by inserting the gluon propagator (8.8) in its 
definition. The $C_0-$term may be dropped as it will not survive the 
subsequent differentiations in eq.(8.2). For the main term, the following 
integral representation is employed:
\begin{equation}
{{\hat R}^2\over \sqrt{1 +A_0 m_q^2 {\hat R}^2}} = {2m_q\sqrt{A_0}\over {2\pi
i}} \oint dR_0 {{\hat R}^2\over {1 +A_0m_q^2 R^2}}, 
\end{equation}
where $R^2$ = ${\hat R}^2-R_0^2$ (Lorentz-invariant). The 4D expression 
${\cal D}(R)$ may now be inferred from its definition in terms of the 3D 
quantity ${\cal D}(\hat R)$:
\begin{equation}
{\cal D}(R)= {\alpha_s(Q^2)\over \pi} {2m_q^2\omega_0^2 \sqrt{A_0} {\hat R}^2
\over {1 + A_0 m_q^2 R^2}}.
\end{equation}
This is as far as one goes by adopting the 3D form (8.8) for 
${\cal D}(\hat R)$. However, it is sufficiently suggestive of the 
extrapolation needed to make it fully covariant, viz. ${\hat R}^2 \to R^2$ 
in the numerator of eq.(8.10), which we adopt in what follows. [On the
other hand, if this replacement it not adopted, then the resulting gluon
condensate will be reduced by a factor $3/4$]. The full propagator in
the Landau gauge is (8.4), where $k_\mu$ is read as $k_\mu=-i \partial_\mu^R$.
 To evaluate the gluon condensate we first note the result:
\begin{equation} 
{\alpha_s \over \pi} <G^a_{\alpha\mu} G^b_{\beta\nu}> = 
[- 2 \partial_\alpha^R\partial_\beta^R {\cal D}_{\mu\nu}^{ab}(R) 
+ 2 \partial_\alpha^R\partial_\nu^R {\cal D}_{\mu\beta}^{ab}(R)] |_{R=0}, 
\end{equation} 
and obtain by straightforward differentiation
\begin{equation}
<g_s^2 G^2> = \sqrt{A_0} 4\pi \alpha_s(4m_q^2)(6m_q\omega_0)^2/\pi^2.
\end{equation}
The remaining question concerns what value of the quark mass $m_q$, i.e. what
flavor, should be employed for evaluating the gluon condensate. The structure
(8.8) does exhibit the desired features of linear confinement and flavor
independence, but the extrapolation of these features in the opposite limit
($R\to 0$), as demanded by eq.(8.14), brings in an ``effective flavor
dependence'' of the final formula (8.15). The heavier the flavor, the more
important is the corresponding mass ($m_q$), {\it vis-a-vis} the $A_0-$term 
in the $q{\bar q}$ potential (8.8). Since, on the other hand, the full 
potential (8.8) fits all the flavor sectors rather well [28], a simple
``weighting'' procedure was chosen in [30], involving only the 3 flavor 
sectors with a nontrivial flavor mass, viz. $s{\bar s}$, $c{\bar c}$, and 
$b{\bar b}$ with equal weights (in the sense of a geometric mean), taking 
account of the $m_q$-dependence $m_q^2 \alpha_s(4m_q^2)$ of eq.(8.15). This 
gives the result
\begin{equation}
<m_q^2 \alpha_s(4m_q^2)> = 13.91 \{m_u^2 \alpha_s(4 m_u^2)\},
\end{equation}
in units of its value in the $(ud)$-region, and its substitution in (8.15) 
yields the final estimate $<g_s^2 G^2>$ = $0.502\, GeV^4$, versus the value 
of $0.47\, GeV^4$ adopted in the QCD sum rule literature [82]. 

\subsection{$<{\bar q}{\cal O}_i q>$ Condensates} 

We now substitute the mass function $m(\hat p^2)$, eqs.(4.33)/(8.9) into 
the general formula (8.1), to derive the various condensates for different
choices of ${\cal O}_i$. As already noted in Sec. 8.1 (in light of color 
gauge invariance), the quantity $p_\mu$ in eqs.(4.33)/(8.9), and everywhere 
else in the following, must be read as $\Pi_\mu$ [43], with
appropriate non-abelian corrections. The formula (8.1) now reads as
\begin{equation}
<{\bar q}{\cal O}_i q>_0 = Tr {-i\over (2\pi)^4}\int d^4\Pi
{{m(\hat \Pi)-i\gamma.\Pi} \over {m^2({\hat \Pi}) + (\gamma.\Pi)^2}}
{\cal O}_i 
\end{equation}
in the absence of external fields. Note that the subtracted part with
$m(\hat \Pi)$=0 in this equation gives no effect on tracing in the absence of
external fields. We first express the denominator in an alternative form:
\begin{equation}
m^2(\hat \Pi)+(\gamma.\Pi)^2={\hat \omega}^2-\Sigma_g
-\Pi_l^2\equiv\Delta-\Sigma_g; \quad \Pi^2=\hat\Pi^2-\Pi_l^2
\end{equation}
\begin{equation}
{\hat \omega}^2=m^2({\hat \Pi})+{\hat \Pi}^2; \quad 
\Sigma_g={1\over 2}g_s {1\over 2} \lambda^a G_{\mu\nu}^a \sigma_{\mu\nu},
\end{equation}
where $\Pi_l$ is the longitudinal component of $\Pi_\mu$,
$d^4\Pi$=$d^3\hat\Pi d\Pi_l$, and the integration must first be carried out
over $\Pi_l$. Because of the presence of the $\Sigma_g$-term in (8.18), 
however, a further "rationalization" of eq.(8.15) is necessary according 
to the identity
\begin{equation}
{1\over {\Delta-\Sigma_g}} \equiv
{1\over {{\hat \omega}^2-\Pi_l^2-\Sigma_g}}={{\Delta+\Sigma_g} \over
{\Delta^2-\Sigma_g^2}}.
\end{equation}
At this stage it is probably adequate to replace $\Sigma_g^2$ in the
denominator of (8.18) by its spin-color-averaged value $<\Sigma_g^2>$:
\begin{equation}
\Sigma_g^2\to <\Sigma_g^2>={1\over 12}<g_s^2 GG> \equiv
\mu^4(=8.48 m_q^4) 
\end{equation}
after the necessary substitutions have been made from (8.14) and (4.21-22).
Thus $<\Sigma_g^2>$ contains a strong signature of the gluon condensate whose
large value introduces some bad analyticity properties in the denominator of
the integrand in (8.15) or (8.18), for purposes of $\Pi_l-$integration, since
the $\hat\omega^2-$term is numerically much smaller than $\mu^2$. It has
been emphasized [30] that this feature has nothing to do with the 3D-4D BSE
treatment, since we have not yet passed the barrier of the orthodox 4D quark
propagator in the background of the gluon field. It is rather a very general
manifestation of the strong spin-color effect of the quark-quark interaction
via the color magnetic field. The problem is not so serious in QED [43] where
the smallness of the coupling constant leaves the counterpart of the $\mu^2$
term well below the positivity limit ( $i.e.$, $\hat\omega^2-\mu^2>0$), but 
the large value of $\mu^2$ in the present (QCD) case tends to invalidate the 
standard analyticity structure of (8.15) for purposes of further integration
with respect to $d^3\hat\Pi$. This problem could not be solved in [30],
but it seems to deserve more serious attention from a wider community. (Taken
literally, it would imply the introduction of a $phase$ in the condensates !)
In the meantime, a conservative view [30] was taken that the maximum allowed 
value of $<\Sigma_g^2>$ (consistent with the positivity of the denominator 
after $\Pi_l$-integration) should not exceed $\hat\omega^4$ for all values of
$\hat\Pi^2$, $i.e.$
\begin{equation}
<\Sigma_g^2>=\sigma^2\leq m_q^4.
\end{equation}
Thus  eq.(8.18) should be understood as
\begin{equation}
{1\over {\Delta-\Sigma_g}} \Rightarrow {{\Delta+\Sigma_g} \over
{\Delta^2-\sigma^2}}; \\    \nonumber 
\quad \Delta\equiv {\hat \omega}^2-{\hat \Pi_l}^2. 
\end{equation}

For the numerator of eq.(8.21) which still carries the spin-dependent quantity
$\Sigma_g$, eq.(8.17), there is no restriction of magnitude for one
$\Sigma_g-$factor only, since it contributes to condensates only after
contracting with another $\Sigma-$factor in eq.(8.15). [However, other 
factors which come from the rationalization of the denominator with higher 
powers of $\Sigma_g$ must be subject to the same restriction]. With
this precaution, eq.(8.15) serves to define two condensates simultaneously,
viz, these with, ${\cal O}_i=1$ and ${\cal O}_i$ $=$
$g_s(\lambda^a/2)G_{\mu\nu}^a\sigma_{\mu\nu}$, where the latter is
expressible in the notation of Ref. [82] as
\begin{equation}
<0|{\bar q}2\Sigma_g q|0>\equiv m_0^2 <\bar qq>_0. 
\end{equation}
To evaluate the integral over $d\Pi_l$, we have
\begin{equation}
{1\over {2\pi i}}\int d\Pi_l {{\Delta;\sigma} \over
{\Delta^2-\sigma^2}} \equiv [I(\sigma);J(\sigma)],
\end{equation}
where 
\begin{equation}
I(\sigma);J(\sigma)= {1\over 2}[{1\over
{\sqrt{{\hat \omega}^2-\sigma}}}\pm  {1\over {\sqrt{{\hat \omega}^2+\sigma}}}]. 
\end{equation}
After collecting the necessary trace factors the final result for the two
condensates is expressible as a simple quadrature ($q=u$ or $d$):
\begin{equation}
<q\bar q>_0 [1;m_0^2]= {3\over \pi^2} \int^\infty_0 {\hat \Pi}^2
\ d\Pi\ m({\hat\Pi})\ [I(\sigma);{{2<\Sigma_g^2>} \over \sigma} J(\sigma)].
\end{equation}
On insertion of the structure (4.33)/(8.9) for the mass function, and 
putting the ``maximum allowed value'' of $\sigma$, viz, $m_q^2$, eq.(8.20), 
the results under CIA are [32a]
\begin{equation}
<q\bar q>_0=(266\ MeV)^3; \qquad  m_0^2=0.130\ GeV^2; 
\end{equation}
these results may be compared with the QCD-SR (input) values [82] of 
$(240 MeV)^3$  and $0.8 GeV^2$ respectively. The corresponding CNPA [41]
result, as worked out in Sect.7.3 with $m({\hat p})$ obtained from Sect.4.4
is $(242 MeV)^3$ for the first item. 
\par
We next calculate three induced condensates $\chi$, $\kappa$, and $J$, due 
to a constant external $e.m.$ field $F_{\mu\nu}$, which are defined as [82]:
\begin{equation}
<{\bar q} \sigma_{\mu\nu} q>_F\equiv e e_q \chi F_{\mu\nu}
<{\bar q}q>_0 ; 
\end{equation}
\begin{equation}
g_s<\bar q (\lambda^a/2) G_{\mu\nu}^a q>_F\equiv e e_q \kappa F_{\mu\nu}
<\bar qq>_0 ;
\end{equation}
\begin{equation}
g_s<{\bar q} (\lambda^a/2) \epsilon_{\mu\nu\alpha\beta} G_{\alpha\beta}^a
q>_F\equiv e e_q \zeta F_{\mu\nu} <\bar qq>_0. 
\end{equation}
In these equations the {\it relative} phases of the induced condensates are
defined with respect to the main condensate $<{\bar q} q>_0$, in accordance
with (8.15), and this feature must be kept systematic track of.
Like the two condensates (8.25), the quantities $\chi$ and $\kappa$ are in a
sense $dual$ to each other, and are best described together. The $e.m.$ field
is introduced through the substitution
\begin{equation}
[{\hat m}+i\gamma.\Pi] \rightarrow [{\hat m}+
i\gamma_\mu(\Pi_\mu-eA_\mu)]
\end{equation}
in the propagator, eq.(8.1), and keeping  only the first order term in 
$A_\mu$. Thus we have to calculate
\begin{equation}
\int Tr S_F (\Pi)(ie\gamma.A) S_F(\Pi)
[\sigma_{\mu\nu};{1\over2}\lambda^a G^a_{\mu\nu}].
\end{equation}
This is facilitated, for a constant $e.m.$ field, by the representation
\begin{equation}
A_\mu=-{1\over 2}x_\nu F_{\mu\nu}; \qquad 
x_\mu=i{\partial\over \partial \Pi_\mu}. 
\end{equation}
The substitution in eq.(8.25) and subsequent trace evaluation is routine but
lengthy. However certain precautions are necessary in the matter of extraction
of two groups of terms, proportional to $\sigma_{\mu\nu}$ and $G_{\mu\nu}$
respectively, {\it before} the trace evaluation, which will survive contraction
with the external $e.m.$ field $F_{\mu\nu}$. Thus,
\begin{equation}
\Pi_\mu \Pi_\nu \Rightarrow {i\over 2} g_s {1\over 2}\lambda_a G^a_{\mu\nu};
\qquad \gamma_\mu \gamma_\nu \Rightarrow i\sigma_{\mu\nu}. 
\end{equation}
In terms like $i\sigma_{\mu\lambda}\Pi_\lambda\Pi_\nu$, additional survivors
come from the symmetrized product $\{\Pi_\lambda,\Pi_\nu\}$ for which we make
the standard isotropy ansatz. In this respect, their association with
(space-like) {\it magnetic} effects makes it more meaningful to do an
effectively 3D averaging, viz. $\Pi_\mu\Pi_\nu$ $\Rightarrow$ ${1\over 3} 
{\hat \Pi}^2 (\delta_{\mu\nu} - {\hat \eta}_\mu {\hat \eta}_\nu)$ where 
${\hat \eta}_\mu$ is a unit vector whose direction need not be specified too
precisely. After this step, the tracing process is straightforward, and we omit
the details. But a useful formula is
\begin{equation}
Tr [{1\over 2}\lambda^a G_{\mu\nu}^a g_s \Sigma_g
\sigma_{\alpha\beta}F_{\alpha\beta}] = {1\over 3} <g_s^2 G^2> F_{\mu\nu}.
\end{equation}
The results for the three quantities $\chi$, $\kappa$, $\zeta$ are [30]:
\begin{equation}
\chi = -3.56\, GeV^{-2}; \quad \kappa = -0.11; \quad 
\zeta= +0.06\, GeV^{-2};
\end{equation} 
where the QCD-SR value for $\chi$ is $(6 \pm 2) GeV^{-2}$ [82]. 

\subsection{Axial Condensates}

  So far there has been no explicit need to subtract the perturbative
contribution $(\hat m=0)$ to the condensates calculated above, since their 
traces are zero. We now consider the $axial$ condensate $({\cal O}_i$=
$i\gamma_\mu\gamma_5)$ in a constant external axial field $A_\mu$, where 
an explicit subtraction is necessary to ensure convergence of the integral. 
This condensate is connected with the axial $isoscalar$ coupling which enters 
the Bjorken sum rule [84] for DIS of polarized electrons on a polarized proton
[85]. It is defined through the relation
\begin{equation}
<\bar qi\gamma_\mu\gamma_5 q>_A= {\cal A}_s A_\mu,
\end{equation}
and its value was calculated in [85b] as $f_\eta^2\approx f_\pi^2$, on the
assumption that the axial  field interacts with the $8^{th}$
component (isoscalar) of the unitary octet current. In the present treatment
it does not need any such extra assumption but can be simply calculated from
eq.(8.1) with $({\cal O}_i=i\gamma_\mu\gamma_5)$, and introducing the axial 
field by the gauge substitution $\Pi_\mu\to \Pi_\mu-\gamma_5 A_\mu$ in the 
propagator, and keeping only the first order term in the expansion. The 
result is expressed by
\begin{equation}
{\cal A}_s A_\mu={-i\over (2\pi)^4} Tr\int d^4\Pi  [S_F(\Pi)\
i\gamma. A \gamma_5\ S_F(\Pi)\ i\gamma_\mu\gamma_5] 
 -[``{{\hat m}=0}'']                   
\end{equation}
where the term under quotes is the value of the main term for ${\hat m}=0$.
Evaluating the trace and using the isotropy condition $<\Pi_\mu
\Pi_\nu>$=$\delta_{\mu\nu}\Pi^2/4$ we obtain
\begin{equation}
{\cal A}_s={-3i\over (2\pi)^4} Tr \int d^4 \Pi [{{{\hat m}^2-\Pi^2/2
+\Sigma_g} \over {(\Delta-\Sigma_g)^2}} +
{{\Pi^2/2-\Sigma_g} \over {(\Pi^2-\Sigma_g)^2}}]
\end{equation}
In this case however it is perhaps not as meaningful to keep track of the
$\Sigma_g$-terms for numerical purposes as for the $e.m.$ case; we shall drop
them at this stage. Then with a simple rearrangement ${\hat m}^2-\Pi^2/2$=
$3\Pi^2/2-\Delta/2$, the $\Delta/2$ term can be combined with the
last term through a Feynman variable $u$ $(0\leq u\leq 1)$ and the pole
integration carried out. The final result is
\begin{equation}
{\cal A}_s ={3\over 4\pi^2}\int_0^\infty {\hat \Pi}^2 d{\hat \Pi}
\int^\infty_0 du\ {\hat m^2} \Bigl[ {3\over {(\hat m^2
+{\hat \Pi}^2)^{3/2}}}+ {1\over {({\hat m}^2 u+{\hat \Pi}^2)^{3/2}}}\Bigr]
\end{equation}
which yields $0.021\ GeV^2$, to be compared with $f_\pi^2\approx 0.018$, or 
perhaps better with $f_\eta^2$ which is the relevant isoscalar quantity [83c] 
having a larger value [13] than $f_\pi^2$. 
\par    
        For a discussion of these results vis-a-vis QCD-SR, see [30]. Since
the spectroscopic linkage of the QCD condensates has been the main theme of
this Section, we should like to end it with the remark that the ({\bf MYTP}-
governed [15]) CIA [16] by itself does not carry information on the dynamics 
of spectroscopy which must be governed by other considerations (non-
perturbative QCD simulated by $DB\chi S$ [4,24]), but it certainly offers a 
broad enough framework to accommodate such dynamics, without having
to look elsewhere. Of course, the importance of spectroscopy as an integral 
part of any `dynamical equation based' approach merely reiterates  a 
philosophy initiated long ago by Feynman et al [25].     
\vspace{1cm}

\section{$qqq$ Dynamics: General Aspects}

The dynamics of baryons as $qqq$ systems represents the third stage of the
three-body problem in its journey from the atomic through nuclear to the
hadronic level of compositeness. The first (atomic) stage had been relatively
free of theoretical ambiguities due to its strong QED foundations in the 
domain of non-relativistic quantum mechanics. In contrast, the second 
(nuclear) stage, although providing the initial stimulus for {\it few-body}
dynamics, has from the outset remained bogged down in a continual empiricity
in the theoretical foundations of strong interaction dynamics. Indeed by the
time the meson exchange picture started being taken seriously for a parallel
treatment of meson-nucleon systems on the lines of electron-photon systems,
the carpet got quietly removed from under its feet, through the slow but
sure realization of its tenuous character born out of the quark compositeness
of the underlying (meson) fields. Indeed the quark-gluon picture which had
taken firm shape by the end of the Seventies, told in no uncertain terms the 
futility of understanding the inter-hadronic forces directly in terms of 
their {it own} species, as if they were elementary fields! On the other hand,
the emergence of nuclear three-body techniques in the Sixties had an instant
impact on the quark-level 3-body problem, thus providing a big boost to its 
development in a language strongly reminiscient of the nuclear 3-body problem,
on the lines of Bethe's Second Principle Theory (see Sec.1.),  except for the
realization of its relativistic character which demands the input dynamics to
to be Bethe-Salpeter-like (albeit with wide variations), rather than 
Schroedinger-like. In this Section we shall give a panoramic view of three 
general aspects governing the dynamics of $qqq$ baryons: i) classification of 
baryonic states [86]; ii) problem of {\it connectedness} in 3-body dynamics
[87]; iii) BS-dynamics for fermionic $qqq$ systems under $DB{\chi}S$, [29b],
in parallel with ${\bar q}q$, Sec.4.3. The  details of topics (ii) and (iii) 
are taken up in Sections {\bf 10} and {\bf 11} respectively.   
\par    
        Yet another type of approach to the $qqq$ problem, as available 
in the literature, concerns parametric representations attuned to 
effective Lagrangians for hadronic transitions to``constituent" quarks, 
with ad hoc assumptions on the hadron-$qqq$ form 
factor [88a], similar (parametric) ansatze for the hadron- quark-diquark 
form factor [88b]; or more often simply direct gaussian parametrizations 
for the $qqq$ wave functions as the starting point of the investigation 
[88c]. Such approaches are often quite effective for the investigations of
some well-defined sectors of hadron physics with quark degrees of freedom,
but are in general much less predictive than dynamical-equation-based 
methods like NJL- Faddeev [89] or BSE-SDE framework [9-12], when extended 
beyond their immediate domains of applicability. 

\subsection{$SU(6) \otimes O(3)$ Classification}

\setcounter{equation}{0}
\renewcommand{\theequation}{9.\arabic{equation}}

        The initial $qqq$ formulation  was provided by a non-relativistic
form of dynamics, and the first systematic classification [86a,b] of $qqq$ 
states proved remarkably successful for the understanding of many details 
of hadronic spectra. On the other hand, the high degree of degeneracy of the 
h.o. model [86a] caused problems on the details of observed states, such as 
the absence of (the relatively low-lying) ${\bf 20}$ states, in favour of 
more restricted types which, in a broad $SU(6) \times O(3)$ classification, 
are all `natural parity' states [86c] 
\begin{equation}
[{\bf 56},(even)^+],[{\bf 70},(odd)^-]; \qquad 
[{\bf 70},(even)^+],[{\bf 56},(odd)^-] 
\end{equation}
while the (complementary) `unnatural parity' states like ${\bf 20}^+$ seemed 
to be missing from the data [13]! The natural parity baryons in turn are
amenable to a simple quark-diquark picture [86d], with diquarks of the types
`scalar-isoscalar' $D_s$ and `(axial)vector-isovector'$ D^a_\mu$, as well 
as (complementary) diquarks of the types (pseudo)vector-isoscalar $D_\mu$ 
and scalar-isovector $D^a_s$ [90a], all of which go to make up the list
(9.1) above. On the other hand the `unnatural' parity baryons require 
diquark ingredients of opposite parity to above, viz., pseudoscalar-isoscalar, 
vector-isovector; vector-isoscalar, and pseudoscalar-isovector, respectively,
to make up a complementary list of $SU(6) \times O(3)$ baryons [29b]:
\begin{equation}
[{\bf 20},(even)^+],[{\bf 70},(even)^-]; \qquad 
[{\bf 70},(odd)^+],[{\bf 56},(even)^-]
\end{equation}              
which have not yet been observed [13, 29b].           
\par 
        Despite the compactness and elegance of the quark-diquark description,
a certain amount of {\it dynamical} 3-body information gets lost due to the 
`freezing' of a quark d.o.f. in the (rigid) diquark structure. While a good
part of the $S_3$ (permutation) symmetry can be recovered by appropriate
$SU(6)$ classification [86d], the dynamical information in the full 3-body 
structure is not entirely retrievable, showing up, e.g., in the $k^2$-
dependence of the e.m. form factor of the $qqq$ baryon. To see more clearly
the interconnection between the two descriptions, let us write down the 
baryon wave function, with proper $S_3$-symmetry, in both the $qqq$ and 
$q-d_q$ notations. To that end, its full wave function $\Psi$, with $S_3$-
symmetry for 3 identical quarks, may be analyzed into its orbital 
$\psi^\alpha$, spin $\chi^\beta$ and isospin $\phi^\gamma$ components, where 
$\alpha$, $\beta$, $\gamma$ superscripts stand for the $S_3$-symmetry types 
[25, 45], which, in the Verde notation [45a], are $(s;m',m'';a)$ for 
symmetric, mixed-symmetric, and antisymmetric respectively. Since only color 
singlet states are being considered, we suppress the (antisymmetric) color 
wave function $C^a$ for brevity, so that the `active' part of the wave 
function is {\it symmetric} [86b]. The full structures of $\Psi$ for 
${\bf 56}$ and ${\bf 70}$ states are [90a]
\begin{eqnarray}
{{N}^d}_{56}       &=& (\chi'\phi'+ \chi"\phi") {\psi}^s/{\sqrt 2}; \quad
{N^q}_{70} =(\phi'\psi'+\phi"\psi"){\chi}^s/{\sqrt 2}; \\  \nonumber
{N^d}_{70}         &=& [(\chi'\phi"+\chi"\phi')\psi'+
(\chi'\phi'-\chi"\phi")\psi"]/2; \\  \nonumber  
{{\Delta}^q}_{56}   &=& \chi^s \phi^s \psi^s; \qquad 
{{\Delta}^d}_{70} =(\chi'\psi'+\chi"\psi"){\phi}^s / {\sqrt 2};
\end{eqnarray}
where the superscripts $d$ and $q$ stand for spin-doublet and spin-quartet
respectively, and the product of the orbital $(\psi)$ and spin $(\chi)$ 
functions for higher $L$-states must be read in the standard sense of
adding angular momenta in terms of C.G coefficients.
For strange baryon ($\Lambda$,$\Sigma$) states, the symmetry is reduced to
$S_2$, due to the higher mass of the $s$-quark, and the corresponding states
have the following representations [90]         
\begin{eqnarray}
{\Lambda}_{56}     &=& \phi'\chi'{\psi}^s/{\sqrt 2}; \quad
{\Lambda}_{70} =\phi'(\chi'\psi" \pm \chi"\psi')/2  \\  \nonumber
{\Sigma}_{56}      &=& \phi"\chi"{\psi}^s/{\sqrt 2}; \quad
{\Sigma}_{70} = \phi"(\chi'\psi' \pm \chi"\psi")/2 \\  \nonumber
{\Lambda}_{70}     &=& =\phi'\psi'{\chi}^s; \qquad 
{\Sigma}_{70} = \phi"\psi"{\chi}^s.
\end{eqnarray}      
\par
        To relate these structures to the quark($q$)-diquark($D$) description,
the $qD$ contents of these wave functions in a lorentz-invariant form may be 
read off from the following correspondence [90a] 
\begin{eqnarray}
\chi'\phi'       & \Leftrightarrow & D_s; \quad  
{\chi"\phi"} \Leftrightarrow {i\gamma_5\gamma_\mu D^a_\mu \tau_a}; \\  \nonumber
{\chi}^s{\phi}^s & \Leftrightarrow & D^a_\mu {\epsilon}^a_\mu; \quad  
\chi'\phi"   \Leftrightarrow  D^a_s \tau^a;     \\    \nonumber
{\chi}^s \phi'   & \Leftrightarrow & D_\mu {\epsilon}_\mu; \quad
\chi" \phi'    \Leftrightarrow  i\gamma_5 \gamma_\mu D_\mu; \\  \nonumber          
\chi^s\phi"      & \Leftrightarrow & D^a_\mu \tau_a \epsilon_\mu
\end{eqnarray}
Here for simplicity, a basis spinor symbol $\Psi$ on the RHS has been 
suppressed for all (baryon) states. However, the additional (Rarita-
Schwinger) spin and isospin symbols needed for several such states to make
up the full baryon structure have been supplied via the unit vectors 
$\epsilon^a_\mu$ and $\epsilon_\mu$ where necessary. [Of course {\it orbital}
functions $\psi$ are needed to make up the spatial overlap for the 
$qD$-pair]. This correspondence may be faithfully substituted in the set
(9.3) to give the precise $qD$ content in $SU(6)$-form to give the different
cases with {\it correct} normalizations. The {\it dynamical} effects are
now entirely contained in the orbital wave function $\psi$. 
\subsection*{9.2 \quad  Connectedness in a 3-Body Amplitude}

The problem of {\it connectedness} [88] in a 3-particle amplitude has been 
in the forefront of few-body dynamics since Faddeev's classic paper [87a] 
showed the proper perspective, by emphasizing the role of the 2-body T-matrix 
as a powerful tool for achieving the goal. The initial stimulus in this 
regard came from the separable assumption due to Mitra [87b] which provided 
a very simple realization of such connectedness via Faddeev's T-matrix 
structure, a result that was given a firmer basis by Lovelace [87c]. An 
alternative strategy for connectedness in a more general $n$-body amplitude 
was provcided by Weinberg [87d] through graphical equations which brought out 
the relative roles of the $T$-and $V$-matrices in a more transparent manner. 
(In particular Weinberg showed that the $T$-matrix was not the only way to 
achieve connectedness). It was emphasized by both Weinberg and Lovelace 
[87c-d] that an important signal for connectedness in the 3-body (or $n$-
body) amplitude is the {\it {absence of any $\delta$-function}} in its
structure, either explicitly or through its defining equation. This signal
is valid irrespective of whether or not the $V$- or the $T$- matrix is
employed for the said dynamical equation. 
\par 
        The above equations were found for a {\it non-relativistic} $n$-
body problem within a basically 3D framework [88d] whose prototype dynamics
is the Schroedinger equation. For the corresponding {\it relativistic}
problem whose typical dynamics may be taken as the Bethe-Salpeter Equation
(BSE) with pairwise kernels within a 4D framework, it should be possible
in principle to follow a logic similar to Weinberg's, using the language of 
Green's functions with corresponding diagrammatic representations [88d],
leading to equations free from $\delta$-functions. However there are other 
{\it physical} issues associated with a 4D support to the BSE kernel of a
confining type, such as contradictions of the spectral predictions [14] 
with data [13]. Indeed, this very issue has been discussed in detail in 
Section 1, culminating in the `two-tier' 3D-4D BSE approach as the central
theme of this article, under the name of Covariant Instantaneity [16] for 
3D support to the BSE kernel [17] which receives formal justification from 
the {\bf MYTP} principle [15]. The principal result of this ansatz is the
{\it {exact interconnection}} between the 3D and 4D forms of the BSE, at 
least for the 4D two-body problem [16]. 
\par
        One may now ask: Does a similar interconnection exist in the
corresponding BS amplitudes for a {\it three-body} system under the same 
conditions of 3D support for the pairwise  BS kernel? The question is of 
great practical value since the 3D reduction of the 4D BSE already provides 
a {\it {fully connected}} integral equation [29b], leading to an approximate
analytic solution (in gaussian form) [29b] for the corresponding 3D wave
function, as a byproduct of its success on the baryon spectra [13]. Therefore
a reconstruction of the 4D $qqq$ wave function in terms of the corresponding 
3D quantities should open up a vista of applications to various types of 
{\it {transition amplitudes}} involving $qqq$ baryons, just as in the two-
body case outlined in Section 4. This exercise is carried out in Section 10,
using Green's function techniques for both the 2- and 3- body systems (the
former for checking against the known results of Section 4). There is however
a big difference between the two systems, born out of the `truncation' of the
Hilbert space due to the 3D support ansatz for the pairwise BSE kernel. Such
truncation, while still allowing an unambiguous reduction of the BSE  from 
the 4D to the 3D level, nevertheless leaves an `undetermined element' in the
{\it {reverse direction}}, viz., from 3D to 4D. This limitation for the
reverse direction is quite general for any $n$-body system  with $n>2$; the
only exception is the case of $n=2$ where both transitions are reversible
without extra assumptions (a sort of degenerate situation). As will be
shown in Sec.10, the extra assumption (in its simplest form) needed to 
complete the reverse transition is facilitated by some 1D $\delta$-function 
which however has nothing to do with connectedness [87]. 

\subsection*{9.3 \quad  Fermionic $qqq$ BSE with $DB{\chi}S$}

We now outline the essential logic of a BSE treatment for a fermionic $qqq$
system, for pairwise kernels with covariant 3D support, under conditions of
$DB{\chi}S$, on closely parallel lines to the ${\bar q}q$ case (Section 4).
In Section 3, the derivation [52] of the equation of motion from an input 
Lagrangian for extended 4-fermion coupling shows that the BSE structure 
(4.2) emerges in the {\it linear} approximation to the $\phi$-field. This 
immediately suggests that the BSE for a $qqq$ system in the same (linear)
approximation must be one with a linear sum over all the three pairs of
interaction, which for spinless quarks reads as [10b]:
\begin{equation}
i(2\pi)^4\Phi(p_1p_2p_3)= \sum_{123}\Delta_{F1}\Delta_{F2} \int d^q_{12}
K(q_{12},q_{12}') \Phi(p_1'p_2'p_3)
\end{equation}                 
where $\Delta_{Fi}$=$-i{\Delta_i}^{-1}$, etc. Under CIA [16], the relative
momenta $q_{ij}$ are `hatted', i.e., they are orthogonal to the total
momenta $P_{ij}$ of the ${ij}$ pairs, as explained in Sec.4. It is however
more convenient for calculational purposes to take all these relative
`hatted' momenta ${\hat q}_{ij}$ to be perpendicular to a common 4-momentum
$P=p_1+p_2+p_3$, instead of to the individual pairs. Technically this amounts 
to the introduction of 3-body forces at the quark level. However the 
difference turns out to be small [10b], which suggests that the 3-body forces
are (expectedly) small. Having thus checked this (small) 3-body effect, we
shall from now on consider a common `hat' symbol for all the three pairs,
i.e., ${\hat q}_{ij;\mu}$=$q_{ij}-q_{ij}.PP_\mu/P^2$. It is this version
for hatted symbols for the $qqq$ problem that we shall consider in the 
next two sections. 

\section{Interlinking 3D And 4D $qqq$ Vertex Fns} 

In this Section we outline a fairly detailed (self-contained) method of 
Green's functions for 2- and 3- particle scattering {\it {near the bound 
state pole}}, for the 3D-4D interconnection between the corresponding wave
functions. For simplicity we consider identical {\it spinless} bosons, with 
pairwise BS kernels under CIA conditions [16], first for the 2-body case for 
calibration, (see Sects.4.1-2), and then for the corresponding 3-body case, 
on the basis of the Green's Fn counterpart of the general structure, eq.(9.6).
   
\subsection{Two-Quark Green's Function Under CIA}
\setcounter{equation}{0}
\renewcommand{\theequation}{10.1.\arabic{equation}}

In the notation and phase convention of Section 4, the 4D $qq$ 
Green's fn $G(p_1p_2;{p_1}'{p_2}')$ near a {\it bound} state 
satisfies a 4D BSE (no inhomogeneous term): 

\begin{equation}
i(2\pi)^4 G(p_1 p_2;{p_1}'{p_2}') = {\Delta_1}^{-1} {\Delta_2}^{-1} \int
d{p_1}'' d{p_2}'' K(p_1 p_2;{p_1}''{p_2}'') G({p_1}''{p_2}'';{p_1}'{p_2}')    
\end{equation}
 where
\begin{equation}
\Delta_1 = {p_1}^2 + {m_q}^2 , 
\end{equation}
and $m_q$ is the mass of each quark. Now using the relative 4- momentum 
$q = (p_1-p_2)/2$ and total 4-momentum $P = p_1 + p_2$ 
(similarly for the other sets), and removing a $\delta$-function
for overall 4-momentum conservation, from each of the $G$- and $K$- 
functions, eq.(10.1.1) reduces to the simpler form    
\begin{equation}
i(2\pi)^4 G(q.q') = {\Delta_1}^{-1} {\Delta_2}^{-1}  \int d{\hat q}'' 
Md{\sigma}'' K({\hat q},{\hat q''}) G(q'',q')
\end{equation}
where ${\hat q}_{\mu} = q_{\mu} - {\sigma} P_{\mu}$, with 
$\sigma = (q.P)/P^2$, is effectively 3D in content (being orthogonal to
$P_{\mu}$). Here we have incorporated the ansatz of a 3D support for the
kernel $K$ (independent of $\sigma$ and ${\sigma}'$), and broken up the 
4D measure $dq''$ arising from (10.1.1) into the product 
$d{\hat q}''Md{\sigma}''$ of a 3D and a 1D measure respectively. We have 
also suppressed the 4-momentum $P_{\mu}$ label, with $(P^2 = -M^2)$, in 
the notation for $G(q.q')$.
\par

        Now define the fully 3D Green's function 
${\hat G}({\hat q},{\hat q}')$ as [47] 
\begin{equation}
{\hat G}({\hat q},{\hat q}') = \int \int M^2 d{\sigma}d{\sigma}'G(q,q')
\end{equation}
and two (hybrid) 3D-4D Green's functions ${\tilde G}({\hat q},q')$,
${\tilde G}(q,{\hat q}')$ as
\begin{equation}
{\tilde G}({\hat q},q') = \int Md{\sigma} G(q,q'); \quad
{\tilde G}(q,{\hat q}') = \int Md{\sigma}' G(q,q');
\end{equation} 
Next, use (10.1.5) in (10.1.3) to give    
\begin{equation}
i(2\pi)^4 {\tilde G}(q,{\hat q}') = {\Delta_1}^{-1} {\Delta_2}^{-1} 
\int dq'' K({\hat q},{\hat q}''){\tilde G}(q'',{\hat q}')  
\end{equation}
Now integrate both sides of (10.1.3) w.r.t. $Md{\sigma}$ and use the result 
\begin{equation}
\int Md{\sigma}{\Delta_1}^{-1} {\Delta_2}^{-1} = 2{\pi}i D^{-1}({\hat q});
\quad D({\hat q}) = 4{\hat \omega}({\hat \omega}^2 - M^2/4);\quad 
{\hat \omega}^2 = {m_q}^2 + {\hat q}^2 
\end{equation}
to give a 3D BSE w.r.t. the variable ${\hat q}$, while keeping the other 
variable $q'$ in a 4D form:
\begin{equation}  
(2\pi)^3 {\tilde G}({\hat q},q') = D^{-1} \int d{\hat q}''  
K({\hat q},{\hat q}'') {\tilde G}({\hat q}'',q')
\end{equation}
A comparison of (10.1.3) with (10.1.8) gives the desired connection between 
the full 4D $G$-function and the hybrid ${\tilde G({\hat q}, q')}$-function: 
\begin{equation}  
2{\pi}i G(q,q') = D({\hat q}){\Delta_1}^{-1}{\Delta_2}^{-1}
{\tilde G}({\hat q},q')
\end{equation}
Again, the symmetry of the left hand side of (10.1.9) w.r.t. $q$ and $q'$ 
allows rewriting the right hand side with the roles of $q$ and $q'$ 
interchanged. This gives the dual form   
\begin{equation}  
2{\pi}i G(q,q') = D({\hat q}'){{\Delta_1}'}^{-1}{{\Delta_2}'}^{-1}
{\tilde G}(q,{\hat q}')
\end{equation}
which on integrating both sides w.r.t. $M d{\sigma}$ gives
\begin{equation}  
2{\pi}i{\tilde G}({\hat q},q') = D({\hat q}'){{\Delta_1}'}^{-1}
{{\Delta_2}'}^{-1}{\hat G}({\hat q},{\hat q}'). 
\end{equation}
Substitution of (10.1.11) in (10.1.9) then gives the symmetrical form
\begin{equation}  
(2{\pi}i)^2 G(q,q') = D({\hat q}){\Delta_1}^{-1}{\Delta_2}^{-1}
{\hat G}({\hat q},{\hat q}')D({\hat q}'){{\Delta_1}'}^{-1}
{{\Delta_2}'}^{-1}
\end{equation}
Finally, integrating both sides of (10.1.8) w.r.t. $M d{\sigma}'$, we 
obtain a fully reduced 3D BSE for the 3D Green's function:
\begin{equation}  
(2\pi)^3 {\hat G}({\hat q},{\hat q}') = D^{-1}({\hat q} \int d{\hat q}''
K({\hat q},{\hat q}'') {\hat G}({\hat q}'',{\hat q}')
\end{equation}
Eq.(10.1.12) which is valid near the bound state pole, expresses the desired 
connection between the 3D and 4D forms of the Green's functions; and 
eq(10.1.13) is the determining equation for the 3D form. A spectral analysis 
can now be made for either of the 3D or 4D Green's functions in the 
standard manner, viz., 
\begin{equation}  
G(q,q') = \sum_n {\Phi}_n(q;P){\Phi}_n^*(q';P)/(P^2 + M^2) 
\end{equation}
where $\Phi$ is the 4D BS wave function. A similar expansion holds for 
the 3D $G$-function ${\hat G}$ in terms of ${\phi}_n({\hat q})$. Substituting
these expansions in (10.1.12), one immediately sees the connection between 
the 3D and 4D wave functions in the form:
\begin{equation}  
2{\pi}i{\Phi}(q,P) = {\Delta_1}^{-1}{\Delta_2}^{-1}D(\hat q){\phi}(\hat q)
\end{equation}
whence the BS vertex function becomes $\Gamma$ = $D \times \phi/(2{\pi}i)$
as found in [16]. We shall make free use of these results, taken as $qq$ 
subsystems, for our study of the $qqq$ $G$-functions in subsections 2 and 3.  

\subsection{3D BSE Reduction for $qqq$ G-fn}
\setcounter{equation}{0}
\renewcommand{\theequation}{10.2.\arabic{equation}}
\par

        As in the two-body case, and in an obvious notation for various 
4-momenta (without the Greek suffixes), we consider the most general 
Green's function $G(p_1 p_2 p_3;{p_1}' {p_2}' {p_3}')$ for 3-quark 
scattering {\it near the bound state pole} (for simplicity) which allows       
us to drop the various inhomogeneous terms from the beginning. Again we 
take out an overall delta function $\delta(p_1 + p_2 + p_3 - P)$ from the
$G$-function  and work with {\it two} internal 4-momenta for each of the 
initial and final states defined as follows [10b]:
\begin{equation}  
{\sqrt 3}{\xi}_3 =p_1 - p_2 \ ; \quad  3{\eta}_3 = - 2p_3 + p_1 +p_2
\end{equation}
\begin{equation}  
P = p_1 + p_2 + p_3 = {p_1}' + {p_2}' + {p_3}'
\end{equation}
and two other sets ${\xi}_1,{\eta}_1$ and ${\xi}_2,{\eta}_2$ defined by 
cyclic permutations from (10.2.1). Further, as we shall consider pairwise
kernels with 3D support, we define the effectively 3D momenta ${\hat p}_i$, 
as well as the three (cyclic) sets of internal momenta 
${\hat \xi}_i,{\hat \eta}_i$, (i = 1,2,3) by [10b]:
\begin{equation}
{\hat p}_i = p_i - {\nu}_i P \ ;\quad  {\hat {\xi}}_i = {\xi}_i - s_i P\  ;
\quad
{\hat {\eta}}_i - t_i P 
\end{equation}
\begin{equation}  
{\nu}_i = (P.p_i)/P^2\  ;\quad s_i = (P.\xi_i)/P^2 \ ;\quad t_i = 
(P.\eta_i)/P^2 \end{equation}
\begin{equation}  
{\sqrt 3} s_3 = \nu_1 - \nu_2 \ ;\quad 3 t_3 = -2 \nu_3 + \nu_1 + \nu_2 \ 
\ ( + {\rm cyclic permutations})
\end{equation}

The space-like momenta ${\hat p}_i$ and the time-like ones $\nu_i$ 
satisfy [10b] 
\begin{equation}  
{\hat p}_1 + {\hat p}_2 + {\hat p}_3 = 0\  ;\quad \nu_1 + \nu_2 + \nu_3 = 1
\end{equation}
Strictly speaking, in the spirit of covariant instantaneity, we should 
have taken the relative 3D momenta ${\hat \xi},{\hat \eta}$ to be in the 
instantaneous frames of the concerned pairs, i.e., w.r.t. the rest frames
of $P_{ij} = p_i +p_j$; however the difference between the rest frames of 
$P$ and $P_{ij}$  is small and calculable [10b], while the use of a common 
3-body rest frame $(P = 0)$ lends considerable simplicity and elegance to 
the formalism.   
\par

        We may now use the foregoing considerations to write down the BSE 
for the 6-point Green's function in terms of relative momenta, on closely 
parallel lines to the 2-body case. To that end note that the 2-body 
relative momenta are $q_{ij} = (p_i - p_j)/2 = {\sqrt 3}{\xi_k}/2$, where 
(ijk) are cyclic permutations of (123). Then for the reduced $qqq$ Green's
function, when the {\it last} interaction was in the (ij) pair, we may use 
the notation $G(\xi_k \eta_k ; {\xi_k}' {\eta_k}')$, together with `hat' 
notations on these 4-momenta when the corresponding time-like components 
are integrated out. Further, since the pair $\xi_k,\eta_k$ is 
{\it {permutation invariant}} as a whole, we may choose to drop the index 
notation from the complete $G$-function to emphasize this symmetry as and 
when needed. The $G$-function for the $qqq$ system satisfies, in the 
neighbourhood of the bound state pole, the following (homogeneous) 4D BSE
for pairwise $qq$ kernels with 3D support:
\begin{equation}  
i(2\pi)^4 G(\xi \eta ;{\xi}' {\eta}') = \sum_{123}
{\Delta_1}^{-1} {\Delta_2}^{-1} \int d{{\hat q}_{12}}'' M d{\sigma_{12}}''
K({\hat q}_{12}, {{\hat q}_{12}}'') G({\xi_3}'' {\eta_3}'';{\xi_3}' {\eta_3}')
\end{equation}
where we have employed a mixed notation ($q_{12}$ versus $\xi_3$) to stress
the two-body nature of the interaction with one spectator at a time, in a 
normalization directly comparable with eq.(10.1.3) for the corresponding 
two-body problem. Note also the connections 
\begin{equation}  
\sigma_{12} = {\sqrt 3}{s_3}/2   ;\quad 
{\hat q}_{12} = {\sqrt 3}{{\hat \xi}_3}/2  ; \quad {\hat \eta}_3 = 
-{\hat p}_3, \quad etc 
\end{equation}  
The next task is to reduce the 4D BSE (10.2.7) to a fully 3D form through a 
sequence of integrations w.r.t. the time-like momenta $s_i,t_i$ applied 
to the different terms on the right hand side, {\it {provided both}} 
variables are simultaneously permuted. We now define the following fully 
3D as well as mixed (hybrid) 3D-4D $G$-functions according as one or more 
of the time-like $\xi,\eta$ variables are integrated out:
\begin{equation}  
{\hat G}({\hat \xi} {\hat \eta};{\hat \xi}' {\hat \eta}') = 
\int \int \int \int ds dt ds' dt' G(\xi \eta ; {\xi}' {\eta}')  
\end{equation}  
which is $S_3$-symmetric.
\begin{equation}  
{\tilde G}_{3\eta}(\xi {\hat \eta};{\xi}' {\hat \eta}') = 
\int \int dt_3 d{t_3}' G(\xi \eta ; {\xi}' {\eta}');
\end{equation}  
\begin{equation}  
{\tilde G}_{3\xi}({\hat \xi}  \eta;{\hat \xi}' {\eta}') = 
\int \int ds_3 d{s_3}' G(\xi \eta ; {\xi}' {\eta}');
\end{equation} 
The last two equations are however {\it not} symmetric w.r.t. the 
permutation group $S_3$, since both the variables ${\xi,\eta}$ are not 
simultaneously transformed; this fact has been indicated in eqs.(10.2.10-11) 
by the suffix ``3" on the corresponding (hybrid) ${\tilde G}$-functions,
to emphasize that the `asymmetry' is w.r.t. the index ``3". We shall term 
such quantities ``$S_3$-indexed", to distinguish them from $S_3$-symmetric 
quantities as in eq.(10.2.9). The full 3D BSE for the ${\hat G}$-function is 
obtained by integrating out both sides of (10.2.7) w.r.t. the $st$-pair 
variables $ds_i d{s_j}' dt_i d{t_j}'$ (giving rise to an $S_3$-symmetric 
quantity), and using (10.2.9) together with (10.2.8) as follows:
\begin{equation}  
(2\pi)^3 {\hat G}({\hat \xi} {\hat \eta} ;{\hat \xi}' {\hat \eta}') = 
\sum_{123} D^{-1}({\hat q}_{12}) \int d{{\hat q}_{12}}'' 
K({\hat q}_{12}, {{\hat q}_{12}}'') {\hat G}({\hat \xi}'' {\hat \eta}'';
{\hat \xi}' {\hat \eta}')  
\end{equation}   
This integral equation for ${\hat G}$ which is the 3-body counterpart of
(10.1.13) for a $qq$ system in the neighbourhood of the bound state pole, 
is the desired 3D BSE for the $qqq$ system in a {\it {fully connected}}
form, i.e., free from delta functions. Now using a spectral decomposition 
for ${\hat G}$ 
\begin{equation}   
{\hat G}({\hat \xi} {\hat \eta};{\hat \xi}' {\hat \eta}')
= \sum_n {\phi}_n( {\hat \xi} {\hat \eta} ;P)
{\phi}_n^*({\hat \xi}' {\hat \eta}';P)/(P^2 + M^2)
\end{equation}   
on both sides of (10.2.12) and equating the residues near a given pole
$P^2 = -M^2$, gives the desired equation for the 3D wave function $\phi$ 
for the bound state in the connected form:
\begin{equation}   
(2\pi)^3 \phi({\hat \xi} {\hat \eta} ;P) = \sum_{123} D^{-1}({\hat q}_{12})
\int d{{\hat q}_{12}}'' K({\hat q}_{12}, {{\hat q}_{12}}'')
\phi({\hat \xi}'' {\hat \eta}'' ;P)
\end{equation}   
Now the $S_3$-symmetry of $\phi$ in the $({\hat \xi}_i, {\hat \eta}_i)$ pair 
is a very useful result for both the solution of (10.2.14) {\it and} for the 
reconstruction of the 4D BS wave function in terms of the 3D wave function 
(10.2.14), as is done in the subsection below.
\subsection{Reconstruction of 4D $qqq$ Wave Function}
\setcounter{equation}{0}
\renewcommand{\theequation}{10.3.\arabic{equation}}
\par

        We now attempt to {\it re-express} the 4D $G$-function given by 
(10.2.7) in terms of the 3D ${\hat G}$-function given by (10.2.12), as the 
$qqq$ counterpart of the $qq$ results (10.1.12-13). To that end we adapt 
the result (10.1.12) to the hybrid Green's function  of the (12) subsystem 
given by ${\tilde G}_{3 \eta}$, eq.(10.2.10), in which the 3-momenta 
${\hat \eta}_3,{{\hat \eta}_3}'$ play a parametric role reflecting the 
spectator status of quark $\# 3$, while the {\it active} roles are played 
by $q_{12}, {q_{12}}' = {\sqrt 3}(\xi_3,{\xi_3}')/2$, for which the analysis 
of subsec.(10.1) applies directly. This gives 
\begin{equation} 
(2{\pi}i)^2 {\tilde G}_{3 \eta}(\xi_3 {\hat \eta}_3; 
{\xi_3}' {{\hat \eta}_3}') 
= D({\hat q}_{12}){\Delta_1}^{-1}{\Delta_2}^{-1}
{\hat G}({\hat \xi_3} {\hat \eta_3}; {\hat \xi_3}' {\hat \eta_3}')
D({{\hat q}_{12}}'){{\Delta_1}'}^{-1}{{\Delta_2}'}^{-1}
\end{equation}
where on the right hand side, the `hatted' $G$-function has full 
$S_3$-symmetry, although (for purposes of book-keeping) we have not 
shown this fact explicitly by deleting the suffix `3' from its 
arguments. A second relation of this kind may be obtained from (10.2.7)
by noting that the 3 terms on its right hand side may be expressed in 
terms of the hybrid ${\tilde G}_{3 \xi}$ functions vide their definitions 
(10.2.11), together with the 2-body interconnection between $(\xi_3,{\xi_3}')$ 
and $({\hat \xi}_3,{{\hat \xi}_3}')$ expressed once again via (10.3.1), but 
without the `hats' on $\eta_3$ and ${\eta_3}'$. This gives
\begin{eqnarray}
({\sqrt 3} \pi i)^2 G(\xi_3 \eta_3; {\xi_3}'{\eta_3}')
&=& ({\sqrt 3} \pi i)^2 G(\xi \eta; {\xi}'{\eta}')\nonumber\\
&=& \sum_{123} {\Delta_1}^{-1}{\Delta_2}^{-1} (\pi i {\sqrt 3})
\int d{{\hat q}_{12}}'' M d{\sigma_{12}}''
K({\hat q}_{12}, {{\hat q}_{12}}'') 
G({\xi_3}'' {\eta_3}'';{\xi_3}' {\eta_3}')\nonumber\\   
&=& \sum_{123} D({\hat q}_{12}) {\Delta_1}^{-1}{\Delta_2}^{-1}
{\tilde G}_{3 \xi}({\hat \xi}_3  \eta_3; {{\hat \xi}_3}' {{\eta}_3}')
{{\Delta_1}'}^{-1} {{\Delta_2}'}^{-1}  
\end{eqnarray}
where the second form exploits the symmetry between $\xi,\eta$ and 
$\xi',\eta'$. 
\par
        At this stage, unlike the 2-body case, the reconstruction of the
4D Green's function is {\it {not yet}} complete for the 3-body case, as 
eq.(10.3.2) clearly shows. This is due to the {\it truncation} of Hilbert 
space implied in the ansatz of 3D support to the pairwise BSE kernel $K$ 
which, while facilitating a 4D to 3D BSE reduction without extra charge, 
does {\it not} have the {\it complete} information to permit the {\it reverse}
transition (3D to 4D) without additional assumptions. 
The physical reasons for the 3D ansatz for the BSE kernel have been discussed 
in detail elsewhere [47], vis-a-vis contemporary approaches. Here we look 
upon this ``inverse" problem as a purely {\it mathematical} one.  
\par
        We must now look for a suitable  ansatz for ${\tilde G}_{3 \xi}$ on 
the right hand side of (10.3.2) in terms of {\it known} quantities, so that 
the reconstructed 4D $G$-function satisfies the 3D equation (10.2.12) exactly,
as a ``check-point" for the entire exercise. We therefore seek a structure 
of the form 
\begin{equation}
{\tilde G}_{3 \xi}({\hat \xi}_3  {\eta}_3; {{\hat \xi}_3}' {{\eta}_3}')
= {\hat G}({{\hat \xi}_3} {\hat \eta}_3; {{\hat \xi}_3}' {{\hat \eta}_3}')
\times F(p_3, {p_3}')    
\end{equation}
where the unknown function $F$ must involve only the momentum of the 
spectator quark $\# 3$. A part of the $\eta_3, {\eta_3}'$ dependence has 
been absorbed in the ${\hat G}$ function on the right, so as to satisfy 
the requirements of $S_3$-symmetry for this 3D quantity [47]. 
\par

        As to the remaining factor $F$, it is necessary to choose its
form in a careful manner so as to conform to the conservation of 
4-momentum for the {\it free} propagation of the spectator between two
neighbouring vertices, consistently with the symmetry between $p_3$ 
and ${p_3}'$. A possible choice consistent with these conditions is
the form :
\begin{equation}
F(p_3, {p_3}') = C_3 {\Delta_3}^{-1} {\delta}(\nu_3 - {\nu_3}') 
\end{equation}
Here ${\Delta_3}^{-1}$ represents the ``free" propagation of quark $\# 3$ 
between successive vertices, while $C_3$ represents some residual effects 
which may at most depend on the 3-momentum ${\hat p}_3$, but must satisfy 
the main constraint that the 3D BSE, (10.2.12), be {\it explicitly} satisfied.
\par

        To check the self-consistency of the ansatz (10.3.4), integrate
both sides of (10.3.2) w.r.t. $ds_3 d{s_3}' dt_3 d{t_3}'$ to recover the 
3D $S_3$-invariant ${\hat G}$-function on the left hand side. Next, in 
the first form on the right hand side, integrate w.r.t. $ds_3 d{s_3}'$ 
on the $G$-function which alone involves these variables. This yields
the quantity ${\tilde G}_{3 \xi}$. At this stage, employ the ansatz 
(10.3.4) to integrate over $dt_3 d{t_3}'$. Consistency with the 3D BSE, 
eq.(10.2.12), now demands 
\begin{equation}
C_3 \int \int d\nu_3 d{\nu_3}' {\Delta_3}^{-1} \delta(\nu_3 - {\nu_3}')
= 1 ; (since dt = d\nu) 
\end{equation}
The 1D integration w.r.t. $d\nu_3$ may be evaluated as a contour 
integral over the propagator ${\Delta}^{-1}$ , which gives the pole 
at $\nu_3 = {\hat \omega}_3/M$, (see below for its definition). Evaluating 
the residue then gives 
\begin{equation}
C_3 = i \pi / (M {\hat \omega}_3 ) ;  \quad
{{\hat \omega}_3}^2 = {m_q}^2 + {{\hat p}_3}^2
\end{equation}
which will reproduce the 3D BSE, eq.(10.2.12), {\it exactly}! Substitution
of (10.3.4) in the second form of (10.3.2) finally gives the desired 3-body 
generalization of (10.1.12) in the form 
\begin{equation}
3 G(\xi \eta; \xi' \eta') = \sum_{123} D({\hat q}_{12}) \Delta_{1F} 
\Delta_{2F} D({{\hat q}_{12}}') {\Delta_{1F}}' {\Delta_{2F}}' 
{\hat G}({\hat \xi_3} {\hat \eta_3}; {\hat \xi_3}' {\hat \eta_3}')
[\Delta_{3F} / (M \pi {\hat \omega}_3)]      
\end{equation}
where for each index, $\Delta_F = - i {\Delta}^{-1}$ is the 
Feynman propagator.
\par

        To find the effect of the ansatz (10.3.4) on the 4D BS 
{\it {wave function}} $\Phi(\xi \eta; P)$, we do a spectral reduction 
like (10.2.13) for the 4D Green's function $G$ on the left hand side of 
(10.3.2). Equating the residues on both sides gives the desired 4D-3D 
connection between $\Phi$ and $\phi$:
\begin{equation}
\Phi(\xi \eta; P) = \sum_{123} D({\hat q}_{12}){\Delta_1}^{-1}{\Delta_2}^{-1}
\phi ({\hat \xi} {\hat \eta}; P) \times 
\sqrt{{\delta(\nu_3 -{\hat \omega}_3/M)} \over{M {\hat \omega}_3 
{\Delta}_3}} 
\end{equation}
{\it defines} the 4D wave fn in terms of piecewise vertex fns $V_i$, as   
\begin{equation}
\Phi(p_1p_2p_3) \equiv {{V_1+V_2+V_3} \over {\Delta_1 \Delta_2\Delta_3}}    
\end{equation}
From (10.3.8-9), we infer the baryon-$qqq$ vertex function $V_3$ corresponding
to the `last' interaction in the $12$-pair  as 
\begin{equation}
V_3 = D({\hat q}_{12}\phi({\hat \xi},{\hat \eta}) \times {\sqrt 
{2\Delta_3\delta({\nu_3}^2 M^2-{\hat \omega}_3^2)}}
\end{equation}
and so on cyclically. (The argument of the 
$\delta$-function inside the radical for $V_3$ simplifies to $p_3^2+m_q^2$). 
This expression is essentially the same as eq.(5.15) of ref.[10b], which
had been obtained from largely intuitive considerations. 
\par
        To account for the appearance of the 1D $\delta$-fn under radical
in (10.3.10), it is explained elsewhere [47] that it has nothing to do with 
connectedness [88] as such, but merely reflects a `dimensional mismatch' due 
to the 3D nature of the pairwise kernel $K$ [16] imbedded in a 4D Hilbert 
space. This in turn is the result of the `contact' nature (in time dimension) 
of the pairwise interaction, somewhat analogous to a Fermi $\delta$-fn 
potential to simulate the effect of the (short range) {\it nuclear} $n-p$ 
interaction in the `molecular' problem of (specular) neutron scattering 
by a hydrogen molecule [91]. As a further self-consistency check, it is 
instructive to compare (10.3.10) with one obtained by taking the limit 
of a point interaction, which amounts to setting $K=Constant$ in the entire
derivation above. This structure [47] which is worked out in Appendix C, is 
free from radicals, and explicitly 4D-invariant, in agreement with the 
so-called NJL-Faddeev (contact [4]) model [92] of 3-particle scattering.

\section{Fermion Quarks: QCD-Motivated $qqq$ BSE}
\par
        We now turn to the more realistic case of fermion quarks for which
we shall draw freely from a relatively recent analysis [29b] of a $qqq$ 
baryon, which is basically a 3-body generalization of subsection 4.3
for the two-body case. For simplicity of description, without sacrificing the
essential physics, we shall specialize to equal mass kinematics (mass=$m_q$).
 
\subsection{3D Reduction of 4D $qqq$ BSE}
\setcounter{equation}{0}
\renewcommand{\theequation}{11.\arabic{equation}}

The starting 4D BSE has the form (c.f. (9.6)):
\begin{equation}
(2\pi)^4 \Psi(p_1p_2p_3)= i S_F(p_1)S_F(p_2) \sum_{123} \int 
d^4 q_{12}' K({\hat q}_{12},{\hat q}_{12}') \Psi(p_1p_2p_3')
\end{equation} 
where the kernels $K_{ij}$ are given by eqs.(4.18-19) for each $ij$ pair, 
except for the Casimir value of the color factor  $F_{12}$ $\equiv $
$\lambda_1.\lambda_2/4$ for the ${\bar 3}$ state of a $qq$ pair (to produce 
a color-singlet baryon), which is just half its `singlet' value for a 
$q{\bar q}$ pair. And of course the $DB{\chi}S$ mechanism is built-in as in 
the two-body case of Sec.4.3. The Gordon reduction of the product of two 
$\gamma_\mu$-matrices [10] also goes through as in Sec.4.3, leading to [29b]:
\begin{equation}
\Phi(p_1p_2p_3) = \sum_{123}{{-i F_{12}} \over {(2\pi)^4 \Delta_1\Delta_2}}
\int d^4q_{12}'V^{(1)}_\mu V^{(2)}_\mu V({hat q}_{12},{\hat q}_{12}'
\Phi(p_1'p_2'p_3)
\end{equation}
where, following the steps of Sec.4.3, the `bosonic' $\Phi$-fn is related to 
the fermionic $\Psi$-fn, as in eq.(4.24), by [10b,29b]:
\begin{equation}
 \Psi(p_i)= {\Pi_1}^3 S_F^{-1}(-p_i) \Phi(p_i); \quad \Delta_i = m_q^2+p_i^2
\end{equation}
while the 4-vectors $V_\mu^{(i)}$ are given by eq.(4.25). 
\par
        Next, for the 3D reduction of eq.(11.2), we need to define the
transverse ${\hat p}_i$ and longitudinal $\nu_i$ components of the 4-momenta
$p_i$, as in Sec.10.2, eqs.(   ), and multiply the pairs of $V_\mu$-fns,
as in Sec.4.3, replacing in the process the longitudinal components $\nu_i$
by their {\it on-shell} values ${\hat \omega}_i/M$, where ${\hat \omega}_i^2$
= $m_q^2+{\hat p}_i^2$  [29b], uniformly from such products [93a]. Now define
the 3D wave function $\psi$ in the as in Sec.10.2, viz., [10b,29b]:
\begin{equation}
\psi({\hat p}_1{\hat p}_2{\hat p}_3) = \int ds_idt_i \Phi(p_1p_2p_3);
\quad {\sqrt 3}s_3=\nu_1-\nu_2; \quad 3t_3 =-2\nu_3+\nu_1+\nu_2 
\end{equation}
The product $ds_idt_i$ is cyclically invariant, so that the definition (11.4)
can be taken over for all the three terms on the RHS of (11.2), with proper
indexing. The rest of the procedure is straightforward, and follows closely
the pattern laid out in the original formulations. Thus one integrates both
sides of (11.2) w.r.t. $ds_3dt_3$, making use of (11.4) as well as the 
measure $d^4q_{12}'$ = $d^3{\hat q}_{12}'Mds_3'{\sqrt 3}/2$ to give on its
RHS $ \int ds_3'dt_3 \Phi'$ = $\psi({\hat p}_1'{\hat p}_2'{\hat p}_3)$. The
additional $ds_3$-integration on the RHS is expressed by the result [10b]:
\begin{equation}
{{\sqrt 3} \over 2} \int {{M ds_3} \over {\Delta_1\Delta_2}} ={{2i\pi} \over
D_{12}}; \quad   
D_{12}= -\Omega_{12} \lambda[(M^2(1-\nu_3)^2,\omega_1^2,\omega_2^2]/
{2M^2(1-\nu_3)^2}   
\end{equation}
\begin{equation}
2/\Omega_{12}= {\hat \mu}_{12}/\omega_1 +{\hat \mu}_{21}/\omega_2;  \quad
{\hat \mu}_{12;21} = {{1-\nu_3}\over 2} \pm {{\omega_1^2-\omega_2^2} \over
{2M^2(1-\nu_3)}}
\end{equation}
where $\nu_3$ has its on-shell value $\omega_3/M$ in the foregoing equations,
as befits a spectator quark in the first term of(11.2). The resultant 3D
reduction of (11.2) now takes the form:
\begin{equation}
\psi({\hat p}_1{\hat p}_2{\hat p}_3)= \sum_{123}{F_{12} \over 
{(2\pi)^3 D_{12}}} \int d^3{\hat q}_{12}'V^{(1)}.V^{(2)} 
V({\hat q}_{12},{\hat q}_{12}') \psi({\hat p}_1'{\hat p}_2'{\hat p}_3)
\end{equation}

\subsection{Reduction to 6D Harmonic Basis}  
 
The next task is to reduce eq.(11.7) to a more transparent form suitable for
numerical treatment. To that end we base our procedure [29b] on the expected
smallness of the $S_3$-invariant quantity $\delta$= $M-\omega_1-\omega_2-
\omega_3$ compared to $\omega_i$ and/or $M$. This gives the crucial result:
\begin{equation}
D_{12}= -4\omega_1\omega_2 \delta+ O(\delta^2); \quad 
-\delta = \omega_1+\omega_2+\omega_3-M
\end{equation}
which ensures that in (11.7), all the three terms on its RHS have a 
{\it {common denominator}} $\delta$ which, when transferred to the LHS,
serves as a natural `energy denominator' for the entire $qqq$ equation.
[Since the terms of $O(\delta^2)$ in (11.8) are fully calculable, any
effect on their omission can be estimated perturbatively if necessary].     
Next, from Sec4.3, the confining part of $V({\hat q}_{12},{\hat q}_{12}')$
is harmonic for $ud$-quarks $(A_0=0)$, so that a perturbative treatment is
possible, based on the (harmonic) confining part of $V({\hat q},{\hat q}')$:
\begin{equation}
V_{con}= {3 \over 4}(2\pi)^3 {\omega_{qq}}^2 [\nabla_{\hat q}^2 +
C_0/\omega_0^2] \delta^3({\hat q}-{\hat q}'); \\
{\omega_{q_1 q_2}}^2 = 4M_{12}{\hat \mu}_{12}{\hat \mu}_{21} \omega_0^2
\alpha_s(M_{12}^2)
\end{equation}
In this formula, the definitions (11.7) for the fractional momenta 
${\hat \mu}_{12, 21}$ conform to their Wightman-Gaerding [56] definitions
for unequal mass kinematics, a la eq.(4.1) of Section 4, since the unequal
masses arise from the mass-shifts $m_q \rightarrow \omega_i$ of the quarks
$(1,2)$ in the presence of the spectator $\#3$. Since such shifts are small,
it is fairly accurate to approximate the fractional momenta as $(1-\nu_3)/2$
each, while $M_{12} \approx M-\omega_3$ only. Now to emphasize the 3D 
character of the various momenta, define the pairwise items: 
\begin{equation}
{\bf L}_{ij} = -i{\bf q}_{ij} \times {\bf \nabla}_{ij}; \quad
{\bf \nabla}_{12}={\bf \nabla}_1-{\bf \nabla}_2; \quad 
2{\bf q}_{12}={\bf p}_1-{\bf p}_2; \quad
{\hat Q}_{12}=4{\bf q}_{12}^2 {\bf \nabla}_{12}^2 +
8{\bf q}_{12}.{\bf \nabla}_{12}+6
\end{equation}      
Also to take full advantage of the HO form (11.9), recast the (small) energy 
denominator $\delta$ in the alternative form [29b]:
\begin{equation}
-2M\delta \approx (\omega_1+\omega_2+\omega_3)^2 -M^2 \leq 
3(\omega_1^2+\omega_2^2+\omega_3^2) \equiv \Delta= 
9m_q^2 + 9(\xi^2+\eta^2)/2 - M^2
\end{equation}
The resulting `Master Equation' (11.7) is in pairwise notation [29b]:
\begin{equation}
\Delta \psi = (W_{con}+ W_{OGE}) \psi  \\ \nonumber
W_{con}= M\omega_0^2 \sum_{123}(1-\nu_3)^2 {\alpha_{12}}^s M_{12} \times
[{\bf \nabla}_{12}^2 +{C_0 \over \omega_0^2} +{1 \over {\omega_1\omega_2}} NCT]
\end{equation}
\begin{equation}
NCT = {{\hat Q}_{12} \over 4}-{C_0 \over \omega_0^2}-{\bf L}_{12} \dot
({\bf \sigma}_1+{\bf \sigma}_2) +{i \over 2} {\bf p}_3 \times   
{\bf \nabla}_{12}\dot ({\bf \sigma}_1-{\bf \sigma}_2)-{\bf \sigma}_1 \dot
{\bf \sigma}_2
\end{equation}
and the $OGE$ term in a `mixed' $({\bf r},{\bf p})$ representation is [29b]:
\begin{equation}
W_{OGE}={4M \over 3}\sum_{123}\alpha_{12}^s[{1 \over r_{12}}+{1 \over 
{\omega_1\omega_2}}({\bf q}_{12}{1 \over r_{12}} \dot {\bf q}_{12} +
\pi \delta^3({\bf r}_{12})(1-2{\bf \sigma}_1\dot {\bf \sigma}_2/3))+ etc]
\end{equation}  
The $OGE$-term is calculated perturbatively in a 6D HO basis given by the
main confining term in (11.12) [29b], which in a common $(\xi,\eta)$ basis 
[29a], now provided by eqs.(10.2.1-6) of Section 10, reads [29b]:
\begin{equation}
\Delta \psi = M \omega_0^2(1-{\bar \nu})^2{\bar \alpha_s}(M-{\bar \omega}) 
[2\nabla_\xi^2+2\nabla_\eta^2+{{C_0 (M^2-3m_q^2+\Delta)} \over 
{2\omega_0^2{\bar \omega}^2}}+ 
{{{\hat Q}_B-8{\bf J}\dot {\bf S}+18} \over {4{\bar \omega}^2}}]\psi
\end{equation}  
where the operators ${\hat Q}_B$, etc are defined in [29a], and the symbols
$(\alpha, \omega, \nu)$ with `bars' represent their `average' values [29b].
From this H.O. equation, the scale parameter $\beta$, analogous to the 2-body
quantity (4.31) may be inferred as [29b]
\begin{equation}
\beta^4= 4M\omega_0^2 {\hat \alpha_s}(1-{\bar \nu})^2(M-{\bar \omega})/9; \quad
{\hat \alpha_s}= 1/[{\bar \alpha}_s^{-1}-2C_0M(1-{\bar \nu})^2/(M-{\bar \omega})] 
\end{equation}
so that the basis function $\psi_0$ in its ground state is $exp[-{1 \over 2}
(\xi^2+\eta^2)/\beta^2]$; similar functions exist for $L$-excited states [40],
providing a basis for perturbative treatment [29a] of the $OGE$ terms (11.14).

\subsection{Complex HO Basis for $qqq$ States}

It is however mathematically simpler to convert eq.(11.15) to a complex basis.
To this end we define the complex dimensionless 3-vectors $z_i$, $z^*_i$, and
their (derivative) conjugate momenta ${\partial}_{z_i}$, as        
\begin{equation}
{\sqrt 2}\beta [z_i; z^*_i] = \xi_i \pm \eta_i; \quad 
{\sqrt 2}\beta^{-1} [\partial_{z_i}; \partial_{z^*_i}] = 
\partial_{\xi_i} \mp \partial_{\eta_i}
\end{equation}
For the construction of angular momenta in complex basis, see Appendix D.
\par
        A more convenient basis for handling the various terms in (11.15)
is provided by the creation/annihilation operator representation [29b, 46b]
defined by {\it two} sets of complex operators
\begin{equation}
{\sqrt 2}a_i=z_i+\partial_{z^*_i}; \quad {\sqrt 2} a^*_i =z^*_i 
+\partial_{z_i}; \quad
{\sqrt 2} a^{\dagger}_i= z^*_i-\partial_{z_i}; \quad
{\sqrt 2}a^{*{\dagger}}_i = z_i- \partial^*_{z_i}
\end{equation}
which satisfy the commutation relations
\begin{equation}
[a_i, a^{\dagger}_j]= [a^*_i, a^{*{\dagger}}] = \delta_{ij}
\end{equation}
with all other pairs commuting. In the next subsection {\bf 11.4}, we
define the number operators $N_c$ and $N^*_c$ which now play the role of 
$N_\xi$ and $N_\eta$, but unlike the latter, the former can be simultaneously 
diagonalized; so their sum $N$ and difference $N_a$ are {\it both} constants 
of motion. Together with certain two-step operators, they form several sets 
of $SO(2,1)$ algebras (described below), which diagonalize the momentum 
dependent operators $Q_B$, etc, in terms of their respective Casimirs [46b], 
so that the solution  of eq.(11.15) takes a simple algegraic form [29b]:
\begin{equation}
F(M,N) \equiv F_{con}(M,N)+F_{OGE}(M.N) = N+3
\end{equation}
where the first term is given by Eq.(4.19) of [29b], while the second term
lends itself to a simple perturbative treatment (see [29b] for details). 
Appendix D gives a summary of the normalized $SU(6) \times O(3)$ structures 
of the 3D $\psi$-fns in the complex basis, which are needed for calculating
the $F_{OGE}$ term of eq.(11.20) above.       

\subsection{$SO(2,1)$ Algebras of Bilinear Operators}
 
We start by defining the number operators $N_c$, $N^*_c$, and the mixed 
quantities $N_m$, $N^*_m$ [46b]
\begin{equation}
N_c = a^{\dagger}_i a_i; \quad N^*_c= a^{*{\dagger}}_i a^*_i; \quad
N_m = a_i a^{*{\dagger}}_i; \quad N^{\dagger}_m = N^*_m = a^*_i a^{\dagger}_i
\end{equation}
and their linear combinations 
\begin{equation}
N = N_c + N^*_c = N_\xi +N_\eta; \quad N_a = N_c - N^*_c
\end{equation} 
Note that both $N$ and $N_a$ are simultaneously diagonal in this (complex)
representation, while in the real $(\xi, \eta)$ basis, only their sum is
diagonal. Next define the two-step operators (and their h.c.'s) [46b]:
\begin{eqnarray}
A           &=&  2a_ia^*_i; \quad C = a_ia_i; \quad C^* = a^*_ia^*_i  \\ \nonumber
A^{\dagger} &=&  2a^{\dagger}_ia^{*{\dagger}}_i; \quad C^{\dagger} = 
a^{\dagger}_ia^{\dagger}_i; \quad C^{*{\dagger}} = a^{*{\dagger}}_i
a^{*{\dagger}}_i
\end{eqnarray}
Now the trio $A$, $A^{\dagger}$ and $N$ form an $S_3$-symmetric set, whose
normalized forms
\begin{equation}
Q_+ = A^{\dagger}/2; \quad Q_- = -A/2; \quad Q_3 = (N+3)/2
\end{equation}
form an $SO(2,1)$ algebra (bounded from below) with the Casimir [93, 46b]
\begin{equation}
u(u+1) = {\bf Q}^2 \equiv  -(AA^{\dagger}+A^{\dagger}A)/8 +(N+3)^2/4
\end{equation}  
where $u(u+1)$ = $+3/4$ for even $N$ and $+2$ for odd $N$, while the 
eigenvalues of $Q_3$ are $-u+k$, (with $k=0,1,2,...$). These imply that 
$u=-3/2$ and $u=-2$ for even and odd $N$ respectively. Similarly the
mixed symmetric set $(C,C^{\dagger},N_c)$ form an $SO(2,1)$ algebra in
the normalized form [46b]
\begin{equation}
Q_{c+} = C^{\dagger}/2; \quad Q_{c-} = -C/2; \quad Q_{c3} = 
{1 \over 2}(N_c + 3/2)
\end{equation}  
with the corresponding Casimir [46b]
\begin{equation}
u_c(u_c+1) \equiv {\bf Q_c^2}= -(CC^{\dagger}+C^{\dagger}C)/8+(N_c+3/2)^2/4
\end{equation}
This spectrum is again bounded from below [93], with the eigenvalues
$Q_{c3}=-u_c+k$, where $u_c=-3/16$ for even $N_c$ and $u_c=+5/16$ for
odd $N_c$. An identical structure holds for the `starred' operators
$(C^*,C^{*{\dagger}},N^*_c)$, with the same eigenvalues. Finally the
trio $(N_m, N^{\dagger}_m=N^*_m, N_a)$ which is $S_3$-antimmetric, satisfy
a `normal' $SO(3)$ algebra [46b]:
\begin{equation}     
[N_m, N_a]=2N_m; \quad [N^{\dagger}_m, N_a]=-2N^{\dagger}_m; \quad
[N_m, N^{\dagger}_m]=-N_a
\end{equation}
with spectra bounded from both above and below. The corresponding Casimir is
\begin{equation}
s(s+1)=(N_m N^{\dagger}_m +N^{\dagger}_m N_m)/2 +N_a^2/4
\end{equation}
The spectrum is here determined from the condition that both $N_c$ and 
$N^*_c$ are non-negative integers. The result is [46b]
\begin{equation}
-N \leq N_a \leq N; \quad s = N/2 
\end{equation}

\subsection{``Exotic'' $qqq$ States}

The comparison of eq.(11.20) with the baryon spectra is described at some 
length in [29b], and it is not our intention here to go into these details
afresh. Instead, we shall end this Section with some qualitative analysis 
on the capacity of this model to identify some exotic baryonic states which 
have for long remained elusive. The main reason for such optimism stems from 
the precise predictions on the `spectroscopic' locations of the states 
on the one hand, and the possibility of making more reliable 
$SU(6) \otimes O(3)$ assignments  for such states on the basis of their 
{\it decay} characteristics which the model also allows within its broad
framework. To see the logic, a good calibration is first provided by the 
fairly accurate location of several `known' states in a parameter-free 
manner; see Table I of [29b] for comparison. With this first check, a more
sensitive test is now a comparison of the alternative $SU(6) \otimes O(3)$
assignments for the mass locations of the same states; see Table II of (29b).      
Specifically the competition is between the ${\bf 56}, {odd}^-$ and 
${\bf 70}, {odd}^-$ assignments for $\Delta$-like states for the same total 
quantum number $N$. The question is clearly of physical interest since         
in the entire history of baryon spectroscopy ${\bf 56},{odd}^-$ states
have suffered from popular perceptions of elusiveness, despite occasional
attempts to the contrary [91]. The analysis in [29b] suggests that the 
${\bf 56}$ assignment has a slight edge over ${\bf 70}$, at least
for a couple of odd-patity states by virtue of `location', but a more
sensitive test requires a more detailed comparative study of the decay
and/or production characteristics that these alternative assignments provide,
vis-a-vis the data (which are still elusive). In this respect it was shown 
in [29b] that the general mechanism of `direct' versus `recoil' modes of 
single-quark transitions [45b], do {\it not} inhibit in any way the 
production of natural parity ${\bf 56}^-$ states w.r.t. the corresponding 
${\bf 70}^-$ states, (perhaps contrary to popular beliefs).    
    
\subsection{CIA vs CNPA for Fermionic $qqq$ Dynamics}

In the foregoing we have mostly described the CIA predictions [29b] on the
baryon spectra. How about the corresponding $qqq$-scenario with the other
{\bf MYTP}-governed CNPA dynamics  whose $q{\bar q}$ counterpart has been
employed in Sections 4-6 ? The reason for avoiding this exercise for the
$qqq$ problem is one of pedagogical necessity. For, from the results of
Sections 4-6, it has been fairly clear that the earlier NPA treatment [40]
based on the old-fashioned NP-language [35] formally provides the same CNPA 
structure of 3D BSE as well as 4D vertex functions for $q{\bar q}$ systems,  
so that a similar $qqq$ structure should be expected. In this respect, the
baryon spectral results [29a] based on the old-fashioned NPA treatment are
already available in detail [40], and the comparison with the CIA treatment
[29b] shows considerable overlap therewith. As for the reconstruction of the
$qqq$ vertex function under CNPA, a closely analogous treatment
akin to Section 10 formally leads to almost identical results, with the
CIA-CNPA correspondence already indicated in Section 4. 
\par
        How about the reconstruction of the 4D $qqq$ vertex function for 
fermion quarks ? Again the treatment, which is analogous for both CIA and 
CNPA, consists in reducing the fermionic structure to an effectively scalar
problem via eq.(11.3) which relates the fermionic BS wave function $\Psi$
to the `scalar' function $\Phi$, fits in smoothly with the treatment
outlined in Section 10 for spin zero quarks, with almost no change, thus
rendering unnecessary another fresh formulation for fermions. As for quark 
loop applications to the $qqq$ problem, the general problem of `Lorentz 
mismatch' of 3D wave functions in a quark-loop integral, that had led us to 
abandon the CIA treatment in favour of CNPA for the $q{\bar q}$ problem 
(see, e.g., Sections 5-6), is also encountered in the $qqq$ case, so that it 
is profitable to adopt CNPA [41] for baryonic transition amplitudes as well.
\par
        The only exceptions are two-loop integrals, as in the self-energy 
problem (Sect.7), or one-loop integrals, as in the vacuum condensate problem 
(see Sect.8), where this pathology is just avoided. Full-fledged baryon-loop 
calculations are still being developed, so such topics are not intended for 
a detailed coverage in this article, except for indicating the results of a
recent calculation of $SU(2)$ $n-p$ mass splitting [95] analogous to the 
treatment of pseudoscalar mass splittings (Sect.7) [32a] by this method.
Thus, using the same value $(4 MeV)$ of the `current' $d-u$ mass difference,
the total $n-p$ mass difference works out as $1.28 MeV$ [95], to be compared
with the experimental value of $1.29 MeV$ [13], except for possible QED 
gauge corrections [63]. On this last item, an indication of the expected
correction is available from its effect on the Kaon e.m. mass difference,
which yields a $\sim 60 \%$ upward revision on its (uncorrected) value of
about $1 MeV$ [32a]; see Appendix C for an estimation of this correction. 
If this value for the kaon case is taken as rough indication of the seme
effect expected for the nucleon case, then (on a proportionate basis) the
QED gauge corrected value for $n-p$ mass difference comes down to $\sim 1MeV$.
For details of this methodology, see [95].              
        
\section{Summary And Conclusions}

In this article an attempt has been made to present a somewhat `less than
conventional' BSE-SDE formalism based on the Markov-Yukawa Transversality
Principle ({\bf MYTP}) [15] on the one hand, and a strongly QCD motivated
4-fermion Lagrangian which generates the BSE-SDE framework by breaking its 
chiral symmetry dynamically ($DB{\chi}S$) [23-27], on the other. The 
{\bf MYTP} mechanism provides an {\it exact} interconnection between the 3D 
and 4D forms of the BSE, so that both can be used interchangeably, a facility 
which does not seem to exist in other alternative 3D BSE formalisms [39], or 
the null-plane formulations--both non-covariant [35] and covariant [36-38]. 
This twin property of the {\bf MYTP}-governed BSE formalism [16], termed 
3D-4D BSE for short, gives rise to a natural `two-tier' description [40], 
the 3D sector (with its relativistic Schroedinger-like BSE) being appropriate 
for making contact with the hadron spectra [13], while the reconstructed 4D 
BSE yields a vertex function which allows the direct use of the language of
Feynman diagrams for evaluating transition amplitudes as 4D loop integrals. 
(This contrasts with other 3D formulations [35-39] which require specialized 
versions of Feynman diagrams [37] for calculating loop integrals).
\par
        At a more quantitative dynamical level, both $q{\bar q}$ and $qqq$ 
hadrons are amenable to a unified treatment, since their respective BSE's 
emanate from a common (input) chiral 4-fermion Lagrangian with a gluon-like 
propagator whose `color-factor' has the right relative strengths for both
systems. And while the 3D-4D structure of the $q{\bar q}$ BSE [16,28], as 
well as the 3D reduction of the $qqq$ BSE [29], have been around for some
time, the missing link of a {\it reconstructed} 4D BS wave function for
the $qqq$ system (only conjectured in [10b]) has now (hopefully) been 
supplied through a formal derivation in {\bf Sect.10} via Green's Function 
techniques [47]. Indeed the main emphasis in this Article has been on
the `second stage' of this two-tier formalism, relating to the calculation
of 4D quark-loop integrals, of which some selected examples have been given
in {\bf Sects.5-7}, to bring out the feasibility of its applications to the 
meson sector. The corresponding applications to baryonic amplitudes via loop
integrals are still being developed, and only a `pilot' example, relating to
$SU(2)$ mass splittings [95], is as yet available. However the scope (and 
feasibility) of such applications is quite substantial [96]. 
\par
        The capacity of this BSE-SDE formalism to relate its parameters 
to the `vacuum condensates' of QCD-SR theory [2] has been sought to be 
brought out in {\bf Sect.8}, wherein it has been shown that several types 
of condensates (both direct and induced) lend themselves with great ease 
to this simple treatment [30], while the corresponding QCD-SR treatments 
[82-85] often need additional ansatze for their evaluation. This facility 
it owes to its (input) gluon propagator on the one hand, and the (derived)
mass function $m({\hat p})$ from the SDE solution [11] on the other. The
two fundamental parameters [11] of the infrared gluon propagator are not 
calculable within this (Bethe's Second Principle oriented) framework, but
they are firmly rooted in spectroscopy, as their contact [28-29] with 
data [13] reveals. Indeed the performance of this spectroscopy-oriented 
BSE-SDE framework in predicting the vacuum condensates, can be directly
attributed to its {\it off-shell} structure. 
\par 
        An inportant (new) aspect of this Study has been a demonstration 
of the powers of {\bf MYTP} extending from its original mandate [15] of 
transversality in terms of `Covaiant Instantaneity' (CIA) [16,17], to a
wider `transversality' on a Covariant Null Plane (CNPA) [41], thus vastly 
enhancing the applicability of this important Principle. In this article we 
have tried to present both CIA and CNPA on very similar lines, but the 
mathematical viability of the latter [41] seems to exceed that of the former 
[16], inasmuch as a CIA treatment of triangle (and higher) loop integrals is 
fraught with problems  of `Lorentz mismatch' of different CIA wave 
functions, leading to ill-defined integrals due to the presence of time-like 
momentum components in the exponential/gaussian factors inside the integrals
concerned [57]. This problem, which has been known since the FKR paper [25],
is properly circumvented in CNPA, except for the (less serious) problem of 
dependence on the `null-plane orientation' which can be tackled through other
means, e.g., a simple device of `Lorentz completion' which yields an 
explicitly Lorentz-invariant structure. This has been illustrated in 
{\bf Sect.5} for the pion form factor which shows the expected high energy 
behaviour as well as very reasonable results [58-62] in both the high and 
low energy regimes. For more general three-hadron amplitudes [31] too,  
similar calculations in {\bf Sect.6} show that the anomalies of ill-defined
4D loop integrals are absent in a CNPA treatment. The only exceptions are 
two-quark loops [32] ({\bf Sect.7}), where both methods, CIA and CNPA, work.  
\par
        Clearly, the {\bf MYTP} is a very powerful Principle which helps 
organize a whole spectrum of phenomena under a single umbrella. It has been
possible to study only a very few (though illustrative) examples to bring
out its powers, but its potential is vast, and warrants many more of such
applications. More importantly, the 3D-4D structure of BS dynamics provided 
by {\bf MYTP} takes in its stride the spectroscopy sector as an integral 
part of the dynamics, as envisaged long ago by Feynman et al [25].   
\par

        A good part of the logic behind this Article was evolved during 
my tenure of an INSA Professorship (1989-94), while the actual contents  
of this Article include both published and unpublished unpublished material
developed subsequently, in my capacity as a free-lance worker (unattached
to any Institution), as part of an ongoing research process.

\section*{Appendix A: Derivation of $F(k^2)$ and $N_H$ for P-meson}

\setcounter{equation}{0}
\renewcommand{\theequation}{A.\arabic{equation}}

In this Appendix we outline the main steps to the derivation of the P-meson
form factor (5.9), as well as the Normalizer (5.8), given in Sec.5 of Text. 
Collecting the various pieces after $p_{2n}$-pole integration, gives for (5.1)
\begin{equation}
F(k^2) = 2(2\pi)^3 N_n(P)N_n(P'){\hat m}_1 \int d^2q_\perp dz_2 P.n 
g(z_2)
e^{-q_\perp^2/\beta^2-f(z_2)/\beta^2} + [1 \Rightarrow 2];
\end{equation} 
\begin{equation}    
f(z_2)= M^2 \eta_k^{-2}[ \theta_kz_2^2-z_2{\hat k}^2{\hat m}_2
+\theta_k {\hat m}_2^2{\hat k}^2/4]; 
\end{equation}
\begin{equation}
 D_n+D_n'=4{\bar P}.n 
[q_\perp^2+ M^2(z_2^2- z_2{\hat k}^2{\hat m}_2/2
+ {\hat m}_2^2 {\hat k}^2/4)/eta_k-\lambda/{4M^2}];
\end{equation} 
\begin{equation}
 g(z_2)= {{D_n+D_n'} \over 4}{{M^2+{\delta m}^2} \over {M^2+k^2/4}}+h(z_2);
\end{equation}
\begin{equation}
h(z_2) = 2{\bar P}.n({\hat m}_2-z_2)[M^2-{\delta m}^2+
{\hat m}_2M^2({\delta m}^2-M^2-k^2/2)/(M^2+k^2/4)]
\end{equation}
The integration over $q_\perp$ and $z_2$ are both routine, the latter with 
a translation $z_2 \rightarrow z_2+{1 \over 2}{\hat m}_2 {\hat k}^2/\theta_k$,
to reduce the gaussian factor to the standard form. Note that, unlike the
conventional (Weinberg) form [39a] of light-front dynamics, the present 4D 
form which permits off-shellness of the internal momenta, does not restrict  
in principle the limits of $z_2$ integration. Thus after the translation, 
the odd-$z_2$ terms can be dropped, and $f(z_2)$ reduces to
\begin{equation}
f(z_2)= M^2 z_2^2\theta_k/\eta_k^2+(M{\hat m}_2{\hat k})^2/{4\theta_k}
\end{equation}
while the $g$-function is a sum of two pieces $g_1+g_2$:
\begin{equation}
g_1 = \eta_k[q_\perp^2 +M^2z_2^2/\eta_k + 
{1 \over 4} M^2 {\hat m}_2^2{\hat k}^2(1+3{\hat k}^2/4)/\theta_k^2 
-\lambda/{4M^2}] (1+{\delta m}^2/M^2); 
\end{equation}
\begin{equation}
g_2 = 2\eta_k (M^2-{\delta m}^2){\hat m}_2 /\theta_k +
2({\delta m}^2-M^2-k^2/2){\hat m}_2^2 {\eta_k}^2 /\theta_k
\end{equation} 
\par
        Before writing the final result for $F(k^2)$, it is instructive at
this stage to infer the normalizer $N_H$ of the hadron, obtained by setting 
$k_\mu=0$, and demanding that $F(0)=1$. This gives after some routine steps: 
\begin{equation}
N_n(P)^{-2} = 2M (2\pi)^3 (P.n/ M)^2 \int d^3{\hat q}
e^{-{\hat q}^2/\beta^2} G(0);
\end{equation}
\begin{equation}
G(0) = [(1+{\delta m}^2/M^2)({\hat q}^2-\lambda/{4M^2})
+2{\hat m}_1{\hat m}_2 (M^2-{\delta m}^2)]           
\end{equation}
where ${\hat q}=(q_\perp, Mz_2)$ is effectively a 3-vector, in conformity
with the requirements of the angular condition [35d,38], which gives a 
formal meaning to its third component $q_3$= $M q.n/P.n$ = $Mz_2$. The 
normalization factor $N_n(P)$ is also seen to vary inversely as $P.n$, 
while the multiplying integral is clearly independent of the NP-orientation
$n_\mu$. To exhibit this $P_n$ independence more explicitly, define a 
`reduced normalizer' $N_H$ which equals $N_n(P) \times P.n/M$ and 
gives for $N_H^{-2}$ the Lorentz-invariant result, eq.(5.8) of Text.
\par    
        Now insert the result $N_n(P)$= $M N_H/P.n$ on the RHS of (A.1), and 
note, via eq.(5.3), that 
\begin{equation}
{M^2 /over {P.nP'.n}} = {M^2 /over {{{\bar P}.n}^2 \eta_k}}; \quad
\eta_k = 1-{\hat k}^2/4.
\end{equation}
One now checks that the factors ${\bar P}.n$ cancel out completely, and the 
evaluation of the gaussian integrals leads after a modest algebra to eq.(5.9) 
of Text, where $G({\hat k})$, after collecting from eqs.(A.6-8), is given by
\begin{equation}
G({\hat k}) = (1+{\delta m}^2/M^2)h({\hat k})
+2(M^2-{\delta m}^2){\hat m}_2/\theta_k +2{\hat m}_2^2 \eta_k \theta_k^{-1}
({\delta m}^2-M^2-k^2/2);
\end{equation}  
\begin{equation}
h({\hat k}) =  (1+{\eta_k}^2/{2\theta_k})\beta^2- \lambda/{4M^2}+  
(M{\hat m}_2{\hat k}/2{\theta_k})^2(1+{3 \over 4}{\hat k}^2); \quad
{\delta m}= m_1-m_2.
\end{equation}
 
\section*{Appendix B: Gauge Corrections to Kaon E.M. Mass} 
\setcounter{equation}{0}
\renewcommand{\theequation}{B.\arabic{equation}}
\par
        We outline here a practical procedure to evaluate the gauge corrections
to the e.m. self-energy of a $q{\bar q}$ system, vide Fig.1b of [18a]. For 
brevity we shall refer to the figures of KL [63] in their notation without 
drawing them anew. Thus Fig,1b of [18a] corresponds to fig 1a of KL [63], 
except for the presence of the hadron lines at the two ends. We shall call 
this simply `1a', with the understanding that the hadron lines are `attached' 
to 1a.  For the actual mathematical symbols (including phase conventions) we 
shall draw freely from [18a], without explanation. In [18a], only 1a of 
KL [63] was calculated, but now one must add 2(a,b,c,d,e) of KL [63], all 
with hadron lines understood at the two ends of each. There is no need to 
calculate 1b or 1c of [63] which are mere e.m. self-energies of single 
quarks (g.i. by themselves), and are routinely absorbed in quark mass 
renormalization (of little significance in this study which has these masses 
as inputs).       
\par    
        A new ingredient is a 4-point vertex in each of 2(a,b,c,d), and 
{\it two} 4-point vertices in 2e, except that the word `point' is now
understood as an extended structure characterized by the hadron-quark vertex 
function $D({\hat q})\phi({\hat q})$ where one must insert a photon line in 
each such $Hq{\bar q}$ blob. Since it is {\it not} a standard point vertex, 
the method [63] of inserting exponential phase integrals with each current is 
not technically feasible; instead we may resort to the simple-minded 
substition $p_i-e_iA(x_i)$ for each 4-momentum $p_i$ (in a mixed $p,x$ 
representation) occurring in the structure of the vertex function, which has
the same physical content, at least up to first order in the e.m. field,
without further comment. This amounts to replacing each ${\hat q}_\mu$ 
occurring in $\Gamma({\hat q})$ = $D({\hat q})\phi({\hat q})$, by
${\hat q}_\mu -e_q A_\mu$, where $e_q = {\hat m}_2 e_1 -{\hat m}_1 e_2$.  
The net result in the first order in $A_\mu$ is a first order correction
to $\Gamma({\hat q})$ of amount $e_q j({\hat q}).A$ defined by 
\begin{equation}
j({\hat q}).A = -4M{\hat q}.A\phi({\hat q}) (1-D({\hat q})/(4M \beta^2))
\end{equation}
(The effect of the hat structure of ${\hat q}$ on the e.m. substitution 
is ignored in this approximate treatment). This effective 4-point vertex 
function is operative at one end in each of 2a,2b,2c,2d of KL [63] and
at both ends of 2e. For the e.m. vertex at the quark lines of 2(a,b,c,d),
we use simply $ie_i \gamma.A$, as in [18a]. The matrix elements can now be
written down on exactly the same lines, and the {\it same} phase convention      
as in [18a], to keep proper track of the gauge corrections with sign. We need
write these down only for 2a and 2e, noting the equalities 2a=2b, as also
2c=2d, and the further substitutions $(1) \rightarrow (2)$ and vice versa
to generate 2c(=2d) from 2a(=2b). The contribution from 2a [63] to the e.m. 
quadratic self-energy of a kaon is expressible as 
\begin{eqnarray}
M^2_{2a} &=& N_H^2 (2\pi)^{-5}e_1 e_q \int  j({\hat q})_\mu D({\hat q}')
\phi({\hat q}')Tr[\gamma_5 D_{F\mu \nu}(k) \nonumber \\
         & & S_F(p_1-{\hat m}_1 k) i e_1 \gamma_\nu 
S_F(p_1') \gamma_5 S_F(-p_2')] d^4q d^4k
\end{eqnarray}
where $p_1'=p_1+{\hat m}_2 k$ and $p_2 = p_2' = p_2-{\hat m}_2 k$ are the
4-momenta of the quarks at the other (right-hand) end, and the photon
propagator in the Landau gauge is $-i(\delta_{\mu\nu} -k_\mu k_\nu/k^2)/k^2$. 
It is now convenient to change the variable from $k_\mu$ to $q_\mu'$, noting 
that $q'= q+{\hat m}_2 k$, which gives  $d^4 k = d^4 q'/{\hat m}_2^4$, etc.
This shows that fig 2a(=2b), where the photon line ends on the heavier quark
$m_1$, gives a bigger contribution than does fig.2c(=2d) which would give
${\hat m}_1^{-4}$ arising from the $d^4k$-measure. Evaluating the traces, and
integrating over the poles of the two time-like momenta $q_0$ and $q_0'$
gives for the sum of the contributions from 2a-2d to the quadratic mass 
difference between ${\bar K}^0$ and $K^-$ as a product of two 3D quadratures 
after some simplifications with factorable approximations a la [18a]:
\begin{eqnarray}
\delta M^2_{2(a-d)} &=& {{6N_H^2 M \delta (e_1 e_q)}\over {(2\pi)^3 
{\hat m}_2^3}} \int d^3{\hat q} \int d^3{\hat q}'{{\phi \phi'}\over
{{\hat q}{\hat q}'\omega_{1k}}}[1- {{D({\hat q})}\over{4M\beta^2}}]\nonumber \\ 
                    & & [({\hat q}^2 (2-4/\pi)-{\hat q}{\hat q}'/3)
(M^2-{\delta m}^2+D({\hat q}'){\omega_1'}^{-1}/2+
D({\hat q}'){\omega_2'}^{-1}/2)\nonumber \\
                    & & + {1 \over 3}{\hat m}_2{\hat q}{\hat q}'
(D({\hat q}'){\omega_2'}^{-1}/2+ M^2-{\delta m}^2)] + [1 \leftrightarrow 2]                  
\end{eqnarray}
Here $\delta (e_i e_q)$ is the ${\bar K}^0$ minus $K^-$ difference
between the indicated charge factors associated with line `i', while 
$\omega_{1,2}'^2 = m_{1,2}^2 + {\hat q}'^2$ and 
$\omega_{1k}^2 = m_1^2 + ({\hat q}-{\hat m}_1 {\hat k})^2$.     
\par
        Next the contribution to $\delta M^2$ arising from fig 2e of KL [63] 
which involves the product of two vertex blobs like (B.1) is given by
\begin{equation}
\delta M^2_{2e} = i N_H^2 (2\pi)^{-5}{e_q}^2 \int d^4q d^4k D_{F\mu\nu}(k)
j({\hat q})_\mu j({\hat q})_\nu 
Tr[\gamma_5 S_F(p_1-{\hat m}_1 k) \gamma_5 S_F(-p_2+{\hat m}_2 k)] 
\end{equation}
This integral is somewhat different in structure from (B.2) in as much as
$k_\mu$ is fully decoupled from either wave function $\phi, \phi'$, both of
which have the same argument ${\hat q}$. This makes it possible to integrate
first over $d^4k$ as well as the time-like component $q_0$ of $q_\mu$ 
neither of which is involved in the vertex function. The relevant integral 
after tracing and rearranging has the form
\begin{eqnarray}
F({\hat q}) &=& 3(-i)^2\int d^4k \int dq_0 k^{-2}
(\delta_{\mu\nu}-k_\mu k_\nu/k^2) \nonumber \\
            & & [{\hat q}^2 -q_0^2 + m_1m_2 
-{\hat m}_1{\hat m}_2 (P-k)^2]/(\Delta_1\Delta_2)
\end{eqnarray}
where $\Delta_i = m_i^2 + (p_i - {\hat m}_i k)^2$. The integral which is
entirely convergent works out after some standard manipulations involving
Feynman techniques as well as differentiation under integral signs as
\begin{equation}
F({\hat q})= 6\pi^3 [m_1m_2+{\hat q}^2+\Lambda][\sqrt{\Lambda}-
\sqrt{\Lambda-{\hat m}_1{\hat m}_2 M^2}]/({\hat m}_1{\hat m}_2 M)^2  
\end{equation}       
where $\Lambda= {\hat m}_1{\hat m}_2 M^2 + D({\hat q})/2M$. And the final
expression for (B.4) in terms of (B.6) is
\begin{equation}
\delta M^2_{2e} = N_H^2 (2\pi)^{-5} {\delta (e_q^2)} \int d^3{\hat q} 
{j({\hat q})}^2 F({\hat q})
\end{equation}  
\par
        Further evaluation of (B.3) and (B.7) can be made a la [18a] in a
straightforward way. The key ingredients are 
\begin{equation}
{\delta {e_1e_q}}= 0.236 e^2 ; \quad {\delta {e_2e_q}}=0.139 e^2; 
\quad {\delta {e_q^2}}= -0.0294 e^2. 
\end{equation}
The break-up of the final results for the diagrams 2(a-e) after dividing the 
results of(B.3) and (B.7) by 2M, since $\delta M^2 = 2M \delta M$, is (in MeV):
\begin{eqnarray}
\delta M_{2a+2b} &=& -0.6996; \quad \delta M_{2c+2d}=+0.1358; \\  \nonumber 
\delta M_{2e}    &=& -0.0481; \quad \delta M_{tot}=-0.612 MeV.
\end{eqnarray}
All these corrections, which reinforce one another due to a complex interplay
of signs, add up to a figure which increases the value $-1.032 MeV$ due to 
Fig. 1(b) of [18a], to $-1.644 MeV$, roughly a $60$ percent (negative) 
increase, which is a rough indication of the type of QED gauge correction to 
be expected from such diagrams.

\section*{Appendix C: A 4D NJL-Faddeev Model}
 
We summarize here the results of a simplified 4D NJL-Faddeev bound state 
problem [47,89], with 3 scalar-isoscalar quarks interacting pairwise in a 
contact fashion, a la NJL [4]. It is merely a special case of 3D-4D-BSE
when its kernel $K$ becomes a constant $\lambda$. For ease of comparison, 
we employ the same notation and phase convention for the various
quantities as in Secs.(4,9), but in view of the bound state nature of the 
problem it is enough to work with the 4D BSE for the wave function only,
without invoking Green's functions. We start with 
the $qq$ problem as a prerequisite for the solution of the $qqq$ problem.

\subsection*{C.1 \quad  $qq$ Bound State in NJL Model}
    
The BSE for the 4D wave function ${\Phi}$ for a $qq$ system may be written
down for the NJL model:
\setcounter{equation}{0}
\renewcommand{\theequation}{C.\arabic{equation}}
\begin{equation}
i(2\pi)^4 \Phi(q_{12}P_{12}) = {\Delta_1}^{-1} {\Delta_2}^{-1} \lambda
 \int d^4{ q_{12}}' \Phi({q_{12}}'P_{12})
\end{equation}   
where $\lambda$ is the strength of the contact NJL interaction for any
pair of (scalar) quarks. The solution of this equation simply reads as [87b]
\begin{equation}
 \Phi(q_{12}P_{12}) = A {\Delta_1}^{-1} {\Delta_2}^{-1}
\end{equation}
When plugged back into (C.1), one gets an  `eigenvalue' equation for the
invariant mass $M_d^2 = -P_{12}^2$ of an isolated bound $qq$ pair in the 
implicit form of a determining equation for $\lambda$: 
\begin{equation}
\lambda^{-1} = -i (2\pi)^{-4} \int d^4q {\Delta_1}^{-1} {\Delta_2}^{-1}
\equiv h(M_d)
\end{equation}
where $\Delta_{1,2}$ = $m_q^2 + q^2 -M_d^2/4 \pm q.P_{12}$, and we have 
indicated the result of integration by a function $h(M_d)$ of the mass 
$M_d$ of the composite bound state (diquark). Unfortunately the integral 
(C.3) is logarithmically divergent, but it can be regularized with a 4D 
ultraviolet cut-off $\Lambda$, together with a Wick rotation, i.e., 
$q_0 \rightarrow iq_0$, which is allowed by the singularities of the two 
propagators. The exact result is:  
\begin{equation}
16 \pi^2 \lambda^{-1} = 1+ \ln (4\Lambda^2 / M_d^2) -
2 {{\sqrt {4m_q^2-M_d^2}} \over{M_d}} {\arcsin (M_d/{2m_q})}
\end{equation} 
under the condition $M_d < 2m_q$. A slightly less accurate but much simpler  
form which is also easier to adapt to the $qqq$ problem to follow, may be 
obtained by the Feynman method of introducing an auxiliary integration 
variable $u$ ($0 < u < 1$) to combine the two propagators, followed by 
a Wick rotation and a translation to integrate over $d^4q$ (ignoring surface 
terms which formally arise due to the logarithmic divergence) :
\begin{equation}
16 \pi^2 \lambda^{-1} = \ln {6\Lambda^2 \over {6m_q^2 - M_d^2}} - 1 
\equiv 16 \pi^2 h(M_d),  
\end{equation}
thus defining a diquark `self-energy' function $h(M)$ where the `on-shell' 
value is $M=M_d$.  Eq.(C.5) also provides a determining equation for the 
NJL strength parameter $\lambda$ in terms of the `diquark' mass $M_d$ and 
the cut-off parameter $\Lambda$, in a clearly 4D invariant form.  
  
\subsection*{C.2 \quad  NJL-$qqq$ Bound State Problem}

We now set up the corresponding NJL-$qqq$ problem under the same $q-q$
contact interaction strength $\lambda$. Using the same notation for the 
various 4-momenta and propagators as listed  in Sec.9, the 4D wave 
function $\Phi(\xi,\eta;P)$ expressed in terms of any of the $S_3$ 
invariant pairs $(\xi_i,\eta_i)$ of internal 4-momenta satisfies the BSE:
\begin{equation}
i(2\pi)^4 \Phi(\xi,\eta;P) = \sum_{123} \lambda
{\Delta_1}^{-1} {\Delta_2}^{-1} \int d^4q_{12}' \Phi({\xi_3}',{\eta_3};P)
\end{equation}
where the arguments of $\Phi$ on the LHS are not-indexed since it is 
$S_3$-symmetric as a whole, while those on the RHS are indexed in order 
to indicate which subsystem is in pairwise interaction (see explanation 
in Sec.9). The solution of this equation may be read off from the observation 
that the integration w.r.t. $q_{12}'= {\sqrt 3}\xi_3'/2$ leaves the respective 
integrals as functions of $\eta_i$ only, where $i= 1,2,3$. Thus [87b]  
\begin{equation}
\Phi(\xi,\eta;P) = \sum_{123}{\Delta_1}^{-1} {\Delta_2}^{-1} F(\eta_3)
\end{equation}
where $F$ is a function of a single variable $\eta_i$. Next, plugging back
the solution (C.7) into the main equation (C.6), gives the following 
integral equation in a single variable $\eta_3$, as a routine procedure 
applicable to separable potentials [87b]:
\begin{equation}
(h(M_d)- h(M_{12})) F(\eta_3) = -i(2\pi)^{-4} \Delta_3^{-1} 
 \int d^4 q_{12}' [F(\eta_2')\Delta_1'^{-1} + (1 \leftrightarrow 2)]
\end{equation} 
Note that the cut-off parameter $\Lambda$ drops out from the LHS, as checked 
by substitution for $h(M)$ from (C.5). This means that the 4D diquark 
propagator  $(h(M_d)-h(M_{12}))^{-1}$ is formally independent of the cut-off 
$\Lambda$, in this simple NJL model.         
\par
        Next, the meaning of the function $F(\eta)$ can be inferred from
an inspection of eq.(C.8), on similar lines to 3D [87b] or 4D [89] studies: 
$F(\eta_3)$ is the 4D `quark(3)-diquark(12)' wave function which
is generated by an exchange force represented by the propagators 
$\Delta_1'^{-1}$ and $\Delta_2'^{-1}$ in the first and second terms on the 
RHS respectively. And the baryon-$qqq$ vertex function $V_3$ corresponding
to a break-up of the baryon into quark(3) and diquark (12) may be identified 
by multiplying this quantity with the product of the inverse propagators 
of quark(3) and diquark(12):
\begin{equation}
V_3 \equiv V(\eta_3) =  \Delta_3 f(\eta_3) F(\eta_3)
\end{equation}     
where the diquark inverse propagator is reexpressed as
\begin{equation}
f(\eta_3) = h(M_d)- h(M_{12}) = (4\pi)^{-2} \ln{{6m_q^2 + \eta_3^2 -4M_B^2/9} 
\over {6m_q^2 -M_d^2}},
\end{equation} 
making use of eq.(C.5) and the kinematical relation 
$\Delta_i = m_q^2+\eta_i^2-M_B^2/9$, where $M_B$ is the mass of the bound 
$qqq$ state, and $i=1,2,3$. The quantity $V_3$ of eq(C.9) may be compared 
directly (except for normalization) with the corresponding `3D-4D-BSE' 
quantity (9.3.10). 

\subsection*{C.3 \quad  Solution of the Bound $qqq$ State Eq.(C.8)}

We now turn to the Lorentz structure of the NJL-$qqq$ equation (C.8), 
as well as an approximate analytic solution for the energy eigenvalues of
the bound $qqq$ states.  To that end we substitute (C.9) in (C.8) to give an 
integral equation for $V(\eta_3)$, with $\eta_2' \equiv \eta$ for short: 
\begin{equation}
V(\eta_3) = -2i(2\pi)^{-4} \int d^4\eta V(\eta) f^{-1}(\eta) \times 
(m_q^2 +\eta^2 - M_B^2/9)^{-1} (m_q^2 + (\eta_3 + \eta)^2 - M_B^2/9)^{-1}
\end{equation}
where the factor $2$ on the RHS signifies equal effects of the two terms
on the RHS of (C.8). For a bound state solution of this equation, with
$M_B < M_d + m_q$, the singularity structures permit a Wick rotation 
$\eta_0 \rightarrow i\eta_0$ which converts $\eta$ into a Euclidean variable
$\eta_E$. This shows without further ado that eq.(C.11) is 4D-invariant
just like its $qq$ counterpart eq.(C.3). This is not quite the same thing as
the old result [13] on O(4)-like spectra with harmonic confinement in the
limit of infinite quark mass [14], since this NJL-Faddeev model of contact
interaction, patterned after similar approaches [89], lacks a confining 
interaction, so that although in principle eq.(C.11) predicts a spectrum 
of bound states at the $qqq$ level (starting with  NJL(contact) pairwise 
interactions), such spectra cannot be a realistic representation of the 
{\it actual} hadron spectra [13].  We now show how this comes about via 
Wick rotation in (C.11).     
\par
        For an approximate analytic solution of eq.(C.11), note that the 
logarithmic function $f(\eta)$ in the integral appearing on the right is 
slowly varying, so that not much error is incurred by taking it out of the 
integral and replacing it with an average value $<f(\eta)>$. It is now 
possible to  `match' both sides with an  effectively constant $V(\eta)$, 
{\it provided} any further logarithmic dependence on $\eta$ is also similarly 
treated for consistency. The integral is now exactly of the type (C.3), i.e., 
logarithmically divergent,  and can be handled successively by Wick rotation, 
Feynman auxiliary variable $u$, and a translation. The result is again of the 
form (C.5), and after cancelling out the factors $V(\eta_3)$ and $V(\eta)$ 
from both sides, the eigenvalue equation reads: 
\begin{equation}
<f(\eta)> = 2 (4\pi)^{-2} [\ln {\Lambda^2 \over 
{<\eta_3^2/6> + m_q^2 -M_B^2/27}} - 1]          
\end{equation}
To simplify this equation, we express all quantities in terms of the $h(M)$
functions given in (C.5) and (C.10), and ignore the difference between 
$\eta = \eta_2'$ and $\eta_3$ inside the logarithms, to give
\begin{equation}
h(M_d)-h(M_{12}) = 2h(M_{12}); \Rightarrow \lambda^{-1} = h(M_d) = 3h(M_{12}) 
\end{equation}
The last equation brings out clearly the fact that the baryon binding comes
about from {\it three} pairs of $qq$ interaction, albeit off-shell, since
the function $<M_{12}^2> = <\eta_3^2> -4M_B^2/9$ still depends on the 
(average) value of $\eta^2$. The qualitative features are thus on expected 
lines, but this oversimplified model is not intended for a realistic fit to 
the nucleon/Delta masses (which at minimum require the introduction of 
spin-isospin d.o.f.), beyond the general feature of a quark-diquark
structure that characterizes an NJL-Faddeev approach [92], as expected from 
any separable potential [87b], of which the NJL model is a special case.  

\subsection*{C.4 \quad  Comparison of NJL-Faddeev with 3D-4D-BSE}     

We end this Appendix with a comparison between the vertex functions (10.3.10) 
and (C.9). The NJL-Faddeev form (C.9) of $V_3$ is  Lorentz invariant, being 
derived from a BSE with a constant kernel, viz the $K=const$ limit of 3D-4D
BSE [16]. Its  quark-diquark form merely reflects the `separable' nature of 
the NJL model [4]. There is no motivation here for a 3D BSE reduction, or
4D reconstruction, since 4D invariance is in-built throughout.  
\par
        In contrast, the vertex function (10.3.10),  obtained from 3D-4D-BSE 
[47],  is merely Lorentz covariant due to the 3D kernel support, but the 
derivation is otherwise more general than NJL-Faddeev, since it is valid for 
any spatial form of the kernel as long as it is 3D in content. This leads to 
an exact 3D reduction of the (4D) BSE  whose formal solution is a 3D wave 
function $\phi({\hat \xi},{\hat \eta})$, a function of {\it two} independent 
3-momenta [10b], in contrast to its NJL counterpart $F(\eta_3)$ in (C.9) 
which is a function of a single 4-momentum $\eta_3$ only. The denominator 
function $D({\hat q}_{12})$ of (10.3.10) similarly is a 3D counterpart of the 
corresponding 4D inverse propagator $f(\eta_3)$ in (C.9). Finally the big 
radical in (10.3.10) corresponds to the inverse propagator $\Delta_3$ in (C.9),
except for its more involved structure, which we now seek to explain.  
\par
        While the `zero extention' in the temporal direction is common to 
both approaches, NJL-Faddeev has also a zero spatial extension, while
3D-4D-BSE has a `normal' spatial extension. Thus any ambiguity in the 
reconstruction of the 4D wave function from the 3D form of the 3D-4D BSE,
vanishes in the $K = const$ limit, so that the same is directly attributable 
to the (mere Lorentz covariant) 3D form of the BSE kernel. Indeed the 1D 
$\delta$-function in (10.3.10) fills up an information gap in the 
reconstruction from a truncated 3D to the full 4D Hilbert space in the 
simplest possible manner, while satisfying a vital self-consistency check by 
reproducing the full structure (10.2.12) of the 3D BSE. This already lends 
{\it sufficiency} to the ansatz (10.3.4) which leads to (10.3.10). As to its 
`necessity', this ansatz has certain desirable properties like on-shell 
propagation of the spectator in between two successive interactions, as well 
as an explicit symmetry in the $p_3$ and ${p_3}'$ momenta. There is a fair 
chance of its uniqueness within some general constraints, but this is still
short of a formal `proof of necessity'.   
\par
        The other question concerns the compatibility of the 1D $\delta$-
function in (10.3.10) with the standard requirement of connectedness [87]. 
Both the $\delta$-function and the $\Delta_{3F}$ propagator appear in 
{\it rational} forms in the 4D Green's function, eq.(10.3.7), reflecting 
a free on-shell propagation of the spectator between two vertex points.
The square root feature in the baryon-$qqq$ vertex function (10.3.10) is a
technical artefact corresponding to an equal distribution of this singulariity 
between the initial and final state vertex points, and has no deeper 
significance. Furthermore, as the steps in Sec.10.2 indicate, the three-body 
connectedness has already been achieved at the 3D level of reduction, so 
the `physics' of this singularity, generated via eq.(10.3.4), must be traced 
to some mechanism other than a lack of connectedness [87] in the 3-body 
scattering amplitude. A plausible analogy is to a sort of (Fermi-like) 
`pseudopotential' of the type employed to simulate the effect of chemical
binding in the coherent scattering of neutrons from a hydrogen molecule 
in connection with the determination of the {\it singlet} $n-p$ scattering 
length [92]. Such $\delta$-function potentials have no deeper significance 
other than depicting the vast mismatch in the frequency scales of nuclear
and molecular interactions. In the present case, the instantaniety in time 
of the pairwise interaction kernel in an otherwise 4D Hilbert space causes a 
similar mismatch, needing a 1D $]delta$-function to fill the gap. And 
just as the `pseudo-potential' in the above example [92] does not have
any observable effect, the singularity under radicals in (10.3.10) will 
{\it not} show up in any physical amplitude for hadronic transitions via 
quark loops, since the Green's functions (10.3.7) involve both the $\delta$-
function and the propagator $\Delta_{3F}$ in {\it rational} forms before 
the relevant quark loop integrations over them are performed.  

\section*{Appendix D: $SU(6) \otimes O(3)$ Wave Fns In Complex Basis}
\setcounter{equation}{0}
\renewcommand{\theequation}{D.\arabic{equation}}    

In this Appendix, we outline a general method ofexpressing the $qqq$ wave 
functions in a complex basis [29b.46b], as an alternative to the `real'
representation given in eqs.(9.1-5). Such a basis gives a compact realization
of the doublet representation of the permutation group $S_3$, with the two
complex vectors ${\bf {z,z^*}}$ substituting for the real pair 
${\bf {\xi,\eta}}$. The action of the permutations $P_{ij}$ on this basis 
in the order $(12);(31);(23)$ is [94a]
\begin{equation}
P_{ij}{\bf z}= [1;e^{2i\pi/3};e^{-2i\pi/3}] {\bf z^*}; \quad
P_{ij}{\bf z^*}= [1;e^{-2i\pi/3};e^{2i\pi}/3]{\bf z}
\end{equation}
Identical doublet representations hold for the orbital $\psi$, spin $\chi$ 
and isospin $\phi$ d.o.f.'s, in the notation of Sections 9-11. To that end,
define the corresponding complex quantities (except for an overall i-factor)
\begin{equation}
{\sqrt 2}[\psi_c; \chi_c; \phi_c] \equiv [\psi"-i\psi'; \chi"-i\chi'; 
\phi"-i\phi']   
\end{equation}
togrther with a second set of complex conjugate relations. Using these 
definitions, the action of the permutation group on the full wave function is
\begin{equation}
P_{ij} [\psi_c \otimes \chi_c \otimes \phi; \psi^*_c \otimes \chi^*_c \otimes
 \phi^*_c] = [\psi^*_c \otimes \chi^*_c \otimes \phi^*_c ; 
\psi_c \otimes \chi_c \otimes \phi]
\end{equation}  
Another important result concerns the action of $P_{ij}$ on any pair of 
component wave functions $(\lambda_c,\mu^*_c)$, where $(\lambda_c,\mu_c)$ 
are any two out of the full set $(\psi,\chi,\phi)$ of three [94a]:
\begin{equation}
P_{ij} [\lambda_c \otimes \mu^*_c; \lambda^*_c \otimes \mu_c] =
[\lambda^*_c \otimes \mu_c; \lambda_c \otimes \mu^*_c]
\end{equation}
As to the quantities $(\psi_s;\psi_a)$, they are $S_3$-singlets by 
themselves, with eigenvalues $\pm 1$ for the $P_{ij}$ operators. The properly
symmetrized $SU(6) \times O(3)$ states, eqs.(9.1-5), are now:
\begin{equation}
{|56>}^q= \psi^s\chi^s\phi^s; \quad 
{|56>}^d= (\chi_c\phi^*_c+\chi^*_c\phi_c)/{\sqrt 2};
\end{equation}
\begin{equation}     
{|70>}^q= \chi^s(\psi_c\phi^*_c +\psi^*_c\phi_c)/{\sqrt 2}; \quad
{|70>}^d= (\psi_c\chi_c\phi_c + \psi^*_c\chi^*_c\phi^*_c)/{\sqrt 2}
\end{equation}
\begin{equation}
{|20>}^q= \psi_a\chi^s\phi_a; \quad 
{|20>}^d= \psi_a (\chi_c\phi^*_c-\chi^*_c\phi_c)/{\sqrt 2}
\end{equation}

\subsection*{D.1 \quad Construction of $\psi$-Fns in Complex Basis}

We now turn to the construction of the orbital $\psi$-functions in terms of
$(z_i,Z^*_i)$, so as to preserve the total angular momentum adapted to the 
complex language.  To that end, the angular momenta (both diagonal and
`mixed') in the complex basis are given by
\begin{equation}
{\bf L_z} = -i {\bf z} \times {\bf \nabla_z}; \quad
{\bf L_{z^*}} = +i{\bf z^*} \times {\bf \nabla{^*_z}} ; \quad
{\bf L_c}=-i{\bf z^*} \times {\bf \nabla_z}; \quad
{\bf L_c^*}=+i{\bf z^*} \times {\bf \nabla_z^*}
\end{equation}
which obey the connections
\begin{equation}
{\bf L}= {\bf L_z} + {\bf L_{z^*}} = {\bf L_\xi}+{\bf L_\eta}; \quad
{\bf L_a}= {\bf L_a}= {\bf L_z} + {\bf L_{z^*}}
\end{equation}
These quantities transform according to eq.(D.1) under the elements $P_{ij}$
of $S_3$.
\par
        To construct angular momentum states of correct $S_3$ symmetry, it is
useful to take those of highest seniority [94b], now expressed in appropriate 
powers of $z_+$ and ${z_+}^*$, and to note that ${\bf z} \dot {bf z^*}$,
$z_+{z_+}^*$ and ${z_+}^3$ are all $S_3$-invariant. The angular momenta 
carried by these basic units are easily checked to be in conformity with
the above (complex) definitions (D.8-9) of the angular momenta. Using these
basic building blocks, the natural parity states of highest seniority [94b]
for a given angular momentum are compactly written in an HO basis as [46b]:
\begin{equation}
{|56^+;70^-;70^+;56^->}= (2z_+{z_+}^*)^{\ell} [1;z_+;z_+^2;z_+^3] 
e^{-2{\bf z}\dot{\bf z^*}}
\end{equation}         
The superscripts $\pm$ on the various states on the LHS {\it correctly} 
describe their parity structures, by noting that ${z_+}^n$ has parity
$(-1)^n$, while ${\bf z} \dot {\bf z^*}$ is a 3-scalar. The $L^P$-values
of the states (D.10) in this order are [46b, 29b]:
\begin{equation}
L^P = (2{\ell})^+; (2{\ell}+1)^-; (2{\ell}+2)^+; (2{\ell}+3)^-;
\end{equation}
while $\ell$ goes through the values $0,1,2,3...$, thus bringing out the
{\it naturalness} of the respective parity structures. 
\par 
        In a similar way it is possible to systematically span all the 
``unnatural'' parity states in the same representation [46b], noting that 
the main carrier of unnatural parity is the axial vector ${\bf \zeta}$ =
$i{\bf z} \times {\bf z^*}$, which is a fully antisymmetric $S_3$-singlet.
The $L^P$-structures of such states of highest seniority, corresponding to
the series (D.10) are [46b,29b]:
\begin{equation}
{|20^+;70^-;70^+;20^->}= {\zeta_+} (2z_+{z_+}^*)^{\ell} 
[1;z_+;z_+^2;z_+^3] e^{-2{\bf z}\dot{\bf z^*}}
\end{equation}
together with the respective $L^P$-values 
\begin{equation}
L^P = (2{\ell}+1)^+; (2{\ell}+2)^-; (2{\ell}+3)^+; (2{\ell}+4)^-;
\end{equation}
thus bringing out the `unnaturalness' of their respective parities. For the
construction of more involved states on these lines, see [46b]. 
\par
        A similar construction is possible for the `spin' wave functions in
the complex basis; see [46b] for details.

\subsection*{D.2 \quad Normalization of Natural and Unnatural Parity Baryons }

We now outline a new method of integration for the normalization of the 
spatial wave functions (D.10) and (D.12) in the 6D $({\bf z}, {bf z^*})$
space, which is rather well-suited to the (complex) variables on hand 
[29b,46b]. The volume measure in this 6D space is expressible in the 
spherical basis as
\begin{equation}
d^6\tau = d^3{\bf z}d^3{\bf z^*} 
=  (dz_+dz_-^*)(dz_-d_+^*)(dz_3dz_3^*)
\end{equation}
where the six elements on the RHS of (D.14) have been rearranged into 3 sets 
of {\it real} 2D volumes, since the three pairs on the RHS, each form complex
conjugate pairs. Now put
\begin{equation}
{\sqrt 2}(z_+;z_-^*) = {\cal R}_1 e^{\pm \vartheta_1}; \quad
{\sqrt 2}(z_-;z_+^*) = {\cal R}_2 e^{\pm \vartheta_2}; \quad   
{\sqrt 2}(z_3;z_3^*) = {\cal R}_1 e^{\pm \vartheta_3};
\end{equation}
Then the volume element (D.14) becomes
\begin{equation}
d^6\tau ={\cal R}_1 d{\cal R}_1 d\vartheta_1 \dot  
{\cal R}_2 d{\cal R}_2 d\vartheta_2 \dot 
{\cal R}_3 d{\cal R}_3 d\vartheta_3 ;
\end{equation}
\begin{equation}
0 \leq {\cal R}_{1,2,3} \leq \inf; \quad 0 \leq \vartheta_{1,2,3} \leq 2\pi;
\quad {\cal R}_1^2 +{\cal R}_2^2 +{\cal R}_3^2= 2 {\bf z} \dot {\bf z}^*
\equiv {\cal R}^2
\end{equation}
Since the phase angles (not quite Euler angles) will not appear in the 
squared modulii of the wave functions, these are integrated out to give
\begin{equation}
d^\tau = \pi^3 d{\cal R}_1^2 d{\cal R}_2^2 d{\cal R}_3^2
\end{equation} 
Now the natural parity sequence (D.10) is compactly expressed as
\begin{equation}
\psi = N_{{\ell}n} (2z_+z_+^*)^{\ell} {z_+}^n e^{-{\cal R}^2/2}
\end{equation}
where $n=0,1,2,3$ for the states (D.10) in sequence, and the normalizer is
\begin{equation}
N_{{\ell}n}^{-2}= \int d^6\tau [{\cal R}_1^2{\cal R}_2^2]^{\ell} 
({\cal R}_1^2/2)^n e^{-{\cal R}^2} 
= \pi^3 \Gamma(\ell+1)\Gamma(\ell+n+1)/2^n,
\end{equation}
which agrees with the result for the $\xi,\eta$-representation [29a].
For the unnatural parity sequence (D.12), the extra $\zeta$-factor gives
\begin{equation}
(\zeta_+\zeta_-^*) = {\cal R}^4/4 -{\cal R}_3^4/4 -{\cal R}_1^2{\cal R}_2^2
\end{equation}
Denoting the corresponding normalizers by ${\tilde N}_{{\ell}n}$, similar
integration now leads to the result
\begin{equation}
{\tilde N}_{{\ell}n}^{-2} = {{\pi^3 \Gamma(\ell+1)\Gamma(\ell+n+1)} \over 
{12(2^n)}} \times [(\ell+n+1)(n+2)+(\ell+1)(\ell+4)]
\end{equation}
Radial excitations can be similarly handled. E.g., one radial excitation
gives an extra multiplicative factor in the normalization integral (D.20)
for natural parity states, giving rise to an extra factor $(2\ell+n+4)$
in (D.20). Further, the reciprocity between the momentum and coordinate 
spaces implied in an HO description as above, allows the same formulation 
to be adapted in the dual space, a result which is useful for evaluating
the $OGE$ corrections to the mass formula (11.20).


\begin{thebibliography}{99}

\bibitem{1}
 A.Chodos et al, Phys.Rev.{\bf D9}, 3471 (1974). 
\bibitem{2}
 (a) M.A.Shifman et al, Nucl.Phys.{\bf B147}, 385 (1979);
 (b) V.L.Chernyak and A.R.Zitnitsky, Phys.Rep.{\bf 112C}, 173 (1984);
 (c) B.L.Ioffe and A.V.Smigla, Nucl.Phys.{\bf B232}, 109 (1984). 
\bibitem{3}
 (a) K.Lane, Phys.Rev.{\bf D10}, 2605 (1974);
 (b) M.K.Volkov, Ann.Phys.(N.Y.){\bf 157}, 282 (1984);
 (c) K.Higashijima, Phys.Rev.{\bf D29}, 1228 (1984).  
\bibitem{4}
 Y.Nambu and G.Jona-Lasino, Phys.Rev.{\bf 122}, 345 (1961).
\bibitem{5}
 (a) E.Witten, Nucl.Phys.{\bf B223}, 422 (1983);
 (b) I.Gasser and H.Leutwyler, Ann.Phys.(N.Y.){\bf 158}, 142 (1984);
 (c) I.Zahed et al, Phys.Rep.{\bf 142C}, 1 (1986)  
 (d) D.Ebert et al, Prog.Part.Nucl.Phys.{\bf 33}, 1-120 (1994).  
\bibitem{6}
 Review: E.V.Shuryak, Phys.Rep.{\bf 115C}, 151 (1984).
\bibitem{7}
 (a) S.G.Matinyan and G.K.Savvidy, Nucl.Phys.{\bf B134}, 539 (1978);
 (b) J.Ambjorn et al, Nucl.Phys.{\bf B152}, 75 (1979);
 (c) Review: H.Dosch, Prog.Part.Nucl.Phys.{\bf 33}, 121-199 (1994).
\bibitem{8}
 (a) T.Goldman and R.W.Haymaker, Phys.Rev.{\bf D24}, 724 (1981);
 (b) G.V.Efimov and M.A.Ivanov, {\it {Quark Confinement Model Of Hadrons}},
 IOP Publishing Ltd., London and Philadelphia, 1993.
\bibitem{9}
 (a) C.D.Roberts et al, Intl.J.Mod.Phys.{\bf A7}, 5607 (1992);
 (b) C.D.Roberts and A.G.Williams, Prog.Part.Nucl.Phys.{\bf 33}, 477 (1994).
\bibitem{10}
 (a) A.N.Mitra, Zeits.f.Phys.{\bf C8}, 25 (1981);
 (b) A.N.Mitra and I.Santhanam, Few-Body Syst.{\bf 12}, 41 (1992);
 (c) R. Barbieri and E.Rimiddi, Nucl.Phys.{\bf B141}, 413 (1978); 
 (d) G.P.Lepage, SLAC-Preprint no.212 (1978).  
\bibitem{11}
 A.N.Mitra and B.M.Sodermark, Int.J.Mod.Phys.{\bf A9}, 915 (1994). 
\bibitem{12}
 Reprint Coll: W.Buchmueller (ed), {\it Quarkonia}, North-Holland, 1992.
\bibitem{13}
 Particle Data Group, Phys.Rev.{\bf D54}, No.1-PartI (1996).
\bibitem{14}
 R.F.Meyer, Nucl.Phys.{\bf B71}, 226 (1974)
\bibitem{15}
 M.A.Markov, Sov.J.Phys.{\bf 3}, 452 (1940);
 H.Yukawa, Phys.Rev.{\bf 77}, 219 (1950)
\bibitem{16}
 A.N.Mitra and S.Bhatnagar, Intl.J.Mod.Phys.{\bf A7}, 121 (1992)
\bibitem{17}
 Yu.L.Kalinowski et al (Pervushin-Group), Phys.Lett.{\bf B231}, 288 (1989);
 Few-Body Syst.{\bf 10}, 87 (1991). 
\bibitem{18}
 V.Kadychevsky, Nucl.Phys.{\bf B6}, 125 (1968)
\bibitem{19}
 (a) H.Sazdjian, J.Math.Phys.{\bf 28}, 2618 (1987);
 (b) J.Bijtebier, Nuovo Cim.{\bf A100}, 91 (1988);
 (c) H.W.Crater and P.Van Alstine, Phys.Rev.{\bf D37}, 1982 (1988)
\bibitem{20}
 R.Dolen, D.Horn and C.Schmid, Phys.Rev.{\bf 166}, 1768 (1968) 
\bibitem{21}
 M.Baker et al, Phys.Rev.Lett.{\bf 11}, 518 (1964);
 Th. A.J. Marris, {\it ibid}.{\bf 12}, 313 (1964).
\bibitem{22}     
 H.D.Politzer, Phys.Rev.Lett.{\bf 30},1346 (1973); \\
 D.J.Gross and F.Wilczek, Phys.rev.{\bf D8},3633 (1973).
\bibitem{23}
 A.Le Yaouanc et al, Phys.Rev.{\bf D29}, 1233 (1984); {\bf D31}, 137 (1985).  
\bibitem{24}
 (a) R.Delbourgo and M.D.Scadron, J.Phys.G {\bf 5}, 1621 (1979); 
 (b) S.Adler and A.C.Davis, Nucl.Phys.{\bf B244}, 469 (1984); 
 (c) H.Pagels et al, Phys.Rev.{\bf D19}, 3080 (1979);
 (d) L.Chang and N.Chang, Phys.Rev.{\bf D29}, 312 (1984). 
\bibitem{25}
 R.P.Feynman et al, Phys.Rev.{\bf D3}, 2706 (1971).  
\bibitem{26}
 C.Izykson and J.-B.Zuber, {\it {Quantum Field Theory}}, Mcgraw-Hill 
 Inc. New York, 1980.   
\bibitem{27}
 Some typical references are: \\
 (a) H.J.Munczek and P.Jain, Phys.Rev.{\bf D46}, 438 (1992);  
 (b) A.G.Williams et al, Ann.Phys.(NY){\bf 210}, 464 (1991);
 (c) K.I.Aoki et al, Phys.Lett.{\bf B266}, 467 (1991). 
\bibitem{28}
 (a) A.Mittal and A.N.Mitra, Phys.Rev.Lett.{\bf 57}, 290 (1986); 
 (b) K.K.Gupta et al, Phys.Rev.{\bf D42}, 1604 (1990).
\bibitem{29}
 (a) A.N.Mitra and D.S.Kulshreshtha, Phys.Rev.{\bf 37}, 1268 (1988); 
 (b) A.Sharma et al, Phys.Rev.{\bf D50}, 454 (1994).
\bibitem{30}
 W.Y.P.Hwang et al, Chinese J. of Phys.{\bf 33}, 397 (1995).
\bibitem{31}
 W.Y.P.Hwang and A.N.Mitra, Few-Body Syst.{\bf 15}, 1 (1993)
\bibitem{32}
 (a) A.N.Mitra, Intl.J.Mod.Phys.{\bf A11}, 5245 (1996).
 (b) A.N.Mitra and K.-C.Yang, Phys.Rev.{\bf C51}, 3404 (1995); 
\bibitem{33}
 (a) A.N.Mitra and D.S.Kulshreshtha, Phys.Rev.{\bf D28}, 588 (1983); 
 (b) A.N.Mitra, A.Pagnamenta and N.N.Singh, Phys.Rev.Lett.
     {\bf D59}, 2408 (1987);
 (c) N.N.Singh et al, Phys.Rev.{\bf D38}, 1454 (1988).
\bibitem{34}
 C.R.Ji and S.Cotanch, Phys.Rev.Lett.{\bf 64}, 1484 (1990).
\bibitem{35}
 (a) P.A.M.Dirac,Rev.Mod.Phys.{\bf 21}, 392 (1949); 
 (b) S.Weinberg, Phys.Rev.{\bf 150}, 1313 (1966); 
 (c) J.Kogut and L.Suskind, Phys.Report.{\bf 8}, 75 (1973);
 (d) H.Leutwyler and J.Stern, Ann.Phys.(N.Y.){\bf 112}, 94 (1978).
\bibitem{36}
 V.A. Karmanov, Nucl.Phys.{\bf B166},378 (1980)
\bibitem{37} 
 R.J.Perry,A.Harindranath and K.Wilson, Phys.Rev.Lett.{\bf 65}, 2959 (1990).
\bibitem{38}
 Review: J.Carbonell et al, Phys.Rep. (1998); to appear.
\bibitem{39}
 (a) E.E.Salpeter, Phys.Rev.{\bf 87}, 328 (1952); 
 (b) A. Logunov and A.N.Tavkhelidze, Nuovo Cimento {\bf 29}, 380 (1963).
 (c) R. Blankenbecler  and R. Sugar, Phys.Rev.{\bf 142}, 105 (1966); 
 (d) F.Gross, {\it ibid} 1448.
\bibitem{40}
 S.Chakrabarty et al, Prog.Part.Nucl.Phys.{\bf 22}, 43-180 (1989).
\bibitem{41}
 A.N.Mitra, LANL Preprint {\bf hep-ph/9812404}, December 1998.
\bibitem{42}
 (a) A.N.Mitra, Proc.Ind.Nat.Sci.Acad.{\bf 54A}, 179 (1988); 
 (b) {\it ibid} {\bf  },     (1996)
\bibitem{43}
 J.Schwinger, Phys.Rev.{\bf 82}, 664 (1951)
\bibitem{44}
 (a) W.E.Caswell and G.P.Lepage, Phys.Rev.{\bf A18}, 810 (1978); 
 (b) S.P.Misra, Phys.Rev.{\bf D18}, 1673 (1978). 
\bibitem{45}
 (a) M.Verde, Handbuch der Physik {\bf 39}, 170 (1957); 
 (b) A.N.Mitra and M.H.Ross, Phys.Rev.{\bf 158}, 1670 (1967); 
 (c) Review: J.M.Richards, Phys.Rep.{\bf C212}, 1 (1992).
\bibitem{46}
 (a) H.Kramer and M.Moshinsky, Nucl.Phys.{\bf 82}, 241 (1966); 
 (b) A.N.Mitra et al, Few-Body Syst.{\bf 19}, 1 (1995);     
 (c) J.Bijtebier, Nuovo Cimento {\bf 81A}, 423 (1985). 
\bibitem{47}
 A.N.Mitra, LANL {\bf hep-th/9803062}, March 1998.    
\bibitem{48}
 Y.Nambu, Phys.Rev.Lett.{\bf 4}, 380 (1960); 
 J.Goldstone, Nuovo Cimento{\bf 19}, 154 (1960).
\bibitem{49}   
 (a) M.Gell-Mann and M.Levy, Nuovo Cimento{\bf 16}, 705 (1960); \\
 (b) T.H.R.Skyrme, Proc.Roy.Soc.{\bf A260}, 127 (1961).
\bibitem{50}
 (a) Y.Nambu, Physica{\bf 15D}, 147 (1985);
 (b) Y.Nambu, in New Theories in Phys, Ed Z.Ajdak, WS (1989);
 (c) Y.Nambu, in Telegdi Festschrift, Ed K.Winter, N.H.Pub Co, (1988);
 (d) Y.Nambu and M.Mukherjee, Phys.Lett.{\bf 209B}, 1 (1988);
 (e) Y.Nambu, U.Chicago Preprint EFI-90-69 (1990). 
\bibitem{51}
 (a) S.Raby et al, Nucl.Phys.{\bf B169}, 373 (1980);
 (b) G.t'Hooft, in Recent Developments in Gauge Theories, Plenum (1979). 
\bibitem{52} 
 H.J. Munczek, Phys.Rev.{\bf D25}, 1579 (1982).
\bibitem{53}
 J.M.Cornwall et al, Phys.Rev.{\bf D10}, 2428 (1974); S.Coleman et al,
 {\it ibid}{\bf D10}, 2491 (1974); H.Kleinert, Phys.lett.{\bf 62B}, 429 (1976);
 V.N.Pervushin et al, Sov.J.Part.Nucl.Phys.{\bf 10}, 444 (1979).   
\bibitem{54}
 D.Atkinson and P.Johnson, Phys.Rev.{\bf D41}, 1661 (1990).   
\bibitem{55}
 H.D. Politzer, Nucl.Phys.{\bf B117}, 397 (1976). 
\bibitem{56}
 A.J.Mcfarlane, Rev.Mod.Phys.{\bf 34}, 41 (1962).
\bibitem{57}
 I.Santhanam et al, Intl.J.Mod.Phys.{bf E2}, 219 (1993).
\bibitem{58}
 (a)E.B.Daley et al, Phys.Rev.Lett.{\bf 45}, 232 (1980); 
 (b)C.Bebec et al, Phys.Rev.{\bf D17}, 1793 (1978).  
\bibitem{59}
 K.G.Chetyrkin et al INR Report no.P-0395, Moscow, 1985.
\bibitem{60}
 G.Farrar and D.Jackson, Phys.Rev.Lett.{\bf 43}, 246 (1979); 
 G.P.Lepage and S.J.Brodsky, Phys Rev.{\bf D22}, 2157 (1980).
\bibitem{61}
 M.Burkardt et al, Phys.Rev.Lett.{\bf 78}, 3059 (1997).
\bibitem{62}
 N.Isgur and C.H.L.Smith, Nucl.Phys.{\bf B317}, 526 (1989); 
 M.Sawicki, Phys.Rev.{\bf D46}, 474 (1992).
\bibitem{63}
 L.S.Kisslinger and Z.Li, Phys.Rev.Lett.{\bf 74}, 2168 (1995).
\bibitem{64}
 H.Pagels and S.Stokar, Phys.Rev.{\bf D20}, 2947 (1979). 
\bibitem{65}
 J.Anjos et al, Phys.Rev.Lett.{\bf 65}, 2630 (1990)
\bibitem{66}
 J.Koerner et al, Z.Phys.{\bf C38}, 511 (1988); M.Bauer et al Z.Phys.
{\bf C42}, (1989); N.Isgur et al, Phys.Rev.{\bf D40}, 1491 (1989);
F.Gilman et al Phys.Rev.{\bf D41}, 142 (1990). 
\bibitem{67}
 K.K.Gupta et al Phys.Lett.{\bf 267B}, 11 (1991).
\bibitem{68}
 J.D.Sullivan, Phys.Rev.{\bf D5}, 1732 (1972).
\bibitem{69}
 W.Y.P.Hwang et al, Phys.Rev.{\bf D45}, 3061 (1992). 
\bibitem{70}
 M.Gell-Mann,R.J.Oakes and B.Renner, Phys.Rev.{\bf 175}, 2195 (1968)
\bibitem{71}
 S.Weinberg, in a Festschrift for I.I.Rabi, ed  L.Motz  (New 
   York Academy of Sciences,1977); p185.
\bibitem{72}
 H.Leutwyler,{\it {Masses of the Light Quarks}}, Bern report 
   {\bf BUTP-94/8}; May 1994.
\bibitem{73}
 K.-C.Yang, Topics in QCD sum Rules, Ph.D. Thesis, Nat Taiwan Univ 1994;
 V.L.Eletsky and B.L.Ioffe , Bern U Preprint BUTP-93/2 (1993).
\bibitem{74}
J.Gasser and H.Leutwyler, Nucl.Phys.{\bf B250},465 (1985)
\bibitem{75}
(a) S.Narison et al, Nucl.Phys.{\bf B212}, 365 (1983);
(b) C.A.Dominguez and E.de Rafael, Ann.Phys.(N.Y) {\bf 174}, 372 (1987);
(c) J.Pashupathy, CTS - preprint (Bangalore) March (1995).
\bibitem{76}
 T.Hatsuda et al, Phys.Rev.Lett.{\bf 66}, 2851 (1991); 
 K.C.Yang et al, Phys.Rev.{\bf D48}, 3001 (1993); 
 V.Eletsky and B.L.Ioffe, Bern Univ. Report BUTP-03/2 (1993).
\bibitem{77}
 A.Nolen and J.Schiffer, Annu.Rev.Nucl.Sci.{\bf 19}, 14 (1969).
\bibitem{78}
 Some recent references on $\rho-\omega$ mixing are: \\ 
 (a) T.Goldman et al, Few-Body Syst.{\bf 12}, 123 (1992);  
 (b) G.Krein et al, Phys.Lett.{\bf B317}, 293 (1993);
 (c) J.Piekarewicz and A.G.Williams, Phys.Rev.{\bf C47}, R2462 (1993);
 (d) T.Hatsuda et al, Phys.Rev.{\bf C49}, 452 (1994).
\bibitem{79}
 (a) R.Abegg et al (TRIUMF), Phys.Rev.{\bf D39}, 2464 (1989);
 (b) S.E.Vigdor et al (Saclay), Phys.Rev.{\bf C46}, 410(1992).
\bibitem{80}
 S.A.Coon and R.C.Barret, Phys.Rev.{\bf C36}, 2189 (1987).
\bibitem{81} 
 K. G. Wilson Phys. Rev.{\bf 179}, 1499 (1969).
\bibitem{82}
 B. L. Ioffe, Acta Phys. Polomica {\bf B16}, 543 (1985)
\bibitem{83} Some typical references are: \\
 B. S. Dewitt, Phys. Rev. {\bf 162}, 1195, 1239 (1967);
 J. Honerkamp, Nucl. Phys. {\bf B48}, 269 (1972);
 R. Kallosh, Nucl. Phys. {\bf B78}, 293, (1974);
 S. G. Matinyan and G. K. Savvidy, Nucl. Phys. {\bf B134}, 539 (1978);
 H. Leutwyler, Nucl. Phys. {\bf B179}, 129 (1981);
 L. F. Abbot, ibid, {\bf B185}, 189 (1982).
\bibitem{84} 
 J.D.Bjorken, Phys.Rev.{\bf 148}, 1467 (1966).
\bibitem{85}
 (a) V.M.Belyaev and Ya.I.Kogan, JETP Lett.{\bf 37}, 730 (1983); 
  C. B. Chiu et al, Phys.Rev.{\bf D32}, 1786 (1985). 
 (b) V.M.Belyaev, B.L.Ioffe and Ya.I.Kogan, Phys.Lett.{\bf 151B}, 290 (1985).
\bibitem{86}
 (a) R.H.Dalitz, Proc.{\bf XIII} Intl.Conf.on HEP, Berkeley, (1966);
 (b) O.W.Greenberg, Phys.Rev.Lett.{\bf 13}, 598 (1964)
 (c) A.N.Mitra, Nuovo Cim.{bf 56A}, 1164 (1968);
 (d) D.B.Lichtenberg, Phys.Rev.{\bf 178}, 2197 (1969).  
\bibitem{87}
 (a) L.D.Faddeev, Sov.Phys.JETP {\bf 12}, 1014 (1961);
 (b) A.N.Mitra, Nucl.Phys.{\bf 32}, 529 (1962);
 (d) C.Lovelace, Phys.Rev.{\bf B135}, 1225 (1964)
 (c) S.Weinberg, Phys.Rev.{\bf B133}, 232 (1964);
\bibitem{88}
 (a) G.Efimov et al, Phys.Rev.{\bf D51}, 176 (1995);
 (b) G.Efimov et al, Few-Body Syst.{\bf 6}, 17 (1989);
 (c) M.A.Ivanov et al, Few-Body Syst.{\bf 21}, 131 (1996);
\bibitem{89}
 S.Huang and J.Tjon, Phys.Rev.{\bf C49}, 1702 (1994); 
 N.Ishii et al, Austr J Phys {\bf 50}, 123 (1997);
 W.Bentz, J. Korean Phys. Soc.{\bf 29} Suppl, 5352 (1996). 
\bibitem{90}
 (a) A.N.Mitra and Anju Sharma, Fortschr.Phys.{\bf 45}, 411-434 (1997); 
 (b) J.T.Londergan and A.N.Mitra, Intl.J.Mod.Phys.{\bf A6}, 2659 (1991)
\bibitem{91}
 R.E.Cutkowsky et al, Phys.Rev.{\bf D20}, 2839 (1979);
\bibitem{92}
 See, e.g., J.M.Blatt and V.F.Weisskopf,{\it {Theoretical Nuclear Physics}},
 John Wiley and Sons Inc, New York, 1952; Chapter II.C, pp 71-80. 
\bibitem{93}
 W.Miller, {\it {Lie Theory and Special Functions}}, Acad.Press, N.Y. (1968);
 G.C.Ghirardi, Nuovo Cim.{\bf A10}, 97 (1972). 
\bibitem{94}
 (a) Yu.A.Simonov, Sov.J.Nucl.phys.{\bf 3}, 461 (1966);
 (b) G.Karl and E.Obryk, Nucl.Phys.{\bf B8}, 609 (1968) 
\bibitem{95}
 A.Sharma and A.N.Mitra, LANL {\bf hep-ph/9707503} July 1997;
\bibitem{96}
 I.Santhanam et al, Few-Body Syst.{\bf 7}, 1 (1990). 

\end{thebibliography}
\end{document}